\def\doctype{0}

\if\doctype0
  \documentclass{article}
  \usepackage{arxiv}
\fi
\if\doctype1 
    \documentclass[Crown,sageh,times]{sagej}
\fi
\usepackage{moreverb,url}
\usepackage[colorlinks,bookmarksopen,bookmarksnumbered,citecolor=red,urlcolor=red]{hyperref}
\usepackage{natbib}
\usepackage{subcaption}
\usepackage{xcolor}
\usepackage{parskip}
\usepackage{pdflscape}
\usepackage{soul}
\if\doctype0
    \usepackage{pdflscape} 
    \usepackage{amsmath}
    \usepackage{graphicx} 
    \usepackage{multirow}
    \usepackage{float}
    \usepackage{geometry}
\fi
\usepackage{cleveref}
\if\doctype0 
\fi
\if\doctype1 
    \newcommand\BibTeX{{\rmfamily B\kern-.05em \textsc{i\kern-.025em b}\kern-.08em
    T\kern-.1667em\lower.7ex\hbox{E}\kern-.125emX}}
\fi

\title{\Large{Data-driven Characterization of Near-Surface Velocity in the San Francisco Bay Area: A Stationary and Spatially Varying Approach}}

\if\doctype0
    \renewcommand{\shorttitle}{Data-driven Characterization of Near-Surface Velocity in SFBA} 
    \newcommand{\headeright}{}
    \newcommand{\undertitle}{Submitted to Earthquake Spectra}
    \hypersetup{
        pdftitle={Data-driven Characterization of Near-Surface Velocity in the San Francisco Bay Area: A Stationary and Spatially Varying Approach},
        pdfauthor={Lavrentiadis G., Seylabi E., Xia F., Tehrani H., Asimaki D. \& McCallen D.},
        pdfkeywords={Sedimentary Velocity Model, San Francisco Bay Area Velocity Model, Site Response Analysis, Wave Propagation, Seismic Hazard Analysis},
    }
\fi
\if\doctype1 
    \def\volumeyear{2024}
    \keywords{Sedimentary Velocity Model; San Francisco Bay Area Velocity Model; Site Response Analysis; Wave Propagation; Seismic Hazard Analysis}
\fi

\if\doctype0
    \author{ 
        \href{https://orcid.org/0000-0001-6546-1340}{\includegraphics[scale=0.06]{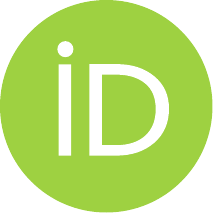}\hspace{1mm}Grigorios Lavrentiadis} \\
        Division of Engineering and Applied Sciences\\
        California Institute of Technology\\
        Pasadena, California, U.S.A. \\
    	\texttt{glavrent@caltech.edu} \\
    	\And
        \href{https://orcid.org/0000-0003-0718-372X}{\includegraphics[scale=0.06]{orcid.pdf}\hspace{1mm}Elnaz Seylabi} \\
        Department of Civil \& Environmental Engineering\\
        University of Nevada, Reno\\
        Reno, Nevada, U.S.A. \\
    	\texttt{elnaze@unr.edu} \\
    	\And
        \href{https://orcid.org/0009-0002-0905-2387}{\includegraphics[scale=0.06]{orcid.pdf}\hspace{1mm}Feiruo Xia} \\
        Division of Engineering and Applied Sciences\\
        California Institute of Technology\\
        Pasadena, California, U.S.A. \\
    	\texttt{fxia@caltech.edu} \\
    	\And
        Hesam Tehrani \\
        Department of Civil \& Environmental Engineering\\
        University of Nevada, Reno\\
        Reno, Nevada, U.S.A. \\
    	\texttt{hsalmanitehrani@unr.edu} \\
    	\And
        \href{https://orcid.org/0000-0002-3008-8088}{\includegraphics[scale=0.06]{orcid.pdf}\hspace{1mm}Domniki Asimaki} \\
        Division of Engineering and Applied Sciences\\
        California Institute of Technology\\
        Pasadena, California, U.S.A. \\
    	\texttt{domniki@caltech.edu} \\ 
    	\And
        \href{https://orcid.org/0000-0003-0386-0324}{\includegraphics[scale=0.06]{orcid.pdf}\hspace{1mm}David McCallen} \\
        Energy Geosciences Division\\
        Lawrence Berkeley National Laboratory\\
        Berkeley, California U.S.A. \\
    	\texttt{dbmccallen@lbl.gov} \\
    }
\fi
\if\doctype1  
    \author{Grigorios Lavrentiadis\affilnum{1}, 
    Elnaz Seylabi\affilnum{2},
    Feiruo Xia\affilnum{1}, 
    Hesam Tehrani\affilnum{2}, 
    Domniki Asimaki\affilnum{1}, 
    and David McCallen\affilnum{3}}
    
    \affiliation{\affilnum{1}California Institute of Technology, Pasadena, California\\
    \affilnum{2}University of Nevada, Reno, Reno, Nevada \\
    \affilnum{3}Lawrence Berkeley National Laboratory, Berkeley, California}
    
    \corrauth{Elnaz Seylabi, University of Nevada, Reno, NV, United States 89557}
    \email{elnaze@unr.edu}
    
    \runninghead{Lavrentiadis et al.}
\fi

\begin{document}
\if\doctype0 \maketitle \fi

\begin{abstract}

This study presents the development of two new sedimentary velocity models for the San Francisco Bay Area (SFBA) to improve the near-surface representation of shear-wave velocity ($V_S$) for large-scale, broadband numerical simulations, with the ultimate goal of enhancing the representation of the sedimentary layers in the Bay Area community velocity model. The first velocity model is stationary and is based solely on $V_{S30}$; the second velocity model is spatially varying and has location-specific adjustments. They were developed using a dataset of 200 measured $V_S$ profiles. Both models were formulated within a hierarchical Bayesian framework, using a parameterization that ensures robust scaling. The spatially varying model includes a slope adjustment term modeled as a Gaussian process to capture site-specific effects based on location. 
Residual analysis shows that both models are unbiased for $V_S$ values up to 1000 m/sec. Along-depth variability models were also developed using within-profile residuals. The proposed models show higher $V_S$ in the San Jose area and Livermore Valley compared to the USGS Bay Area community velocity model by a factor of two or more in some cases. Goodness-of-fit (GOF) comparisons using one-dimensional linear site-response analysis at selected sites demonstrate that the proposed models outperform the USGS model in capturing near-surface amplification across a broad frequency range. Incorporating along-depth variability further improves the GOF scores by reducing over-amplification at high frequencies. 
These results underscore the importance of integrating data-driven models of the shallow crust, like the ones presented here, in coarser regional community velocity models to enhance regional seismic hazard assessments.

\end{abstract}

\if\doctype1 \maketitle \fi

\section{Introduction}

Three-dimensional earthquake ground motion simulations were long constrained to low frequencies due to computational cost \citep{olsen2006strong,graves2010broadband,bielak2010shakeout}; over the last decade, however, the continuous growth of high-performance computing systems has made higher frequency simulations an increasingly realistic target \citep{taborda2013ground,taborda2014ground,bielak2016verification,pinilla2024,mccallen2024regional}.
This fact has motivated researchers to refine existing source, geology, and velocity models to capture source, path, and site effects of deterministic broadband ground motion simulations over a finer resolution \citep[e.g.,][]{shi2013rupture,taborda2012earthquake,roten2014expected,elnaz,savran2016model}. 
In particular, the literature suggests that small-scale variations in the shallow velocity structure causing localized scattering and nonlinear sediment behavior may significantly affect the amplitude and frequency content of ground motions, especially in the frequency range of engineering interest. This emphasizes the need for a more accurate representation of the shallow crustal layers in large-scale numerical models. 

Weathered rocks and sedimentary deposits in the shallow crust and their transition to the stiffer bedrock are frequently represented by smooth functions of shear wave velocity with depth \citep[e.g.,][]{small2017scec}, which are typically determined independently from the deeper velocity structure, and are sometimes based on shallow geotechnical and geophysical investigations. 
\citet{ely2010VS30} tackled this issue by proposing a smooth geometric function, the so-called geotechnical layer, parameterized by the time-average shear wave velocity at the top 30m ($V_{S30}$), to represent the shear-wave velocity profile ($V_S$) at the top 350 m. 
\citet{Shi2018} used a set of nearly a thousand profiles in California to develop a sediment velocity model (SVM) up to a depth-to-shear-wave-velocity of 1 km/sec ($Z_{1.0}$) using an analytic expression which is a function of $V_{S30}$. Among other differences from \citet{ely2010VS30}, the \citet{Shi2018} model and its variants did not predefine a fixed depth of convergence between sediments and basement rock, allowing for impedance contrasts that respect the basement rock geometry and enabling the simulation of realistic basin edge effects, as demonstrated by \cite{taborda2016}. More recently, \citet{marafi2021generic} proposed a velocity model for the Cascadia by using 218 profiles and modifying the model by \citet{Shi2018} to incorporate both $V_{S30}$ and $Z_{1.0}$ as input parameters to represent better the deeper velocity structure for the region of interest. 

The current USGS 3-D Geologic and Seismic Velocity Models of the San Francisco Bay Area (SFBA) region comprises a high-resolution model of 290 km by 140 km by 45 km volume surrounded by a coarser grid spanning 650 km by 330 km by 45 km volume. The finest grid size in the detailed seismic velocity model is 100 m in the horizontal direction and 25 m in the vertical direction \citep{scienceplan}, which inevitably limits the capabilities of earthquake simulations to capture site effects. The initial version (SFVM v.05.0.0) was designed to simulate wave propagation for the 1906 (Mw=7.9) San Francisco and the 1989 (Mw=6.9) Loma Prieta earthquakes, as well as a series of hypothetical (Mw=7.9) earthquake scenarios on the San Andreas fault \citep{aagaard2008ground_a, aagaard2008ground_b}. Subsequent evaluations revealed systematic biases in the synthetic waveforms produced by the model, which were addressed through updates to the SFVM v.08.3.0 to enhance accuracy and consistency. The updated version has since been used in various studies focusing on the effects that earthquake source processes, coupled with 3D path and site effects, have on the spatial variability of ground motions \citep{hirakawa2022evaluation}. The last version of SFVM is version v.21.1, released recently based on the final modifications on v.08.03. This final version includes adjustments made to improve the accuracy of ground-motion predictions in the SFBA region. These adjustments were made by analyzing geologic structures and experimenting with synthetic motions to match observed ones \citep{hirakawa2022evaluation}.

In this study, we present two new sedimentary velocity models intended to refine the representation of the shallow crustal velocity structure in the USGS velocity model for the Bay Area. Our models use measured Vs profiles in the region to reduce biases in the upper few hundred meters.
The first model is stationary, namely providing the median shear-wave velocity profile as a function of $V_{S30}$ independent of location. The second model is spatially varying, namely accounts for the profile's location in addition to the site's $V_{S30}$.
Compared to state-of-the-art sediment velocity models: (i) our models are explicitly constrained based on  $V_{S30}$ and are based on robust scaling relationships that improve model performance, particularly outside the range of available data, and (ii) our spatially varying model accounts for location-based systematic differences in velocity scaling.

The remainder of the paper is organized as follows:
\hyperref[sec:data]{Section: Velocity Profile Datasets} provides an overview of the $V_{S}$ profiles in the Bay Area region that were used in the model development.  
\hyperref[sec:model_dev]{Section: Model Development} describes the formulation and statistical framework employed for both the stationary and spatially varying velocity models.  
\hyperref[sec:results]{Section: Results} presents the trained models, evaluates their data fit, and addresses issues pertinent to the interface between our SVMs and the USGS SFBA (i.e. where does the SVM stop and the USGS SFBA start) as well as their along-depth variability.  
\hyperref[sec:evaluation]{Section: Evaluation} explores the implementation of the proposed models to enhance the near-surface velocity representation of the USGS SFBA velocity model and evaluates their relative ability to capture site amplification under the assumption of 1D site response.  
\hyperref[sec:conclusions]{Section: Conclusions} highlights the key findings of this study. We should also note here that this study is a continuation and, as such, supersedes the sedimentary velocity model presented in \cite{tehrani2023}.

\section{Velocity Profile Datasets} \label{sec:data}

\begin{figure} 
    \centering
    \includegraphics[trim = 0in .5in 0in 0.5in,clip,width= 0.6\textwidth]{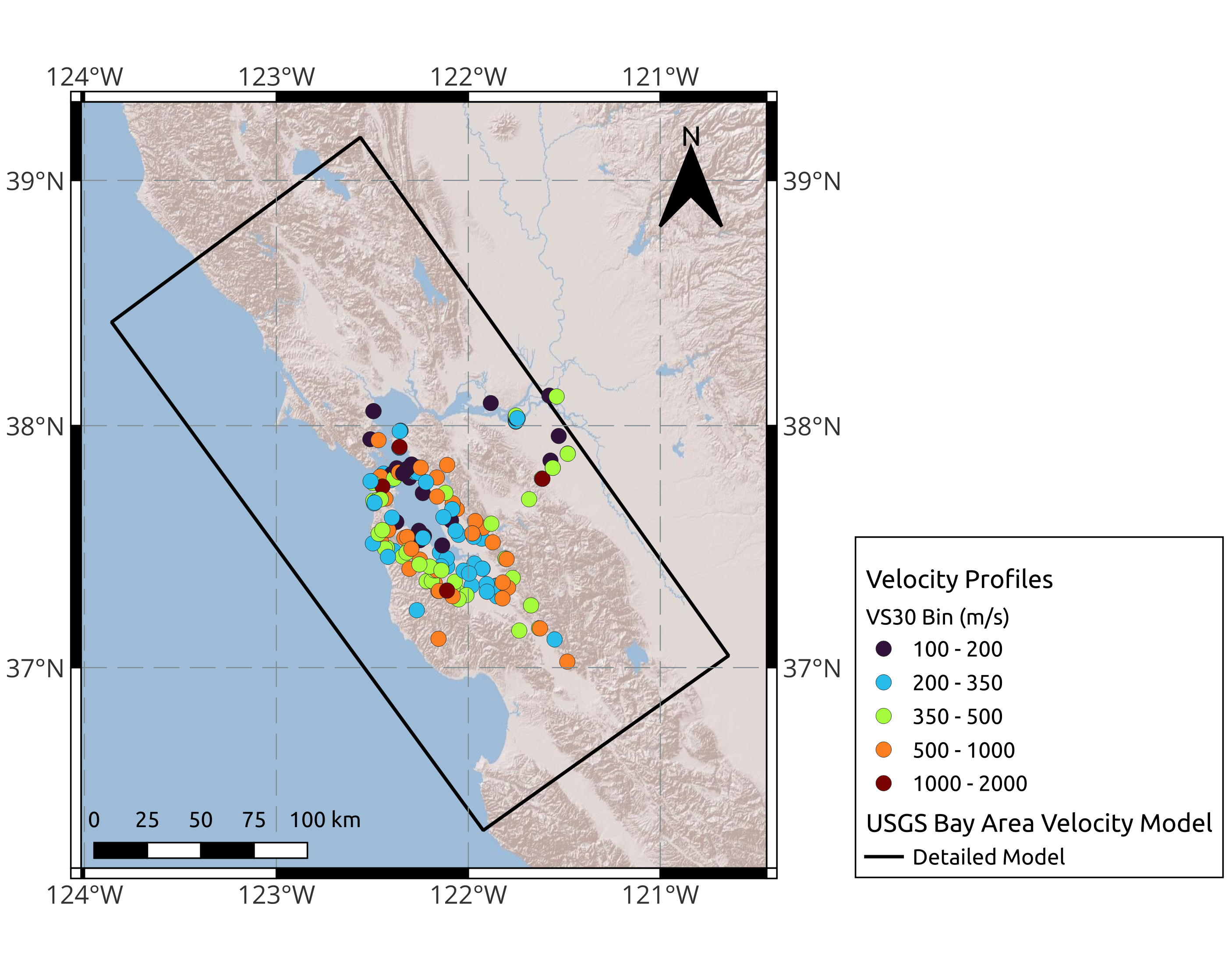}
    \caption{Locations of the compiled profiles used to develop the proposed $V_S$ models.}
    \label{fig:location_profiles}
\end{figure}

The Vs profiles were compiled from five sets of profiles within the SFBA, including 13 profiles from \citet{yong2013arra}, 92 profiles from \citet{boore2003compendium}, 29 profiles collected by California Geological Survey, three profiles measured by LeRoy Crandall and Associates, and 63 profiles obtained from the web portal shear-wave velocity profile database (VSPDB) \citep{kwak2021web}. 
The composite dataset comprises 200 profiles with locations shown in \Cref{fig:location_profiles}. The profiles from \citet{yong2013arra} were derived using noninvasive methods, those from \citet{kwak2021web} were based on both invasive and noninvasive methods, and the rest were obtained using invasive techniques \citep[e.g., suspension logging, cross-hole, and down-hole tests,][]{Shi2018}. Previous studies have indicated that noninvasive shear-wave velocity profiling techniques can produce similar results to invasive techniques \citep{boore1998comparing, boore2008comparisons, brown2002comparison,rix2002near, stephenson2005blind, bas2022p}. Therefore, all profiles are used to develop the proposed models.

\begin{figure} 
    \centering
    \includegraphics[width= 0.5\textwidth]{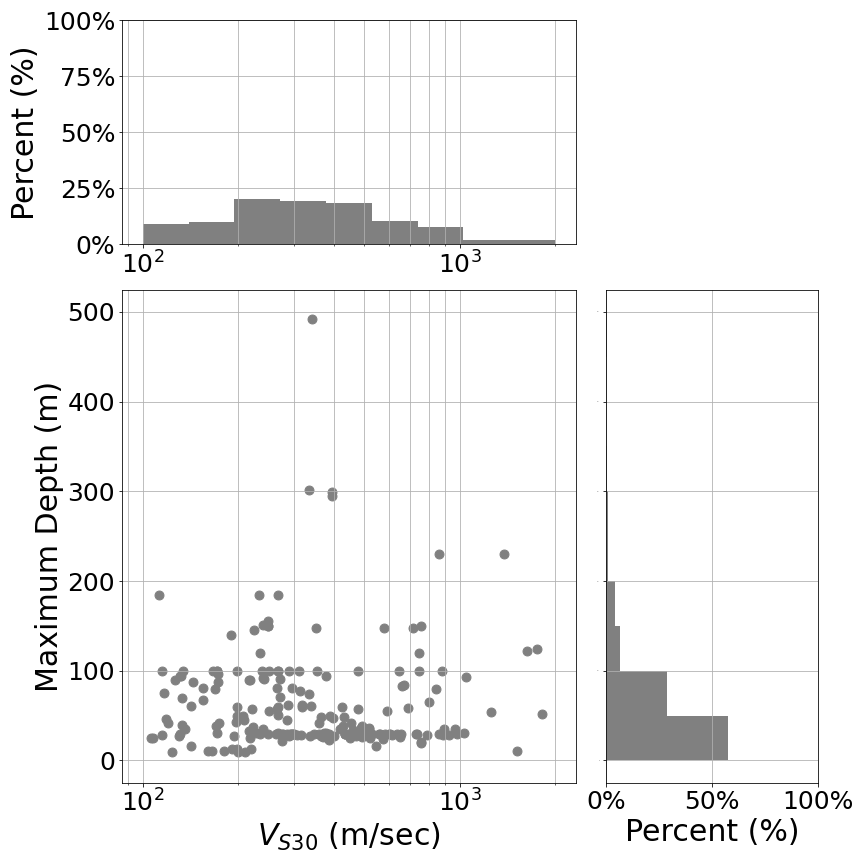}
    \caption{Distributions of $V_{S30}$ and maximum depth of the profiles used in the model development.}
    \label{fig:data_distribution}
\end{figure}

The $V_{S30}$ and maximum depth distribution of the compiled profiles are shown in \Cref{fig:data_distribution}.
Most profiles span a $V_{S30}$ range between 140 and 740 m/sec, with the minimum $V_{S30}$ being 105 m/sec and the maximum being 1825 m/sec.
The maximum depth of most profiles is less than 100 m, with 60\% being less than 50 m deep and 30\% being between 50 and 100 m deep.

\section{Model Development}\label{sec:model_dev}

This section outlines the parameterization of the median velocity profile, followed by a discussion of the regression methods used for the stationary and spatially varying models.

\subsection{Parametrization}

The analytical expression for the median velocity profile is based on \citet{Shi2018} (Equation \ref{eq:velocity_model}). The selected relationship maintains a constant shear-wave velocity $V_{S0}$ down to a depth $z^*$ of $2.5\,\mathrm{m}$, and it follows a power function below that depth. The parameter $k$ controls the slope, while the parameter $n$ dictates the curvature of the velocity profiles. For $n=1$, the profile is linear; for $n>1$, it is convex; and for $n<1$, it is concave.

\begin{align} \label{eq:velocity_model}
    \overline{V}_S(z) = 
    \begin{cases}
        V_{S0}                      & \textrm{for} ~ z \leq z^* \\   
        V_{S0} (1+k(z-z^*))^{1/n}   & \textrm{for} ~ z > z^*
    \end{cases}
\end{align}

An added feature of our model, compared to previous studies, is that $V_{S0}$ is evaluated explicitly, constraining the time-average shear wave velocity of the top 30 meters of the median profile to match the target $V_{S30}$:
\begin{equation} \label{eq:vs0_con}
    \frac{30}{\int_{0}^{30} \frac{1}{\overline{V}_{S}(z)}\text{d}z } = V_{S30}
\end{equation}
By integrating this equation analytically and solving for $V_{S0}$, we obtain:
\begin{align} \label{eq:vs0}
    V_{S0} = V_{S30}
    \begin{cases}
        \frac{ \left( 1 + (30 - z^*) k \right)^{1/n} + 2.5 k \left( 1 - \frac{1}{n} \right) - 1 }{ z^* k \left( 1 - \frac{1}{n} \right) } & \textrm{for} ~ n \neq 1 \\
        \frac{ z^* + \frac{1}{k} \ln( 1 + k (30 - z^*)) }{30} & \textrm{for} ~ n = 1
    \end{cases}
\end{align}
This constraint ensures that the $V_{S30}$ of the generated profiles matches the input $V_{S30}$ exactly while reducing the number of scaling relationships that need to be determined, resulting in a more stable regression.

The scaling parameters $n$ and $k$ are next modeled as functions of $V_{S30}$. The parameter $n$ is represented by a sigmoid function ($S(x) = 1/(1+\exp(-x))$) that asymptotically approaches unity for low $V_{S30}$, and $s_2$ for high $V_{S30}$ (Equation \ref{eq:n_scl}). On the other hand, $k$ uses a combination of a sigmoid and hinge function ($H(x) = x \ln(1 + \exp(x))$), which converges to $\exp(r_1)$ for low $V_{S30}$, and increases with an $r_3$ slope in log-log space for high $V_{S30}$ (Equation \ref{eq:k_scl}).
These relationships were based on trends observed in $k$ and $n$ in \citet{Shi2018, marafi2021generic} (Figure S1 in the electronic supplement). The sigmoid and hinge functions were preferred over the high-order polynomials used in previous studies because they achieve more robust scaling near the data limits, eliminating the large oscillations typically caused by polynomial fits.
To minimize trade-offs, $k$ and $n$ were constrained to share the same sigmoid scaling ($V_{S30scl} = (\ln(V_{S30}) - V_{S30ref} ) / V_{S30w}$), with $V_{S30ref}$ controlling the midpoint and $V_{S30w}$ the width of the transition between low and high $V_{S30}$. The common transition parameters ensure that $V_{S0}$ follows a monotonically increasing scaling with $V_{S30}$.

\begin{equation} \label{eq:n_scl}
    n = 1 + s_2 ~ S( V_{S30scl} )
\end{equation}

\begin{equation} \label{eq:k_scl}
    k = \exp\left[ r_1 + r_2  S( V_{S30scl} ) + r_3 V_{S30~w} H( V_{S30scl} ) \right];
\end{equation}

Both the stationary and spatially varying velocity models were formulated as hierarchical Bayesian models. In Bayesian statistics, the uncertainty of the model parameters ($\mathbf{\theta}$) before observing the data is expressed by the prior distribution ($p(\mathbf{\theta})$). The uncertainty of the model parameters informed by the available data is quantified by the posterior distribution ($p(\mathbf{\theta}|y,\mathbf{x})$), where $y$ is the response variable (here, $\ln(V_S)$) and $\mathbf{x}$ are the conditional variables (here, $z$ and $V_{S30}$). The influence of the data is captured by the likelihood function, which is defined by the probability density function of the velocity model:
\begin{equation}
    \mathcal{L}(\mathbf{\theta}) = pdf(y|f(\mathbf{x},\mathbf{\theta}),\sigma^2)
\end{equation}
where $f(\mathbf{x},\mathbf{\theta})$ represents the median velocity profile in log space, and $\sigma$ is the standard deviation of the model residuals.

Bayes's rule provides the means for calculating the posterior distribution based on the prior distribution and likelihood function as:
\begin{equation}
    p(\mathbf{\theta}|y,\mathbf{x}) \propto \mathcal{L}(\mathbf{\theta}) p(\mathbf{\theta})
\end{equation}
which we evaluated numerically using the computer software STAN \citep{Stan2023}.

Based on preliminary exploratory analyses, the residuals between the measured and median velocity profiles for both models were defined in log space, assuming they follow a normal distribution:
\begin{equation}
    \epsilon = \ln(V_S) - \ln(\overline{V}_S)
\end{equation}
\begin{equation}
    \epsilon \sim \mathcal{N}(0, \sigma)
\end{equation}
where $V_S$ are the shear wave velocity observations, $\overline{V}_S$ represents the median profile values at the same depths, and $\sigma$ is the standard deviation in log space. The assumption that the residuals are log-normally distributed is also supported by previous studies, including \citet{Shi2018} and \citet{marafi2021generic}.

The following sections detail the development of the two models. \hyperref[sec:st_model_dev]{Subsection: Stationary Model} outlines the development of the stationary model, while \hyperref[sec:sv_model_dev]{Subsection: Spatially Varying Model} describes the development of the spatially varying model.

\subsection{Stationary Model} \label{sec:st_model_dev}

The stationary velocity model is formulated as follows:
\begin{equation}
    y_{ij} = f_{stat}(\mathbf{x}_{ij}, \mathbf{\theta}_{i}) + \epsilon_{ij}
\end{equation}
where $y_{ij}$ is the $j^{th}$ velocity observation, in log space, of the $i^{th}$ profile. The term $f_{stat}$ represents the median prediction of the stationary model, and $\epsilon_{ij}$ is the model's misfit. 
The input parameters for $f_{stat}$ include the depth to the middle of the layer ($\mathbf{x}_{ij}$) and the array of conditional variables for the profile in question, which for the stationary model is solely $V_{S30}$.

Next, the prior distributions in the regression of the stationary model are defined.

The midpoint of the $V_{S30}$ scaling, $V_{S30ref}$, is modeled with a normal distribution that has a mean of $5.7$ and standard deviation of $0.1$:
\begin{equation}
    V_{S30ref} \sim \mathcal{N}(5.7, 0.1)    
\end{equation}
In linear space, the median value of the midpoint is $300$ m/sec while the 16th and 84th percentiles are $270$ and $330$, respectively. 

The width of the transition was modeled using a gamma distribution with shape and scale parameters of $2.0$ and $0.5$, respectively:
\begin{equation}
    V_{S30w} \sim Gamma(2.0, 0.5)    
\end{equation}
A gamma distribution was chosen as the prior because its support is on the positive side of the real line, ensuring that $V_{S30w}$ remains physically meaningful. The prior mean is one, with the 16th to 84th percentile ranging from $0.35$ to $1.68$.

The prior distribution for the change in curvature as a function $Vs30$ is a log-normal distribution with a mean of $2.0$ in logspace and a standard deviation of $0.3$:
\begin{equation}
    s_2 \sim \mathcal{LN}(2.0, 0.3)
\end{equation}
A negative $s_2$ would imply a concave median profile as $V_{S30}$ increases (i.e., as velocity profiles become stiffer), which was deemed unphysical based on simple overburden pressure effects on material stiffness and, by extension, shear wave velocity. 
The log-normal distribution has a strictly positive support, resulting in median $V_S$ profiles that are convex for the entire $V_{S30}$ range.
The median, 16th, and 84th percentiles for this prior distribution are $7.40$, $5.48$, and $9.96$, respectively.

An uninformative prior is used for $r_{1stat}$, which controls the slope of $V_S$ at low $V_{S30}$. The coefficient $r_{1stat}$ corresponds to $r_1$ in Equation \ref{eq:k_scl} for the stationary model. It is modeled using a normal distribution with a mean of zero and a standard deviation of $5.0$:
\begin{equation}
    r_{1stat} \sim \mathcal{N}(0.0, 5.0)
\end{equation}
A zero value for $r_{1stat}$ represents a unit slope for $V_S$ with depth (i.e., $\exp(r_{1stat})$), while the 16th and 84th percentiles correspond to slopes of 0.001 and 150, respectively, covering a broad range of plausible values.

The rate of slope change for profiles with intermediate $V_{S30}$ is controlled by $r_{2stat}$, which follows a log-normal distribution with a log-mean of $0.5$ and a standard deviation of $0.5$:
\begin{equation}
    r_{2stat} \sim \mathcal{LN}(0.5, 0.5)
\end{equation}
Similar to $r_{1stat}$, $r_{2stat}$ corresponds to $r_2$ in Equation \ref{eq:k_scl} for the stationary model. A prior with positive support ensures an increasing $V_S$ slope for higher $V_{S30}$, consistent with findings from previous studies \citep{asimaki2014, marafi2021generic}.

The coefficient $r_3$ controls the rate of change at high $V_{S30}$. An exponential prior distribution is used to penalize unnecessary model complexity \citep{Simpson2017, Lavrentiadis2021}:
\begin{equation}
    r_3 \sim Exp(2.0)
\end{equation}
The exponential distribution has most of its mass near zero, ensuring a minimal rate of slope change unless there is significant evidence in the empirical data supporting a more complex model.

The standard deviation for the misfit is modeled with a log-normal distribution, with a log-mean and standard deviation of $-1.0$ and $0.6$, respectively:
\begin{equation}
    \sigma \sim \mathcal{LN}(-1.,0.6)
\end{equation}
The 16th to 84th percentile spans the range from $0.2$ to $0.65$, covering the range of standard deviations of previous models. 

\subsection{Spatially Varying Model} \label{sec:sv_model_dev}

One limitation of the stationary model is that the standard deviation may be overestimated due to the incorporation of any systematic differences between velocity profiles with similar $V_{S30}$. 
The spatially varying model aims to address this by separating the within-profile variability from the systematic effects. It accomplishes this task by including a spatially varying slope adjustment ($\delta Br$), which is a function of the profile's location (Equation \ref{eq:vs_svar}).
No spatially varying adjustment was included in the curvature of the profiles, $n$, due to the small dataset size, which does not allow for the differentiation of such an effect from a slope adjustment.
Additionally, no spatially varying adjustment was included in $V_{S0}$ as it is fully determined based on $V_{S30}$, $k$, and $n$ parameters (Equation \ref{eq:vs0_con}). 

\begin{equation} \label{eq:vs_svar}
    y_{ij} = f_{svar}(\mathbf{x}_{ij},\mathbf{\theta}_{i}, \delta B r_{i}) + \epsilon_{ij}  
\end{equation}

The mean $k - V_{S30}$ scaling is adjusted from the stationary model to account for the systematic effects (Equation \ref{eq:r_adj}). Specifically, $\delta r_{1svar}$ modifies the slope of the $V_S$ profiles at low $V_{S30}$ relative to the stationary model, while $\delta r_{2svar}$ adjusts the slope of the $V_S$ profiles at intermediate to high $V_{S30}$.

\begin{align} \label{eq:r_adj}
\begin{split}
    r_{1svar} &= r_{1stat} + \delta r_{1svar} \\   
    r_{2svar} &= r_{2stat} + \delta r_{2svar}
\end{split}
\end{align}

The updated $k - V_{S30}$ scaling, which includes both the mean and profile-specific adjustments, is presented in Equation \ref{eq:k_scl_svar}. The scaling terms for low and intermediate $V_{S30}$, $r_1$ and $r_2$, correspond to the adjusted values $r_{1svar}$ and $r_{2svar}$, respectively, while the scaling term $r_3$, for high $V_{S30}$, remains fixed at its stationary model value. A new $r_3$ term was not estimated for the spatially varying model due to the limited number of stiff profiles in the dataset.

\begin{equation} \label{eq:k_scl_svar}
    k = \exp\left[ r_1 + r_2  S( V_{S30scl} ) + r_3 V_{S30~w} H( V_{S30scl} ) + \delta B_r \right]
\end{equation}

The $\delta r_{1svar}$, and $\delta r_{2svar}$ adjustments are modeled with normal priors with zero mean and $0.2$ standard deviation (Equation \ref{eq:dr_prior}). 
The zero mean ensures that the $k - V_{S30}$ scaling remains the same as the stationary model unless there is significant evidence from the data to suggest otherwise.

\begin{align} \label{eq:dr_prior}
\begin{split}
    \delta r_{1svar} &\sim \mathcal{N}(0.0, 0.2) \\   
    \delta r_{2svar} &\sim \mathcal{N}(0.0, 0.2) 
\end{split}
\end{align}

The spatially varying term, $\delta B_r$, is modeled as a Gaussian Process with a zero mean and a kernel function, $\kappa_{\delta B_r}(\vec{t}_{s},\vec{t}^\prime_{s})$. The zero mean ensures that, away from the profiles in the training dataset, the slope of the generated profiles remains consistent with the average slope for the given $V_{S30}$. The kernel function encodes the site-specific information, allowing it to learn the length scale and the magnitude of the spatial variability in the slope of the $V_S$ profiles.
\begin{equation}
    \delta B_r \sim \mathcal{GP}\left(0, \kappa_{\delta B_r}(\vec{x}_{s},\vec{x}_{s^\prime}) \right)
\end{equation}
The kernel is modeled with a negative exponential kernel function:
\begin{equation}
    \kappa_{\delta B_r}(\vec{x}_{s},\vec{x}_{s^\prime}) = \omega_{\delta Br}^2 \exp\left( \frac{ |\vec{x}_{s} - \vec{x}_{s^\prime}| }{ \ell_{\delta Br} } \right)
\end{equation}
where $\omega_{\delta Br}$ controls the size, $\ell_{\delta Br}$ controls the length scale of the spatial variability, and $|\vec{x}_{s} - \vec{x}_{s^\prime}|$ is the euclidean distance between $s$ and $s^\prime$ profiles.

The length scale of variability is modeled using an inverse gamma prior distribution with shape and scale parameters of $2.0$ and $50.0$, respectively:
\begin{equation}
    \ell_{\delta B_r} \sim InvGamma(2.0, 50.0)
\end{equation}
The inverse gamma distribution has positive support, ensuring the length scale remains physically meaningful. The 16th to 84th percentile spans from $15$ to $70$ km.

The scale of the spatial variability is modeled using a centered half-normal distribution with a standard deviation of $0.02$:
\begin{equation}
    \omega_{\delta Br} \sim \mathcal{T}(0,)\mathcal{N}(0.0, 0.02)
\end{equation}
The zero lower truncation limit ensures the standard deviation remains positive, while the zero mean and small standard deviation concentrate most of the prior distribution's mass near zero. This allows for spatially varying curvature only if the data provide significant support.

\section{Regression Results} \label{sec:results}

This section presents the results of the stationary and spatially varying models, followed by the along-depth correlation model and the SVM termination criteria.

\subsection{Stationary Model}

Table \ref{tab:st_coeffs} summarizes the estimated coefficients of the stationary model. Coefficients $V_{S30ref}$, $V_{S30w}$, $r_1$, and $\sigma$ are estimated with the smallest uncertainty, followed by $r_2$ and $s_2$, which have wider posterior distributions. The coefficient $r_3$ has the largest uncertainty. The terms $r_2$, $r_3$, and $s_2$ control the slope and curvature scaling for stiff profiles (i.e., large $V_{S30}$), where the training dataset is scant. This suggests that future data collection efforts focused on stiff sites will have the greatest impact on refining the $V_S(z)$ model scaling. 
Figure S2 in the electronic supplement displays the trace plots and full posterior distributions of the Markov chains, indicating good model convergence.

\begin{table}[htbp!]
\small
\caption{Posterior distributions of stationary model coefficients}
    \centering
    \begin{tabular}{|l||c|c|c|c|c|c|c|}
        \hline
         Coefficient          & $V_{S30ref}$ & $V_{S30w}$ & $r_1$ & $r_2$ & $r_3$ & $s_2$ & $\sigma$ \\
         \hline
         Mean                 & 6.5045 & 0.4368 & -2.2960 & 5.4669 & 0.4236 & 7.1685 & 0.3759 \\
         Median               & 6.4990 & 0.4354 & -2.2986 & 5.3966 & 0.3886 & 7.0741 & 0.3759 \\
         $5^{th}$ Percentile  & 6.3505 & 0.3866 & -2.4135 & 3.9247 & 0.0335 & 5.7509 & 0.3686 \\
         $95^{th}$ Percentile & 6.6780 & 0.4916 & -2.1700 & 7.2551 & 0.9193 & 8.9065 & 0.3834 \\ \hline
    \end{tabular}
    \label{tab:st_coeffs}
\end{table}

\begin{figure}[htbp!]
    \centering
    \begin{subfigure}[t]{0.32\textwidth}
        \caption{} \label{fig:st_model_scl_k}
        \includegraphics[height = 0.95\textwidth]{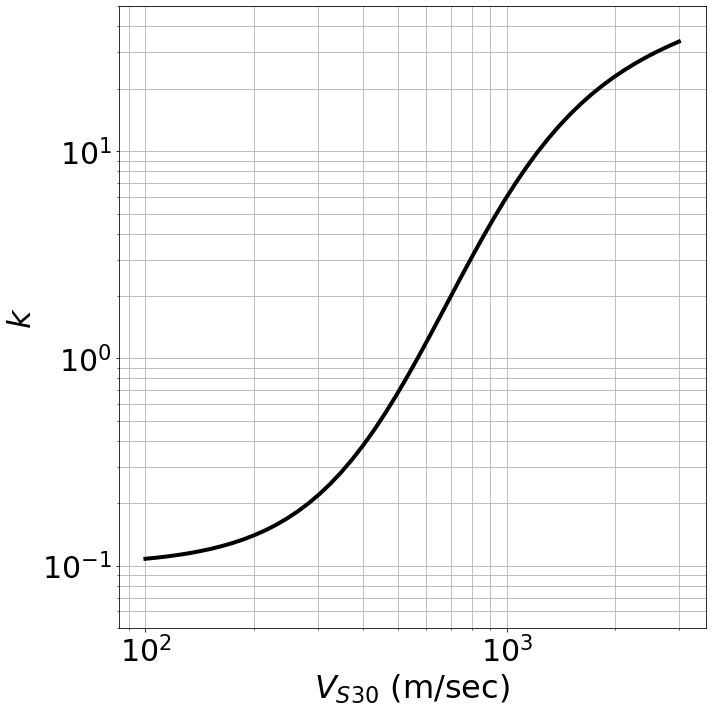}
    \end{subfigure} 
    \begin{subfigure}[t]{0.32\textwidth}
        \caption{} \label{fig:st_model_scl_n}
        \includegraphics[height = 0.95\textwidth]{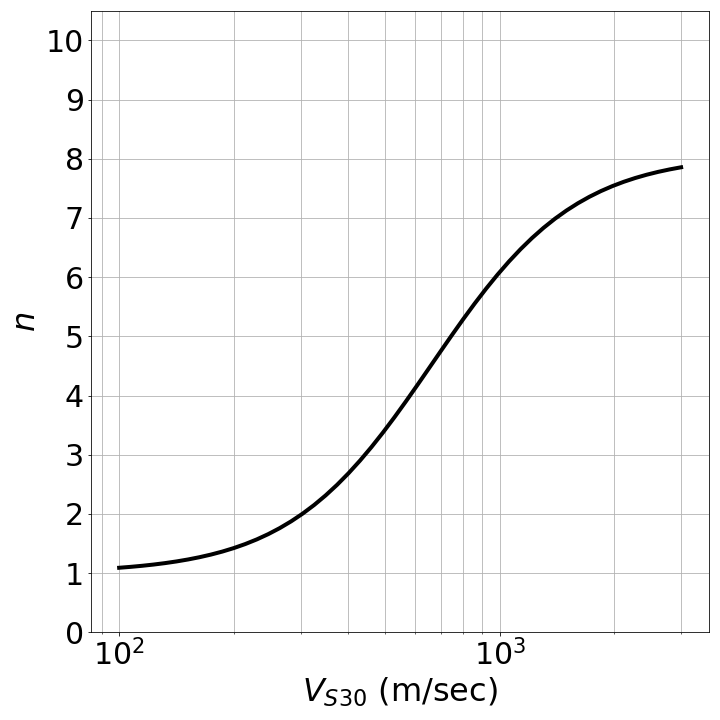}
    \end{subfigure}  
    \begin{subfigure}[t]{0.32\textwidth}
        \caption{} \label{fig:st_model_scl_vs0}
        \includegraphics[height = 0.95\textwidth]{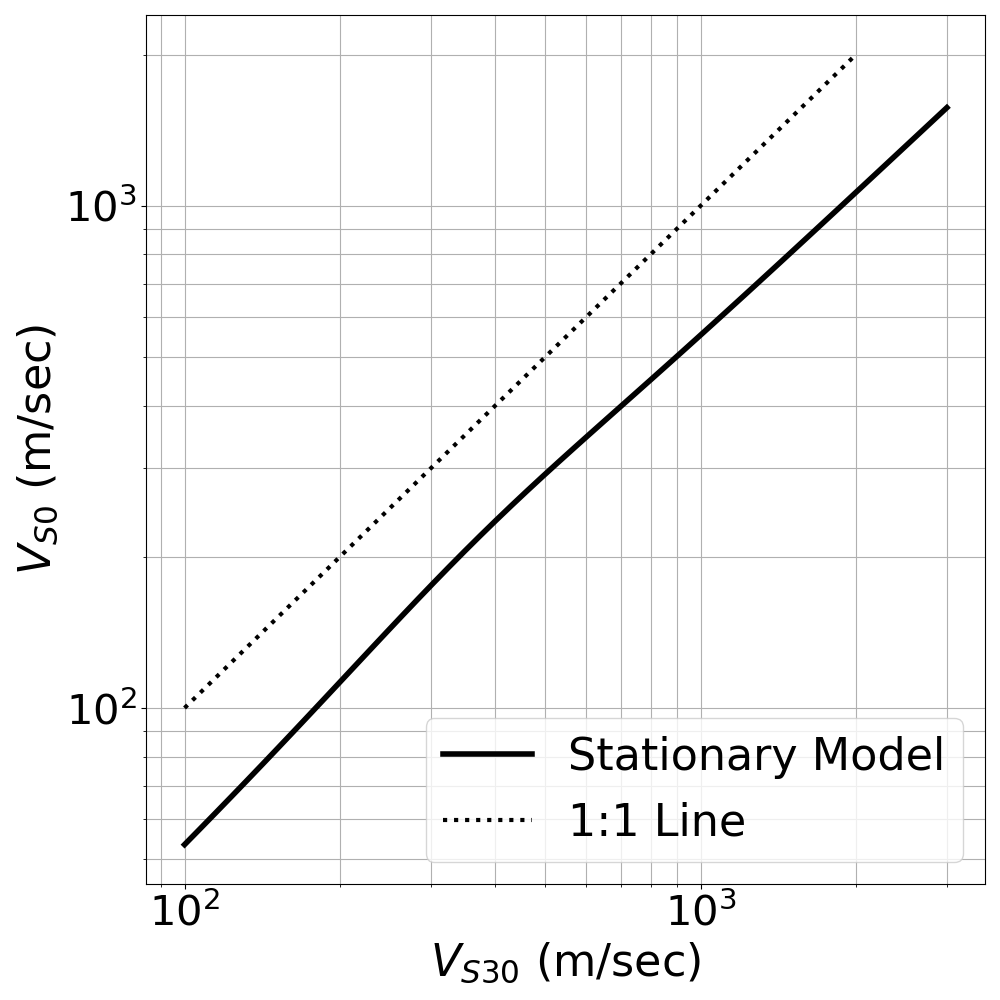}
    \end{subfigure}  
    \caption{Scaling relationships for the stationary model; subfigure (a) $V_{S30}$ versus $k$ scaling, subfigure (b) $V_{S30}$ versus $n$ scaling, and subfigure (c) $V_{S30}$ versus $V_{S0}$ scaling.}
    \label{fig:st_model_scl}
\end{figure}

The scaling relationships for $k$, $n$, and $V_{S0}$ of the stationary model are presented in Figure \ref{fig:st_model_scl}. Subfigure \ref{fig:st_model_scl_k} shows the slope scaling as a function of $V_{S30}$. For soft profiles ($V_{S30} < 200$ m/sec), the median profile converges asymptotically to a $0.1$ slope. As $V_{S30}$ increases, the median profiles become steeper. The rate of slope change is highest for intermediate $V_{S30}$ values ($300 < V_{S30} < 800$), while for $V_{S30}$ above $1000$ m/sec, the rate of change decreases.
Subfigure \ref{fig:st_model_scl_n} presents the curvature scaling as a function of $V_{S30}$ where similar trends to $k-V_{S30}$ scaling are observed. At low $V_{S30}$ values, the profiles have little curvature, $n$ asymptotically reaches unity, while the curvature of the median profiles increases for larger $V_{S30}$. 
For very stiff profiles, the curvature of the profiles asymptotically converges to $8$.
The $V_{S0} - V_{S30}$ scaling as defined by the $k$ and $n$ scaling relationships is presented in Subfigure \ref{fig:st_model_scl_vs0}. 
The derived relationship follows a marginally concave trend in log-log space and is approximately a factor of two below the one-to-one line. 

\begin{figure}
    \centering
    \begin{subfigure}[t]{0.32\textwidth}
        \caption{}  \label{fig:st_model_res_depth}
        \includegraphics[height = 0.95\textwidth]{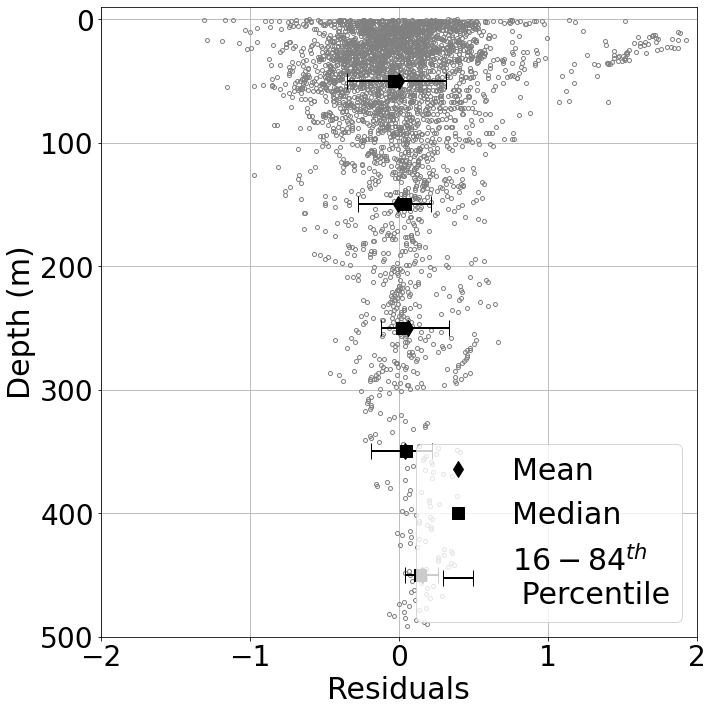}
    \end{subfigure} 
    \begin{subfigure}[t]{0.32\textwidth}
        \caption{} \label{fig:st_model_res_vs30}
        \includegraphics[height = 0.95\textwidth]{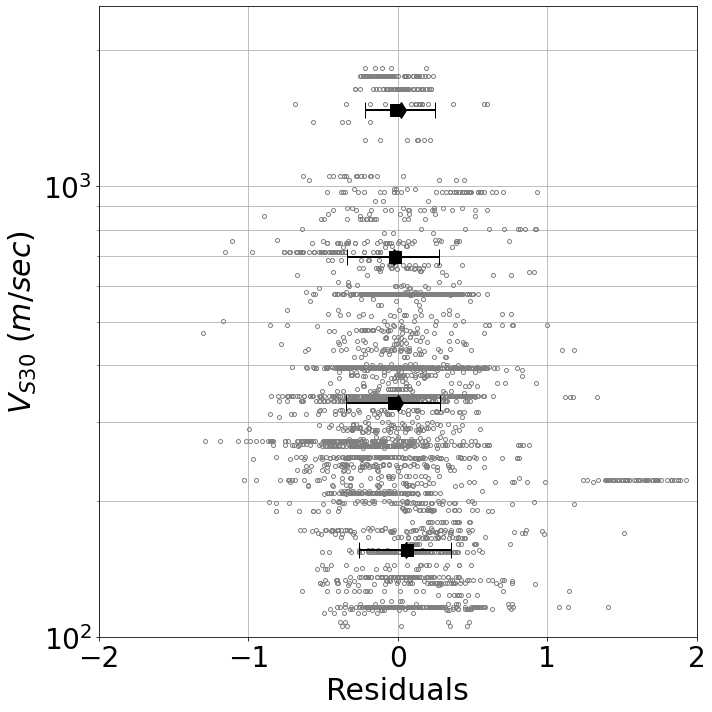}
    \end{subfigure}    
    \begin{subfigure}[t]{0.32\textwidth}
        \caption{} \label{fig:st_model_res_vs}
        \includegraphics[height = 0.95\textwidth]{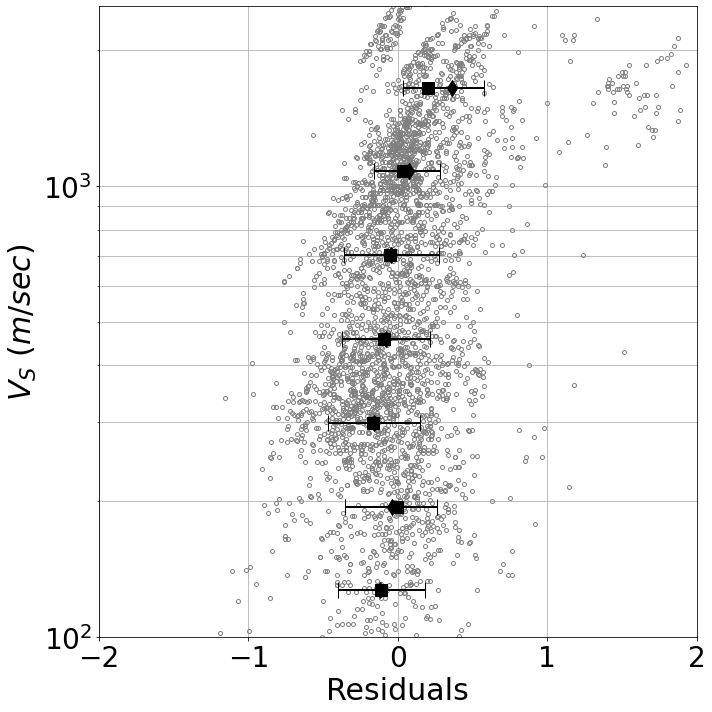}
    \end{subfigure}    
    \caption{Stationary model residuals versus depth on subfigure (a),  versus $V_{S30}$ on subfigure (b), and versus $V_S$ on subfigure (c). Error bars indicate the median, $16^{th}$, and $84^{th}$ percentiles.}
    \label{fig:st_model_res}
\end{figure}

The stationary model residuals, arranged as a function of depth, $V_{S30}$, and shear-wave velocity, are presented in Subfigures \ref{fig:st_model_res_depth}, \ref{fig:st_model_res_vs30}, and \ref{fig:st_model_res_vs}, respectively. Additionally, the along-depth residuals for different $V_{S30}$ and profile depth bins are provided in the electronic supplement (Figures S3 and S4). 
The near-zero mean of the binned residuals across all figures suggests that the model is unbiased for $V_S < 1000$ m/sec. A slight deviation in the binned residuals is observed at higher shear-wave velocities, as shown in Subfigure \ref{fig:st_model_res_vs}. This behavior can be attributed to higher shear-wave velocities corresponding to bedrock layers beneath the Bay Area sedimentary layers.
Accurately modeling these stiff layers would require an additional branch in the median velocity model (Equation \ref{eq:velocity_model}) to capture the $V_S - z$ scaling within the rock layers. Nonetheless, since the proposed model primarily aims to represent the basin layers, this misfit occurs outside the scope of the current model.

Regarding the model's misfit, the approximately equal spacing between the 16th–50th and 50th–84th percentiles across all bins suggests that a symmetric distribution, such as a normal distribution, is appropriate. Additionally, the constant width of the 16th to 84th percentile range in bins with a large number of data points indicates that using a constant standard deviation with respect to depth and $V_{S30}$ is reasonable.

t
\subsection{Spatially Varying Model}

Table \ref{tab:svar_coeffs} summarizes the coefficients for the spatially varying model. The coefficients $r_1$ and $r_2$ are adjusted relative to those in the stationary model, with $r_1$ being approximately $0.3$ units smaller and $r_2$ about $0.54$ units larger than their counterparts in the stationary model. The standard deviation of the spatially varying model is roughly $25\%$ smaller than the stationary model, highlighting the importance of properly capturing profile-specific effects.
The magnitude of spatial variability for $\delta B_r$ is $0.31$ units, with a correlation length of $1.91$ km. Figure S5 in the electronic supplement presents the trace plots and full posterior distributions for the slope adjustments and hyperparameters of the spatially varying term, indicating good model convergence, similar to the stationary model.

\begin{table}[htbp!]
\small
\caption{Posterior distributions of spatially varying model coefficients}
    \centering
    \begin{tabular}{|l||c|c|c|c|c|c|c|c|c|}
        \hline
         Coefficient          & $V_{S30ref}$ & $V_{S30w}$ & $r_1$ & $r_2$ & $r_3$ & $s_2$ & $\sigma$ & $
         \ell_{\delta Br}$ (km) & $\omega_{\delta Br}$\\
         \hline
         Fixed Value          & 6.4990 & 0.4355 &     -   &    -   & 0.3897 & 7.0713 &    -   &    -   &    -   \\
         Mean                 &    -   &    -   & -2.6097 & 5.9316 &    -   &    -   & 0.2807 & 1.9471 & 0.3159 \\
         Median               &    -   &    -   & -2.6102 & 5.9329 &    -   &    -   & 0.2807 & 1.9104 & 0.3156 \\
         5th Percentile  &    -   &    -   & -2.7168 & 5.6428 &    -   &    -   & 0.2746 & 1.4690 & 0.2860 \\
         95th Percentile &    -   &    -   & -2.5012 & 6.2179 &    -   &    -   & 0.2869 & 2.5526 & 0.3458 \\ \hline
    \end{tabular}
    \label{tab:svar_coeffs}
\end{table}

\begin{figure}[htbp!]
    \centering
    \begin{subfigure}[t]{0.29\textwidth}
        \caption{} \label{fig:sv_model_scl_k}
        \includegraphics[height = 0.95\textwidth]{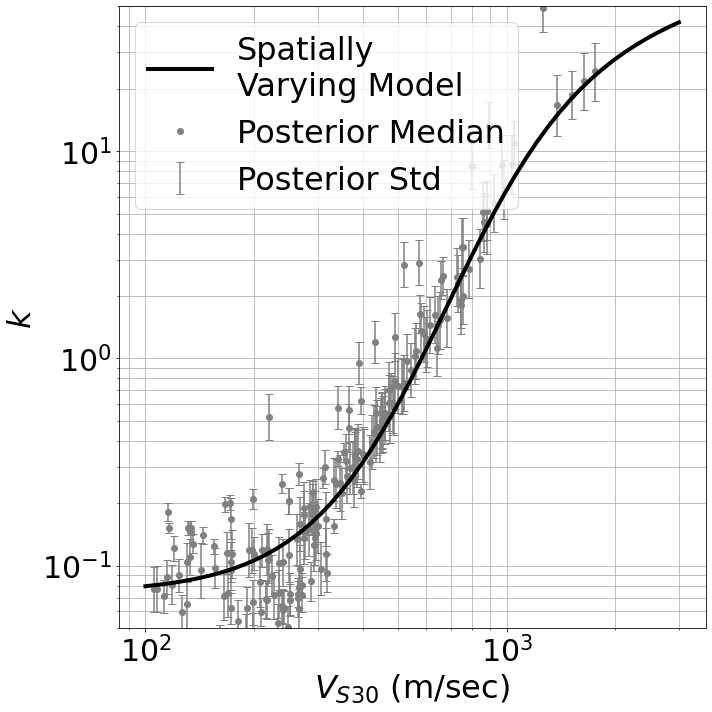}
    \end{subfigure} 
    \begin{subfigure}[t]{0.29\textwidth}
        \caption{} \label{fig:sv_model_scl_n}
        \includegraphics[height = 0.95\textwidth]{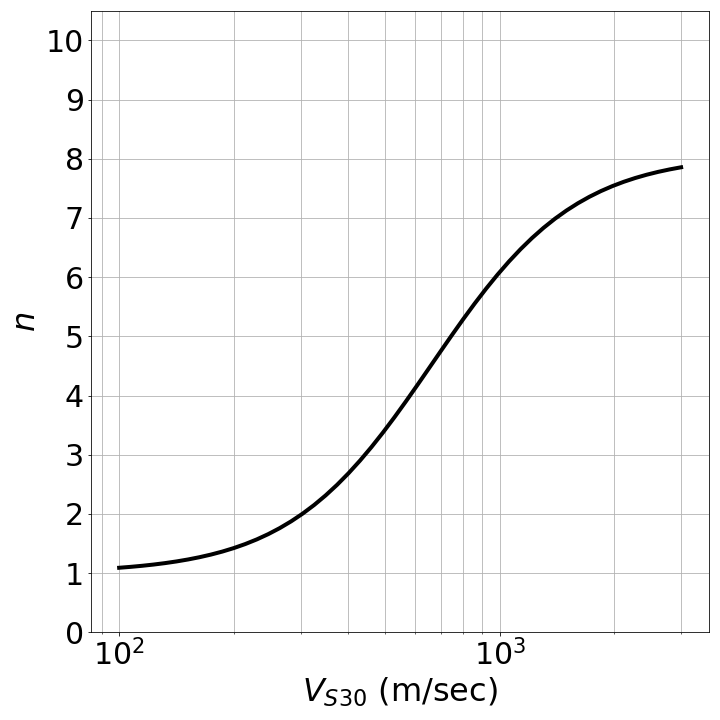}
    \end{subfigure}  
    \begin{subfigure}[t]{0.29\textwidth}
        \caption{} \label{fig:sv_model_scl_vs0}
        \includegraphics[height = 0.95\textwidth]{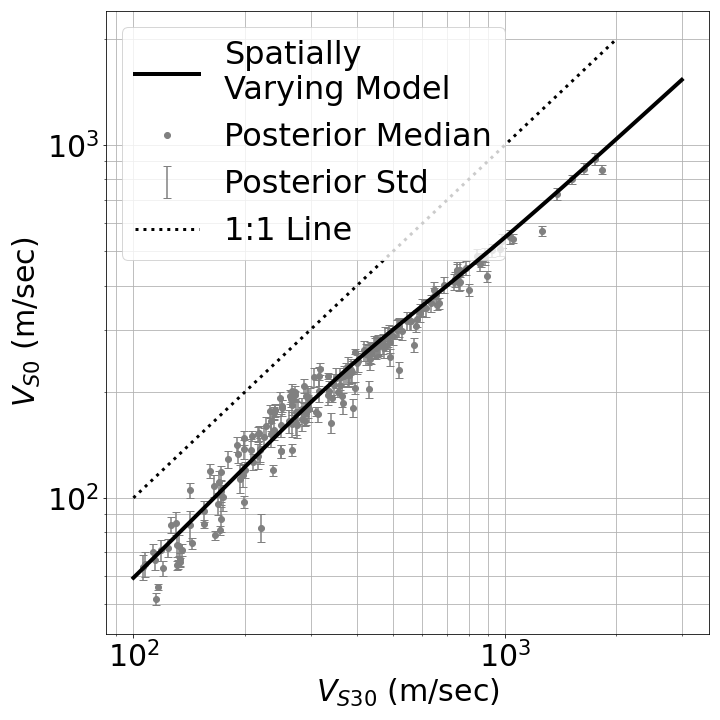}
    \end{subfigure}  
    \caption{Scaling relationships for the spatially varying model; subfigure (a) $V_{S30}$ versus $k$ scaling, and subfigure (c) $V_{S30}$ versus $V_{S0}$ scaling.}
    \label{fig:sv_model_scl}
\end{figure}

Figure \ref{fig:sv_model_scl} presents the scaling relationships for the spatially varying model. Subfigure \ref{fig:sv_model_scl_k} shows the slope scaling as a function of $V_{S30}$. The profile-specific slope estimates are represented by circular markers, with error bars indicating the posterior standard deviation. Overall, these estimates align well with the global scaling relationship.
Subfigure \ref{fig:sv_model_scl_n} presents the curvature scaling versus $V_{S30}$, which is kept the same as the stationary model. Subfigure \ref{fig:sv_model_scl_vs0} displays the shear-wave velocity at the top of the profiles as a function of $V_{S30}$. The profile-specific $V_{S0}$ adjustments reflect the deviation of the profile-specific slope from its scaling relationship to satisfy the $V_{S30}$ constraint.

\begin{figure}[htbp!]
    \centering
    \begin{subfigure}[t]{0.32\textwidth}
        \caption{} \label{fig:sv_model_res_depth}
        \includegraphics[height = 0.95\textwidth]{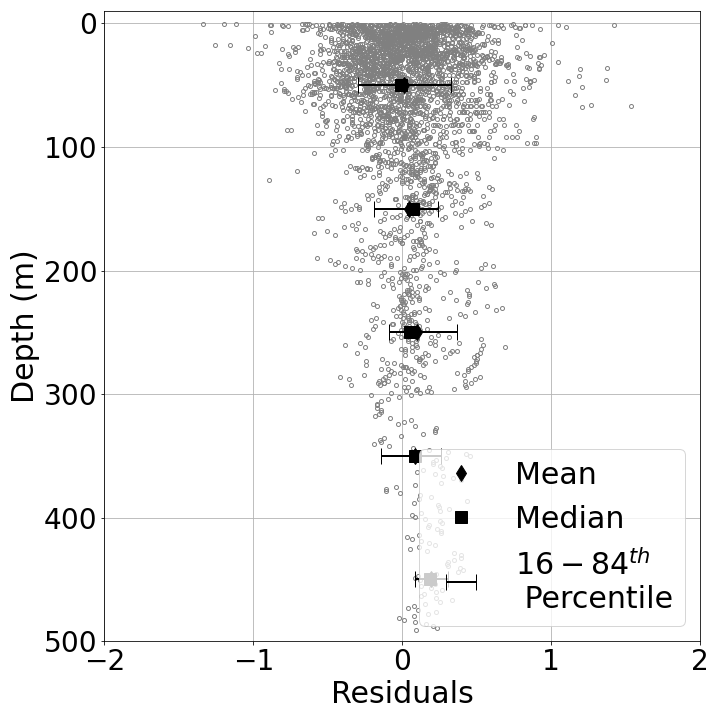}
    \end{subfigure} 
    \begin{subfigure}[t]{0.32\textwidth}
        \caption{} \label{fig:sv_model_res_vs30}
        \includegraphics[height = 0.95\textwidth]{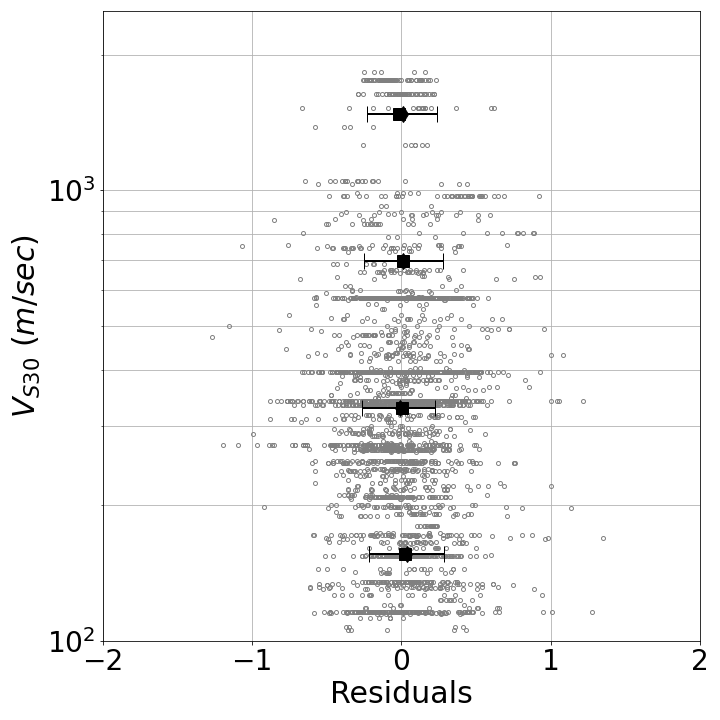}
    \end{subfigure}  
    \begin{subfigure}[t]{0.32\textwidth}
        \caption{} \label{fig:sv_model_res_vs}
        \includegraphics[height = 0.95\textwidth]{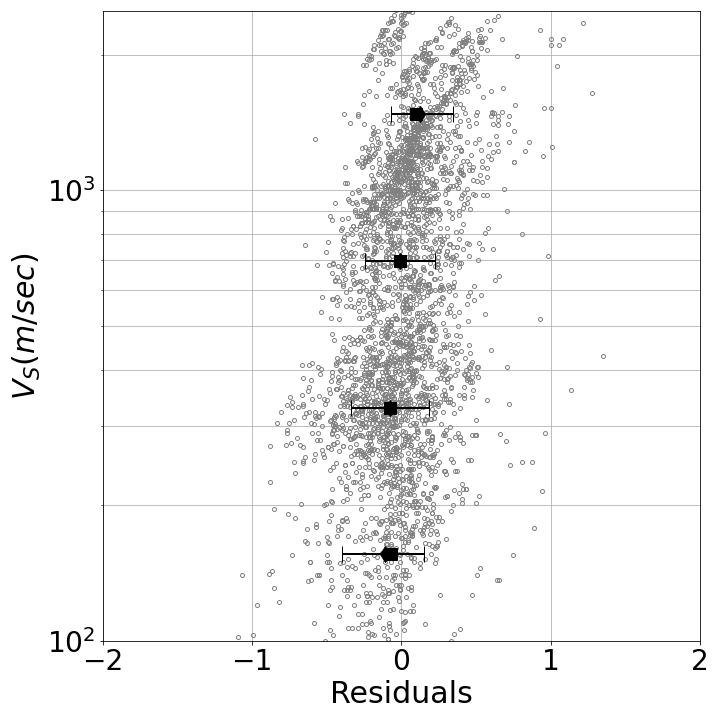}
    \end{subfigure}  
    \caption{Spatially varying model within profile residuals versus depth on subfigure, (a) versus $V_{S30}$ on subfigure (b), and versus $V_S$ on subfigure (c); error bars indicate the median, $16^{th}$, and $84^{th}$ percentiles.}
    \label{fig:sv_model_res}
\end{figure}

The within profile residuals are summarized in Figure \ref{fig:sv_model_res}; they are depicted versus depth in subfigure \ref{fig:sv_model_res_depth}, versus $V_{S30}$ in subfigure \ref{fig:sv_model_res_vs30}, and versus $V_{S}$ in subfigure \ref{fig:sv_model_res_vs}.
Furthermore, they are depicted as a function of depth binned by $V_{S30}$ and profile length in the electronic supplement (Figures S6 and S7). 
The mean of the binned residuals is close to zero, indicating that the model is unbiased with respect to depth, $V_{S30}$, and layer $V_S$. Although there is still some bias at higher $V_S$ values ($V_S > 1000$ m/sec), it is smaller than the stationary model. 
Additionally, similarly to the stationary model, the nearly constant distance from the mean to the 16th and 84th percentiles, as well as the constant width between these percentiles, suggest that a symmetric distribution with a constant standard deviation is appropriate.

\begin{figure}[htbp!]
    \centering
    \begin{subfigure}[t]{0.48\textwidth}
        \caption{}
        \includegraphics[height = 0.78\textwidth]{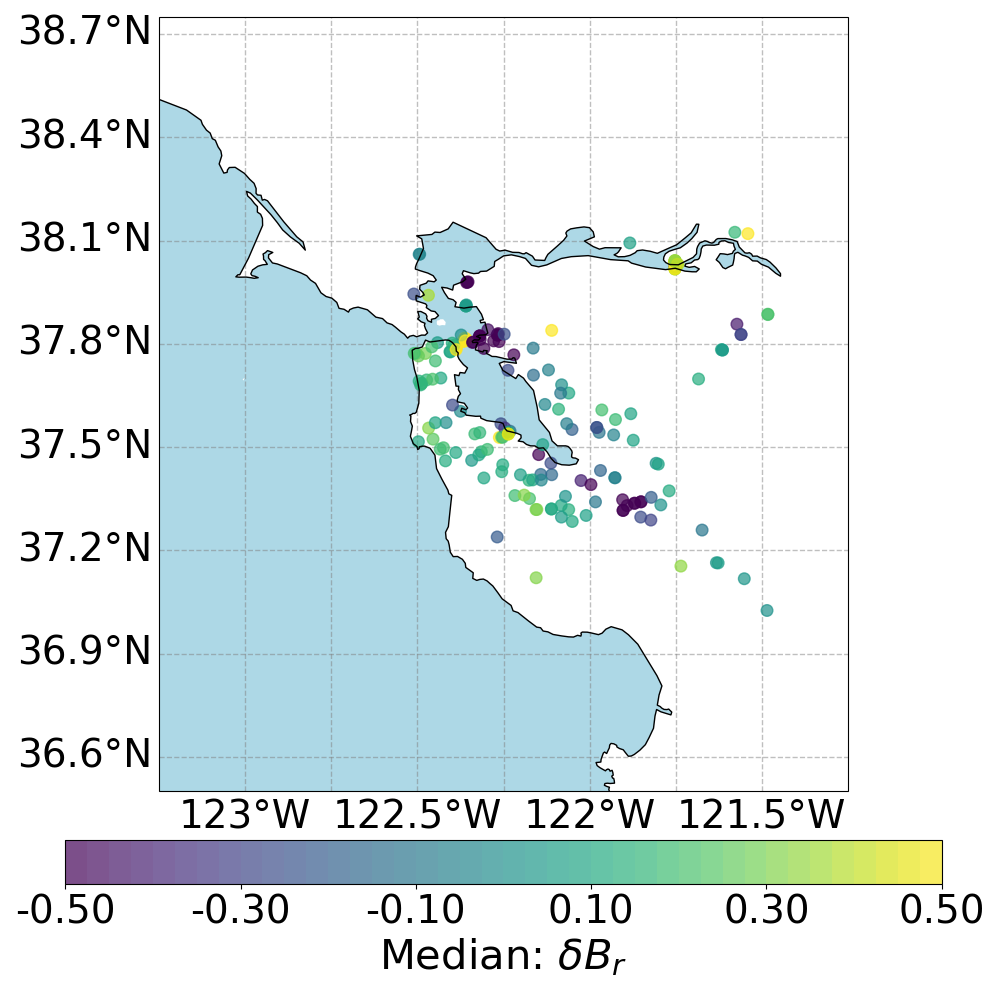}
        \label{fig:sv_model_dBr_med}
    \end{subfigure} 
    \begin{subfigure}[t]{0.48\textwidth}
        \caption{}
        \includegraphics[height = 0.78\textwidth]{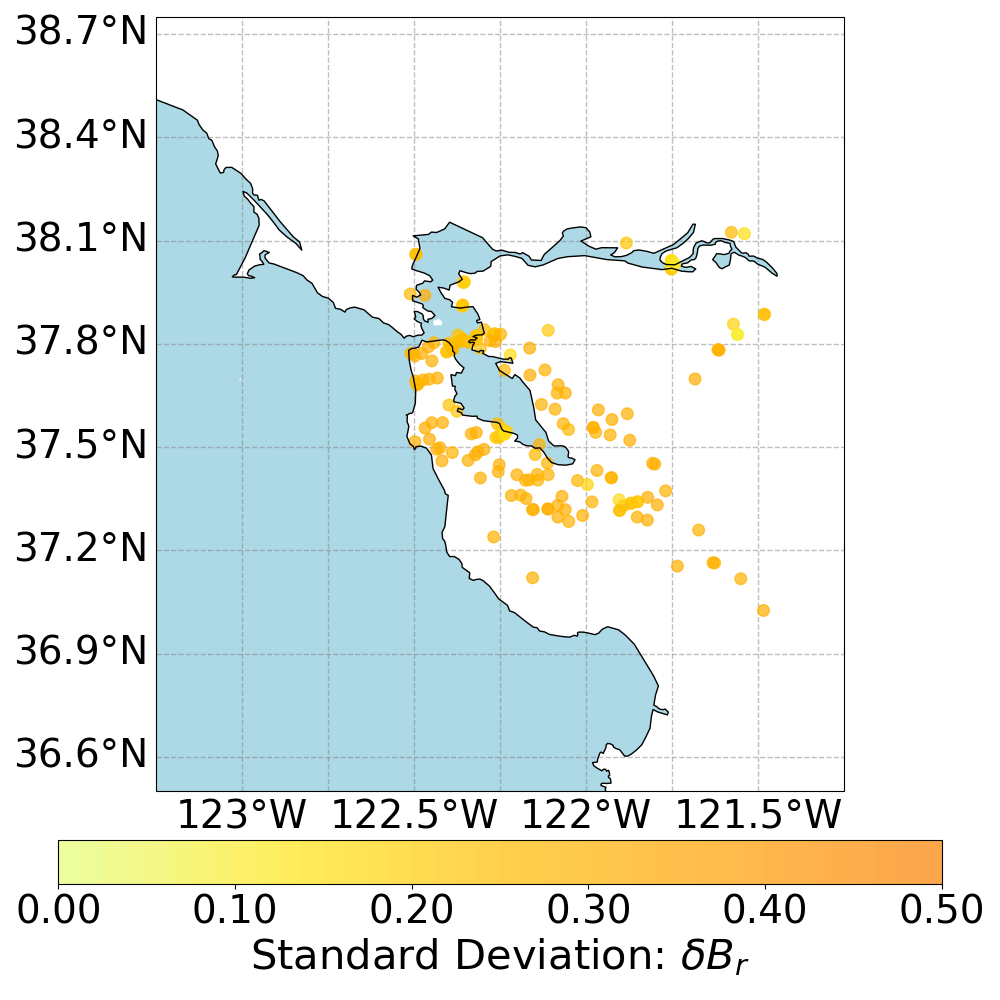}
        \label{fig:sv_model_dBr_unc}
    \end{subfigure}  
    \caption{Spatially varying slope adjustment term, $\delta B_r$, (a) median value, (b) uncertainty.}
    \label{fig:sv_model_dBr}
\end{figure}

The spatial variability of the slope adjustment is shown in Figure \ref{fig:sv_model_dBr}. Subfigure \ref{fig:sv_model_dBr_med} presents the median estimate of $\delta B_r$, while Subfigure \ref{fig:sv_model_dBr_unc} presents its posterior uncertainty. 
Around the Bay Area estuary, $\delta B_r$ is negative, indicating a gentler slope (i.e., a slower increase of $V_S$ with depth). In contrast, inland, in the San Francisco peninsula, San Mateo, and East Bay, $\delta B_r$ is positive, corresponding to steeper profiles. Due to the relatively short correlation length, the uncertainty in $\delta B_r$ does not display significant spatial variability.

Aggregating $\delta B_r$ by surface geological unit does not reveal any systematic trends in the average values (see Figure S8 in the electronic supplement). This outcome is expected, as surface geology is correlated with $V_{S30}$, which influences the median slope (Equation \ref{eq:k_scl_svar}). Instead, the surface geology appears to impact the variability of the slope adjustment within each class. The highest variability is observed in artificial fill, alluvium (Quaternary and Holocene) deposits, and offshore profiles, while older Pleistocene and Pliocene units exhibit lower variability. An exception is seen with profiles in the Franciscan Complex, which also display significant variability.
Future studies should evaluate a more complex correlation structure that accounts for adjustments within and between geological units.

\subsection{Along-Depth Correlation} \label{sec:depth_correlation}

\begin{figure}[htbp!]
    \centering
    \begin{subfigure}[t]{0.48\textwidth}
        \caption{}
        \includegraphics[height = 0.78\textwidth]{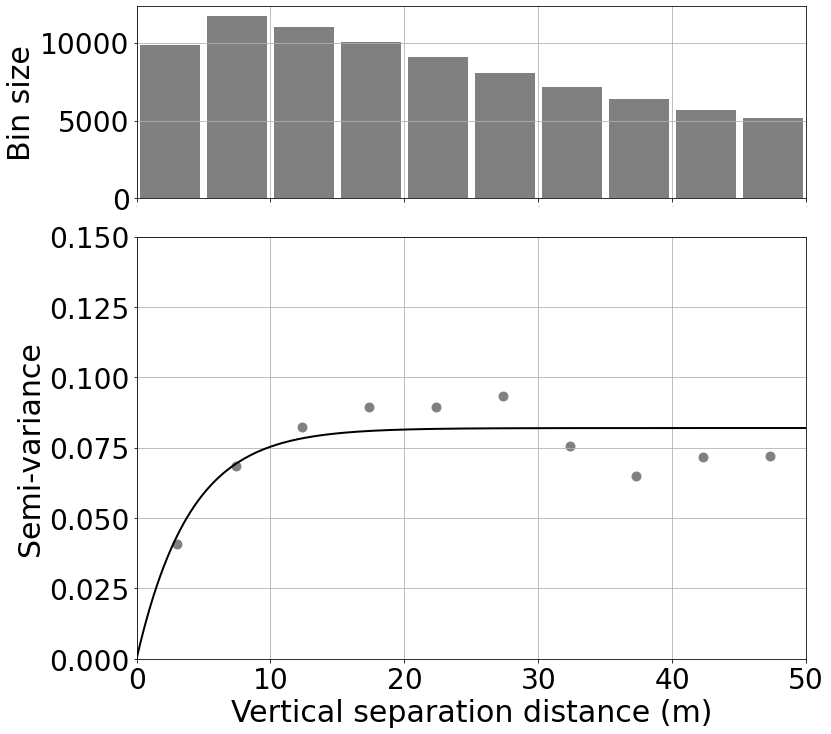}
    \end{subfigure} 
    \begin{subfigure}[t]{0.48\textwidth}
        \caption{}
        \includegraphics[height = 0.78\textwidth]{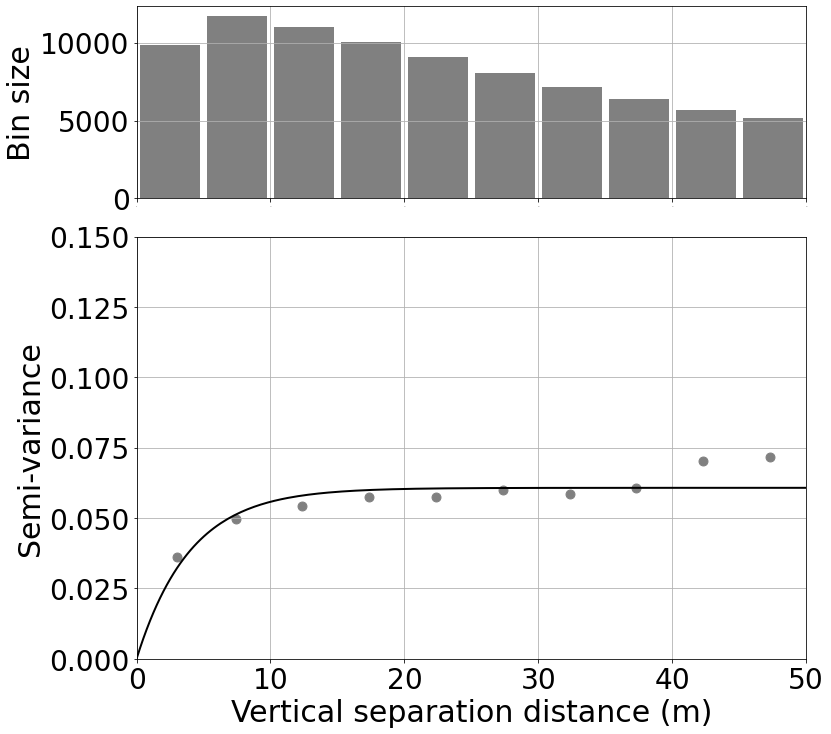}
    \end{subfigure}  
    \caption{Along-depth semivariogram for (a) stationary model, (b) spatially varying model; circular marker represent the empirical semivariance while the solid line represent the semivariance model fit.}
    \label{fig:semivar_depth}
\end{figure}

The along-depth correlations for both models were determined through semivariogram analysis using an exponential covariance. Table \ref{tab:semivar_depth_coeffs} summarizes the estimated along-depth correlation length ($r$) and semivariance ($s$) for the stationary and spatially varying models while the raw semivariograms and model fits are shown in Figure \ref{fig:semivar_depth}. 
In both models, the estimated correlation lengths are approximately $10$ m. As expected, the semivariance of the spatially varying model is smaller than that of the stationary model.

To reduce the computational cost, the along-depth correlation structure was determined separately from the lateral spatially varying component. Capturing both correlation structures in a single regression would require manipulating covariance matrices the size of the total number of velocity layers. 
In the current implementation, the spatially varying term ($\delta B_r$) is modeled with a covariance matrix based on the number of velocity profiles, while the semivariogram analysis computes the semivariance of the model residuals within each profile. This approach effectively produces a pseudo-3D SVM, where the two horizontal dimensions are decoupled from the vertical one. The two orders of magnitude length-scale difference between the correlation length in the horizontal plane (2km) and vertical direction (10m) support the presence of a sedimentary depositional environment and further strengthen our assumption of decoupling the two directions. Still, 
Future studies should focus on capturing the full 3D correlation structure, especially in transition zones such as alluvium fans and basin edges where the ratio of correlation lengths will be closer to one, using more efficient regression methodologies and denser velocity observations.

\begin{table}[htbp!]
\small
	\caption{Along-depth semivariogram parameters; Correlation length ($r$) and semi-variance ($s$).}
	\centering
    \label{tab:semivar_depth_coeffs}
	\begin{tabular}{|l|l|c|c|}
	    \hline
	    Model      &  Coefficient      & $r$ (m) &  $s$   \\ 
	    \hline
	    Stationary & Estimate          & 11.9293 & 0.0820 \\
                   & Standard error    & 2.9375  & 0.0037 \\ \hline
	    Spatially  & Estimate          & 11.9778 & 0.0607 \\
	    Varying    & Standard error    & 2.1545  & 0.0020 \\ \hline
     \end{tabular}
\end{table}

\subsection{The Sediment Velocity Model - Basement Rock Interface} \label{sec:termination}

Determining the appropriate conditions for transitioning from the sediment velocity models to the underlying basement rock models, in this case, the USGS SFBA community velocity model, has been an ongoing challenge in previous studies. For example, \citet{ely2010VS30} prescribed a fixed depth where the two models connected seamlessly; \citet{Shi2018}, on the other hand, prescribed an upper limit of $V_{S30}$ applicability on their SVM, beyond which the community velocity model was valid, allowing for impedance contrasts to form at the interface. In this work, attempts to identify a specific depth or a shear-wave velocity at which the empirical data and the SFBA model converged were unsuccessful (Figures S9 and S10 in the electronic supplement).

For this reason, we recommend following the sedimentary velocity model until a shear-wave velocity of $1000$ m/sec is reached or until the rock layer underlying the basin is encountered. To prevent any negative impedance contrast--where the $V_S$ of the shallow layers exceeds that of the underlying layers--the maximum $V_S$ value between the SVM and USGS SFBA models should be adopted. Beyond these thresholds, the USGS SFBA model should be applied.

\section{Model Evaluation} \label{sec:evaluation}

The model evaluation is divided into two subsections: first, the horizontal cross-sections of the proposed models are compared with the USGS model; next, we evaluate their impact on site amplification through a one-dimensional site-response analysis.

\begin{landscape}

\begin{figure}[htbp!]
    \centering
    \begin{minipage}[c]{0.3\textwidth}
        \begin{subfigure}[t]{1\textwidth}
            \caption{}
            \includegraphics[height = 0.95\textwidth]{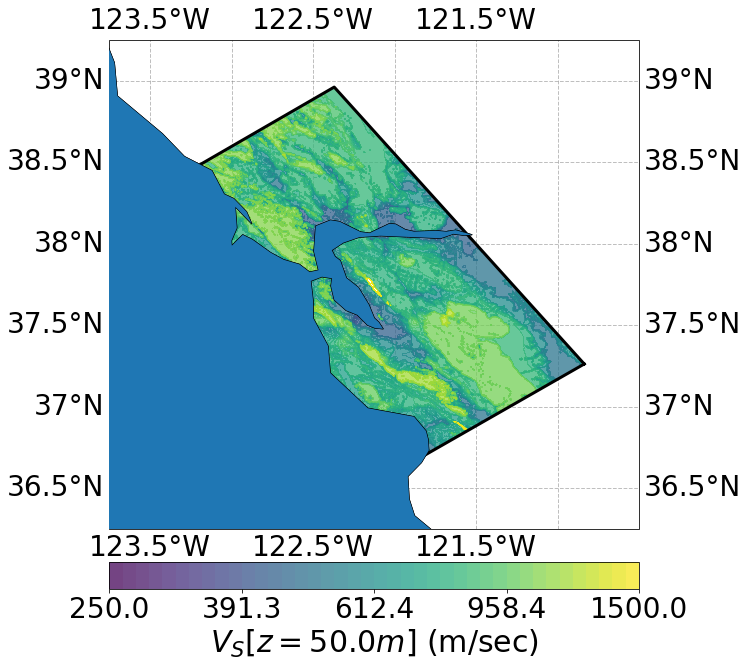} 
            \label{fig:vel_model_50m_stat}
        \end{subfigure} 
    \end{minipage}
    \begin{minipage}[b]{0.3\textwidth}
        \begin{subfigure}[t]{1\textwidth}
            \caption{}
            \includegraphics[height = 0.95\textwidth]{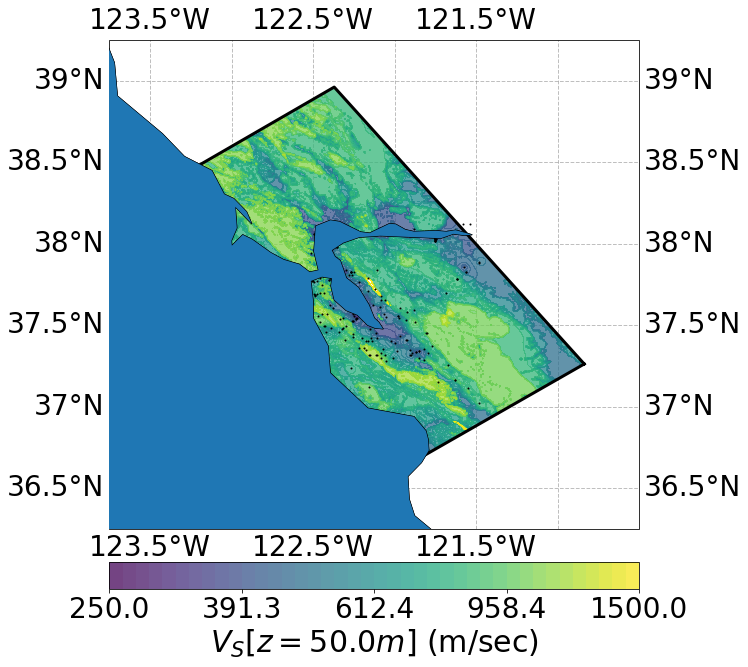}
            \label{fig:vel_model_50m_svar_med_con}
        \end{subfigure} \\
        \begin{subfigure}[t]{1\textwidth}
            \caption{}
            \includegraphics[height = 0.95\textwidth]{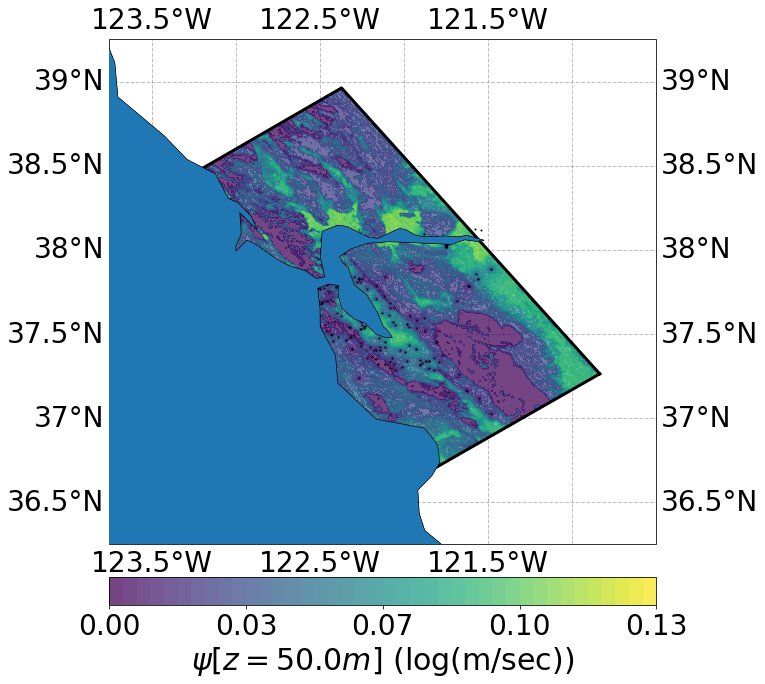}
            \label{fig:vel_model_50m_svar_unc_con}
        \end{subfigure}  
    \end{minipage}
    \begin{minipage}[b]{0.3\textwidth}
        \begin{subfigure}[t]{1\textwidth}
            \caption{}
            \includegraphics[height = 0.95\textwidth]{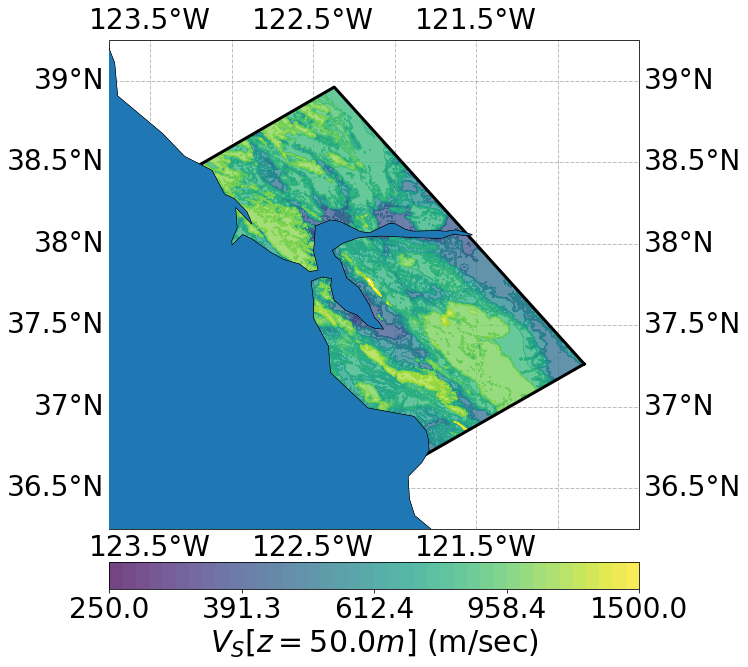}
            \label{fig:vel_model_50m_svar_med_uncon}
        \end{subfigure} \\
        \begin{subfigure}[t]{1\textwidth}
            \caption{}
            \includegraphics[height = 0.95\textwidth]{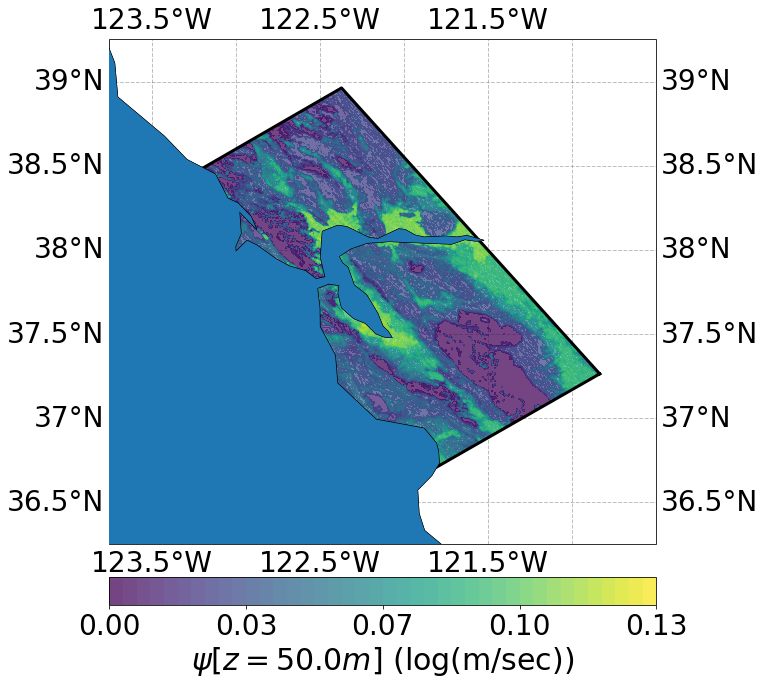}
            \label{fig:vel_model_50m_svar_unc_uncon}
        \end{subfigure}  
    \end{minipage}
    \begin{minipage}[c]{0.3\textwidth}
        \begin{subfigure}[t]{1\textwidth}
            \caption{}
            \includegraphics[height = 0.95\textwidth]{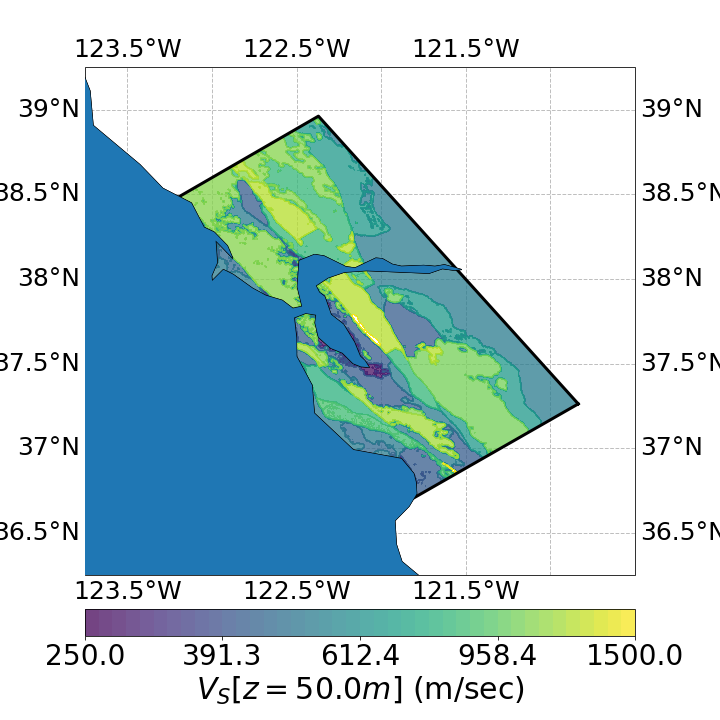}
        \end{subfigure} 
    \end{minipage}
    \caption{Velocity model cross session at a depth of $50~m$: 
    (a) mean $V_S$ of the stationary model, 
    (b) mean $V_S$ of spatially varying model conditioned on available velocity profiles, 
    (c) $V_S$ uncertainty of spatially varying model conditioned on available velocity profiles, 
    (d) mean $V_S$ of the unconditional spatially varying model, 
    (e) $V_S$ uncertainty of unconditional spatially varying model conditioned on available velocity profiles, 
    (d) $V_S$ of USGS San-Francisco Bay Area velocity model.}
    \label{fig:vel_model_50m}
\end{figure}

\end{landscape}

Figure \ref{fig:vel_model_50m} presents a comparison of the cross-section at 50m depth across four models: the stationary model, the spatially varying model conditioned on the available data, the spatially varying model without any data conditioning, and the USGS SFBA model. Additional cross-sections at 10 m and 100 m depth are provided in the electronic supplement (Figures S11 and S12). The $V_{S30}$ input for both the stationary and spatially varying models is obtained from \cite{wills2015next}.

The most notable differences between the proposed models and the USGS model occur in the South Bay and Livermore Valley, where both the stationary and spatially varying models suggest higher shear-wave velocities. Differences in the South Bay are more pronounced at shallower depths, while in Livermore, they become more significant at greater depths. Comparisons at $10$, $50$, and $100$ m depth between the stationary and spatially varying models with the USGS SFBA velocity model are presented in the electronic supplement (Figures S14 to S16). These differences could be the underlying factor explaining the findings by \cite{pinilla2024}, who reported over-amplification of seismic waves in these regions based on 3D simulations when compared to observations; although, a new set of 3D simulations with the proposed models would be required to test the hypothesis. 

An additional observation is that, as Subfigures \ref{fig:vel_model_50m_stat}, \ref{fig:vel_model_50m_svar_med_con} and \ref{fig:vel_model_50m_svar_med_uncon} show, in regions with limited profile observations, the stationary and spatially varying models are similar; however, near measured velocity profiles, the conditional spatially varying model smoothly deviates from the stationary and unconditional spatially varying model to capture site-specific slope adjustments.
Additionally,  Subfigures \ref{fig:vel_model_50m_svar_unc_con} and \ref{fig:vel_model_50m_svar_unc_uncon} display the $V_S$ uncertainty from $\delta B_r$ in the spatially varying models. In both cases, the largest uncertainty occurs in regions with low $V_{S30}$ values due to the shallower $V_S - z$ slopes. Subfigure \ref{fig:vel_model_50m_svar_unc_con} also demonstrates how including the velocity profiles in the conditional model reduces uncertainty compared to the unconditional model for sites in the vicinity of velocity observations.

\subsection{Site Response Analysis}

We lastly conducted a series of linear site-response analyses to evaluate the proposed models' performance in terms of predicting near-surface amplification relative to the USGS SFBA velocity model under 1D site response assumptions. We specifically performed four sets of site-response analyses: one using the measured velocity profiles as the ground truth and three others using the stationary, spatially varying, and USGS models evaluated at the same locations as the observed profiles.

The comparison between the alternative models was based on the goodness-of-fit (GOF) measure proposed by \cite{shi2017stiffness}, which considers peak-ground acceleration (PGA), peak-ground velocity (PGV), peak-ground displacement (PGD), Arias intensity, significant duration, Fourier amplitude, and spectral acceleration to assess the fit between the models and ground truth conditions. A score of zero indicates a perfect match between the model and reference conditions; a positive score indicates an overestimation of intensity measures (IMs) relative to the reference conditions, while a negative score indicates an underestimation of IMs.
An ensemble of Ricker wavelets was used as input outcrop motion for the site-response analyses. The generated ground motion has a flat frequency content from $0.1$ to $10.0~\text{Hz}$, allowing for comparison of the velocity profiles over a wide range of frequencies (Figures S17 and S18 in the electronic supplement).
Comparisons for the stationary and spatially varying models were performed both with and without along-depth variability, sampled with the covariance described in \hyperref[sec:depth_correlation]{Subsection: Along-Depth Correlation}. An example of the profiles used in the site-response analysis is provided in Figure S19 in the electronic supplement.

\begin{figure}[htbp!]
    \centering
    \begin{subfigure}[t]{0.32\textwidth}
        \caption{}
        \includegraphics[width = 1\textwidth]{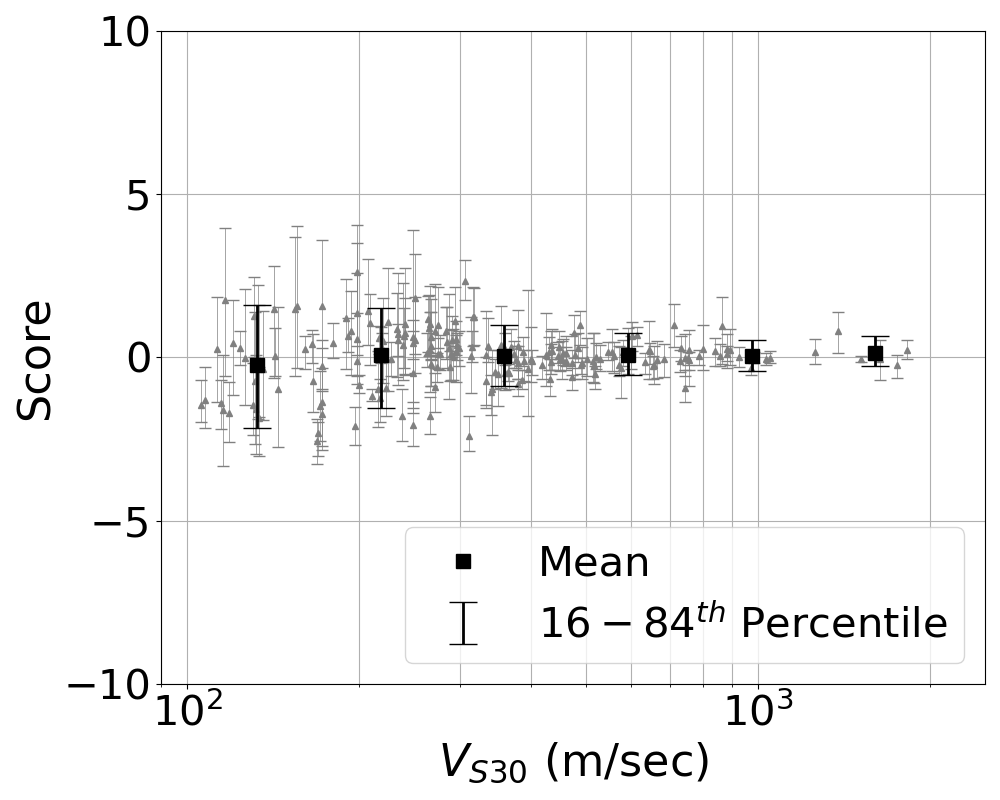}
    \end{subfigure} 
    \hfill
    \begin{subfigure}[t]{0.32\textwidth}
        \caption{}
        \includegraphics[width = 1\textwidth]{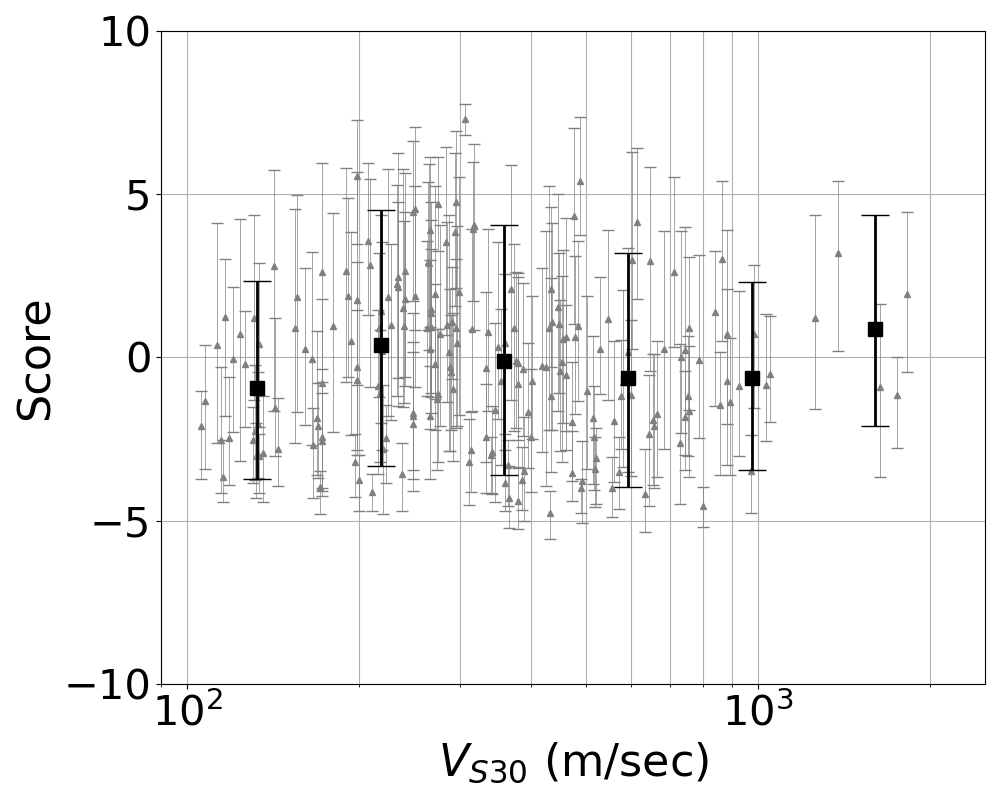}
    \end{subfigure}  
    \hfill
    \begin{subfigure}[t]{0.32\textwidth}
        \caption{}
        \includegraphics[width = 1\textwidth]{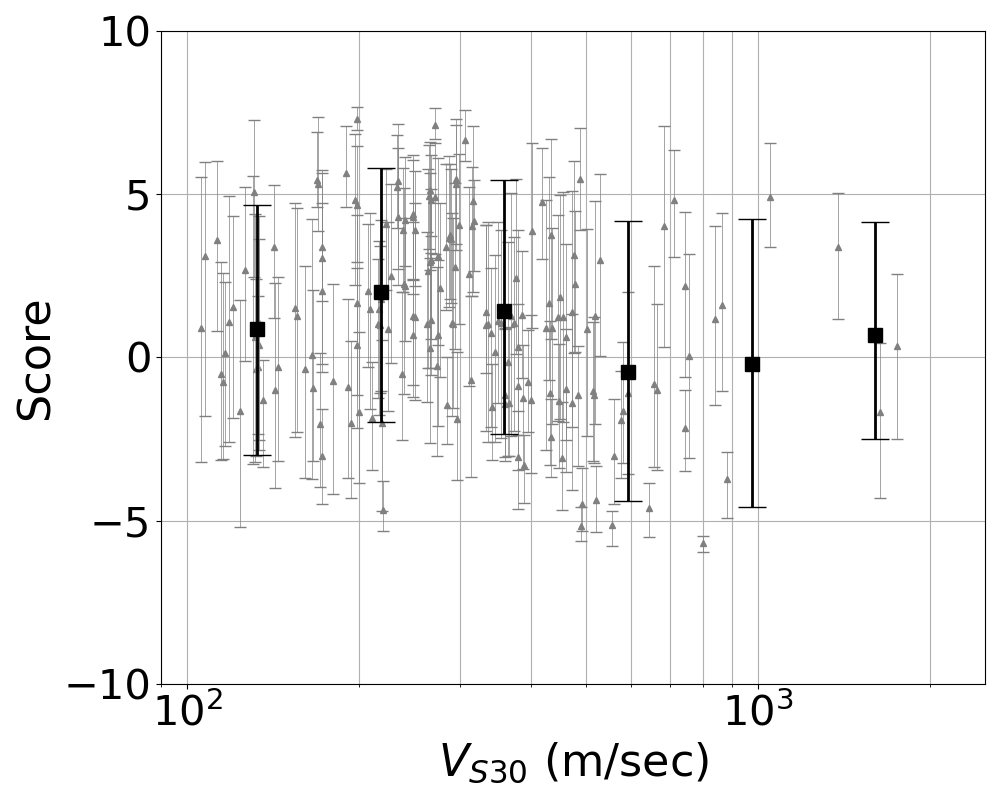}
    \end{subfigure}  
    \caption{Average goodness-of-fit (GOF) scores for the stationary model profiles with along-depth variability; gray dots represent the GOF scores of individual profiles, black squares indicate the mean GOF values for each $V_{S30}$ bin, and error bars show the 16th to 84th percentile range; 
    (a) Frequency bin: $f \in [0.01, f_P)~\text{Hz}$,
    (b) Frequency bin: $f \in [f_P, 2 f_P)~\text{Hz}$,
    (c) Frequency bin: $f \in [2 f_P, 10)~\text{Hz}$.}
    \label{fig:sra_gof_svar_drlz}
\end{figure}

\begin{figure}[htbp!]
    \centering
    \begin{subfigure}[t]{0.32\textwidth}
        \caption{}
        \includegraphics[width = 1\textwidth]{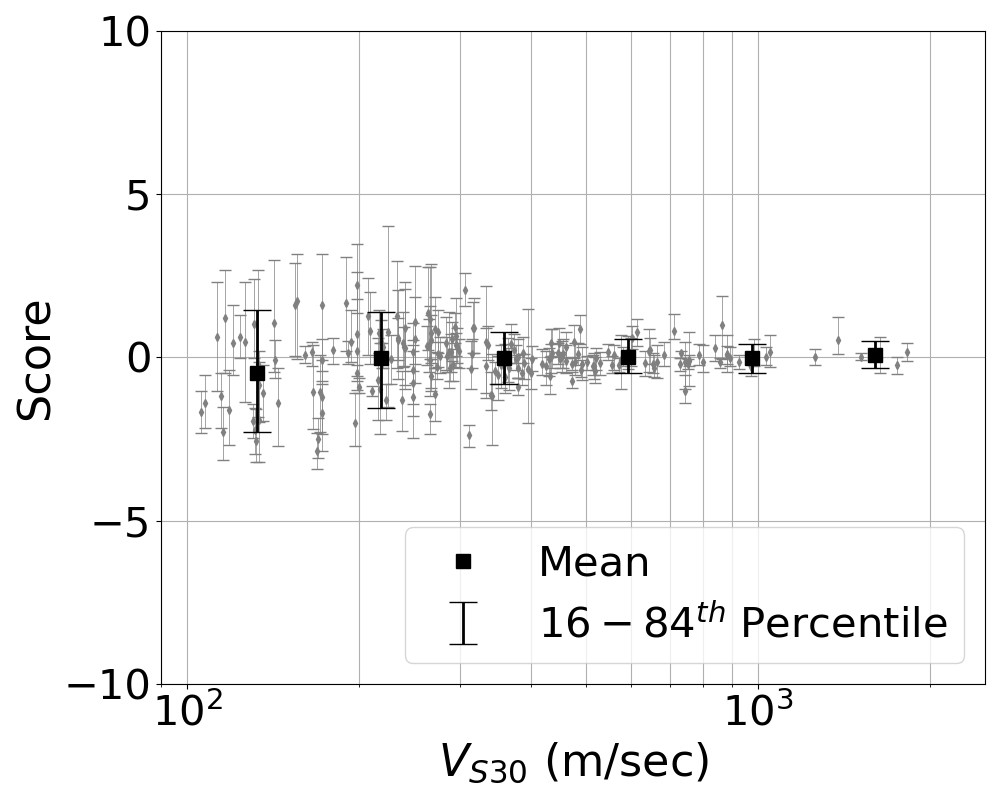}
    \end{subfigure} 
    \hfill
    \begin{subfigure}[t]{0.32\textwidth}
        \caption{}
        \includegraphics[width = 1\textwidth]{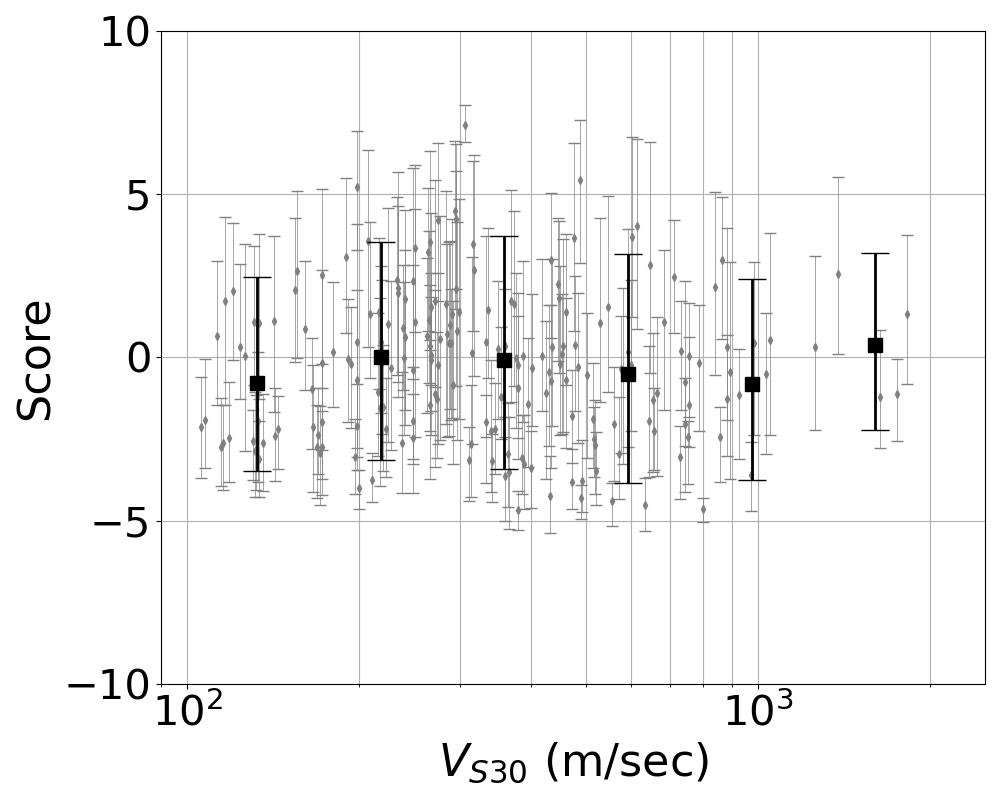}
    \end{subfigure}  
    \hfill
    \begin{subfigure}[t]{0.32\textwidth}
        \caption{}
        \includegraphics[width = 1\textwidth]{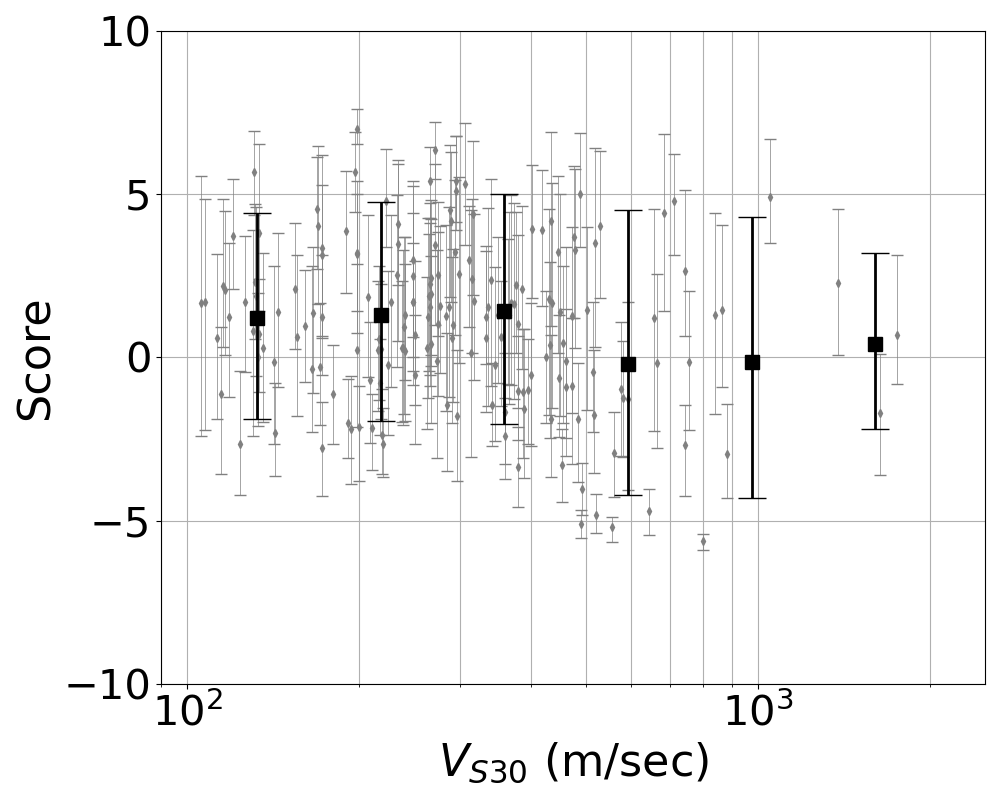}
    \end{subfigure}  
    \caption{Average goodness-of-fit (GOF) scores for the spatially varying model profiles with along-depth variability and uncertainty in the slope adjustment factor $\delta B_r$; gray dots represent the GOF scores of individual profiles, black squares indicate the mean GOF values for each $V_{S30}$ bin, and error bars show the 16th to 84th percentile range; 
    (a) Frequency bin: $f \in [0.01, f_P)~\text{Hz}$,
    (b) Frequency bin: $f \in [f_P, 2 f_P)~\text{Hz}$,
    (c) Frequency bin: $f \in [2 f_P, 10)~\text{Hz}$.}
    \label{fig:sra_gof_svar_srlz_drlz}
\end{figure}

\begin{figure}[htbp!]
    \centering
    \begin{subfigure}[t]{0.32\textwidth}
        \caption{}
        \includegraphics[width = 1\textwidth]{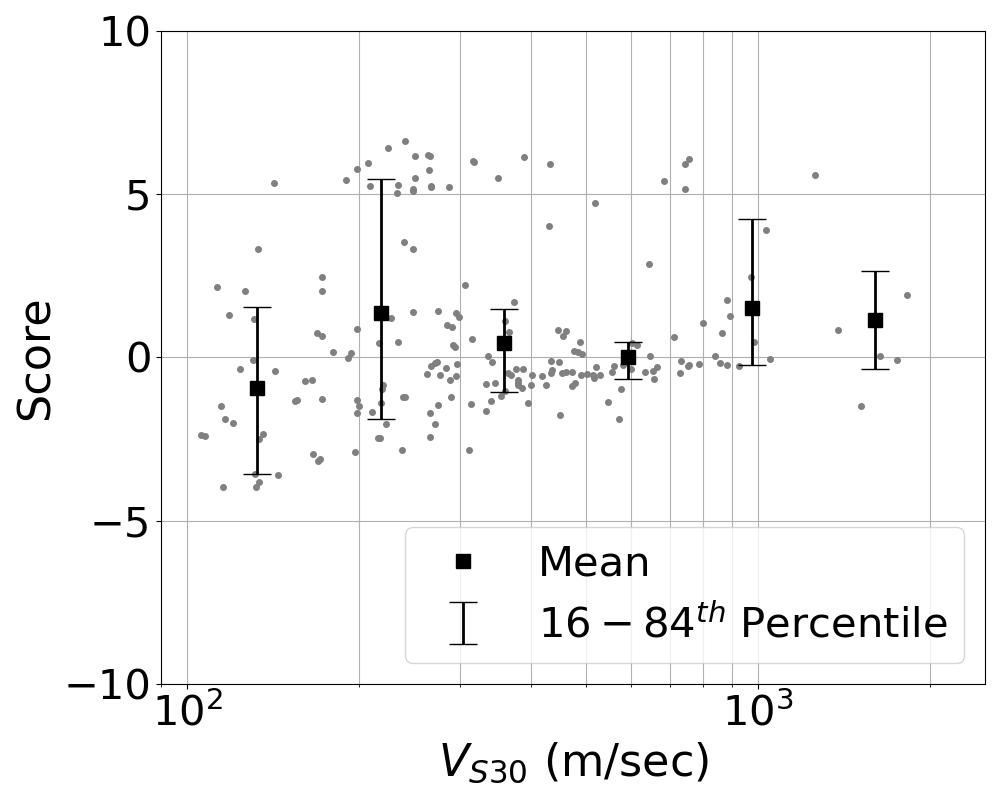}
    \end{subfigure} 
    \hfill
    \begin{subfigure}[t]{0.32\textwidth}
        \caption{}
        \includegraphics[width = 1\textwidth]{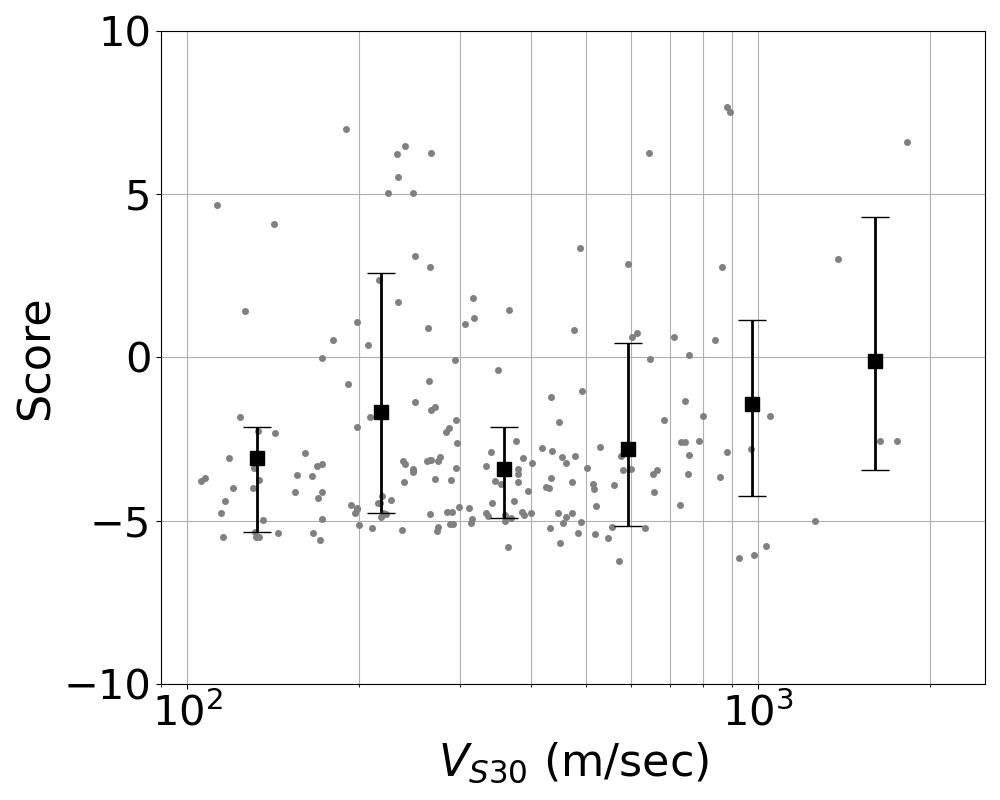}
    \end{subfigure}  
    \hfill
    \begin{subfigure}[t]{0.32\textwidth}
        \caption{}
        \includegraphics[width = 1\textwidth]{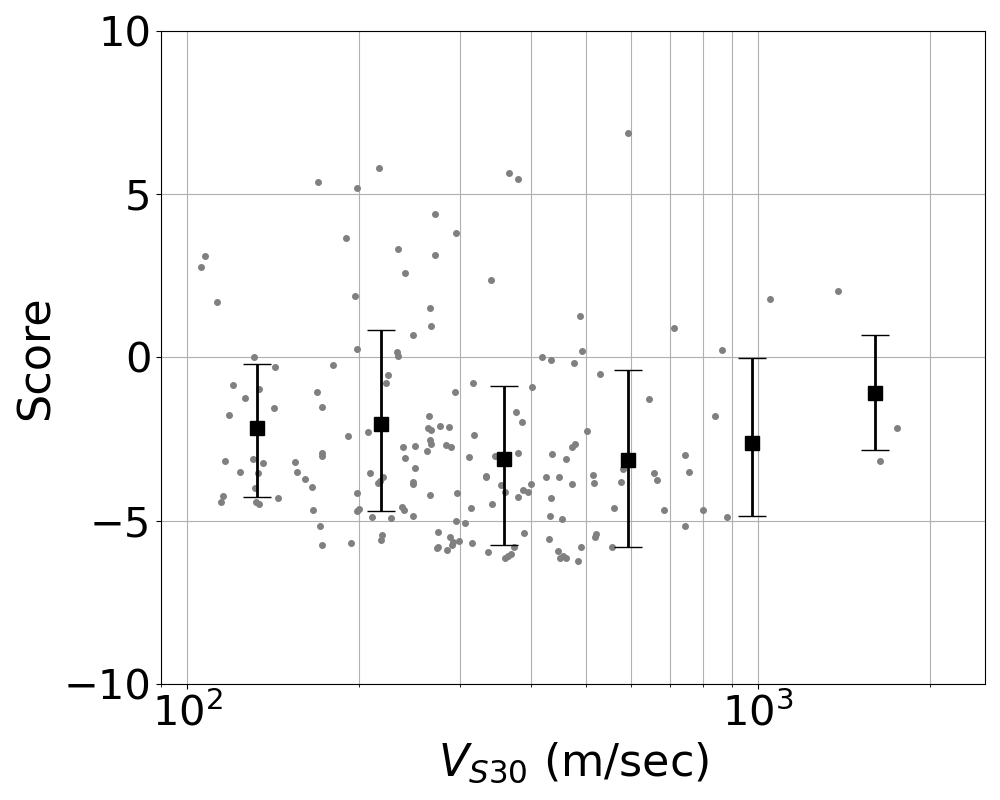}
    \end{subfigure}
    \caption{Average goodness-of-fit (GOF) scores for USGS SFBA velocity profiles; gray dots represent the GOF scores of individual profiles, black squares indicate the mean GOF values for each $V_{S30}$ bin, and error bars show the 16th to 84th percentile range; 
    (a) Frequency bin: $f \in [0.01, f_P)~\text{Hz}$,
    (b) Frequency bin: $f \in [f_P, 2 f_P)~\text{Hz}$,
    (c) Frequency bin: $f \in [2 f_P, 10)~\text{Hz}$.}
    \label{fig:sra_gof_usgs}
\end{figure}

Figures \ref{fig:sra_gof_svar_drlz}, \ref{fig:sra_gof_svar_srlz_drlz}, and \ref{fig:sra_gof_usgs} present the GOF scores for the stationary, spatially varying, and USGS SFBA models, respectively, relative to the ground truth profile measurements. The comparison is made across three frequency bins: $0.01$ Hz to $f_P$, $f_P$ to $2f_P$, and $2f_P$ to $10$ Hz, where $f_P$ represents the frequency corresponding to the first fundamental mode of each profile, calculated using the quarter-wavelength approximation:

\begin{equation} \label{eq:fp}
    f_P = \left(4 \sum_{i=1}^n \frac{dz_i}{V_{Si}} \right)^{-1}
\end{equation}

\noindent where $V_{Si}$ and $dz_i$ are the shear-wave velocity and thickness of the $i^{th}$ layer, respectively.
The frequency bins are defined in terms of $f_P$ rather than fixed values to account for the varying heights and stiffness of different profiles, which result in different resonant frequency ranges. The quarter-wavelength approximation was chosen over a full wave propagation definition as it is simpler to compute in forward applications.

Figure \ref{fig:sra_gof_svar_drlz} shows that the stationary model captures the mean amplification (i.e., the moving average of the GOF score is near zero) in the $0.01$ to $f_P$ frequency range across the full range of $V_{S30}$ values. It also recovers the average amplification in the $f_P$ to $2 f_P$ frequency bin. This comparison, at the same time, highlights that along-depth variability has a more pronounced impact in amplifying high frequencies shown with the wider 16-84th percentiles.

Similar trends are observed in the GOF comparisons for the spatially varying model (Figure \ref{fig:sra_gof_svar_srlz_drlz}). Note that this model results in a reduction of the GOF range by approximately $10\%$ for both the $0.01$ to $f_P$ and $f_P$ to $2 f_P$ frequency bins. 

Higher modes, represented by the $2 f_P$ to 10 Hz frequency bin, show an overestimation of amplification for soft to medium stiffness profiles ($V_{S30} < 400~\text{m/sec}$) in both the stationary and spatially varying models. However, for stiffer profiles, the spatially varying model achieves a better fit.
Since the proposed models are approximations of the background medium velocity, it is expected that they are unable to accurately capture modal shapes beyond the first mode.

A similar comparison, excluding along-depth variability, is provided in the electronic supplement (Figures S20 and S21). Comparing the two cases shows that incorporating along-depth variability increases the range of GOF scores within each profile in the medium- and high-frequency bins while also reducing over-amplification at high frequencies. This reduction is due to the scattering of high frequencies caused by small perturbations of the velocity profiles, demonstrating the importance of considering the along-depth variability in site-response analyses intended to develop statistical models.

The same evaluation using the USGS SFBA velocity profiles is shown in Figure \ref{fig:sra_gof_usgs}, which shows a wider scatter of GOF scores at low frequencies and an underestimation of near-surface amplification in the medium and high-frequency ranges. This underestimation is expected, given the coarseness of the USGS SFBA model (with a vertical resolution of $25$ m in the top $400$ m), which limits its ability to capture the softer layers near the surface. 

These results suggest that modifying the shallow layers using the proposed stationary or spatially varying model could potentially enhance ground motion predictions for a wide range of frequencies of engineering interest. Incorporating along-depth variability will further improve these predictions in the high-frequency regime. Three-dimensional validation simulations are necessary, as mentioned above, to validate this hypothesis. 

\section{Conclusions} \label{sec:conclusions}

Two data-driven velocity models conditioned on $V_{S30}$ were developed in this study using a set of velocity profiles collected from the San Francisco Bay Area. The first model follows a stationary assumption where $V_{S30}$ fully characterizes the median scaling. The second model relaxes this assumption by incorporating a spatially varying slope adjustment based on the profile's location, allowing for better capture of profile-specific effects.

The median profile scaling adopts the functional form by  \cite{Shi2018}, which defines the median $V_S$ versus depth as a function of the shear wave velocity at the surface ($V_{S0}$), slope ($k$), and curvature ($n$). In the stationary model, the parameters $k$ and $n$ are expressed as functions of $V_{S30}$, while in the spatially varying model, $k$ also varies with the profile's location. A sigmoid function is used for the $n - V_{S30}$ scaling, and a composite sigmoid and hinge function is used for the $k - V_{S30}$ scaling, with both $k$ and $n$ relationships sharing the same sigmoid scaling. This formulation minimizes trade-offs between slope and curvature while providing a more robust extrapolation beyond the data range than polynomial fits. New to previous studies, the $V_{S0} - V_{S30}$ scaling is computed analytically to satisfy the $V_{S30}$ constraint, reducing the number of unknown coefficients and ensuring that the input $V_{S30}$ matches that of the generated profiles.

The spatially varying model is derived from the stationary model to maintain consistent scaling in the absence of site-specific information. The slope adjustment ($\delta B_r$) is modeled as a zero-mean Gaussian process with a negative exponential kernel function, ensuring that the $V_S - z$ scaling smoothly transitions to capture profile-specific effects near measured profiles while following the global scaling elsewhere.

Residual analysis indicates that both the stationary and spatially varying models are unbiased with respect to depth and $V_{S30}$, with a small deviation observed at high $V_S$ values (above $1000$ m/sec). This deviation likely occurs when profiles extend into the underlying rock layers, which would require a different $V_S - z$ scaling. As the proposed models are intended for sedimentary layers, this misfit lies outside the scope of the current model. The profile-specific adjustments in the spatially varying model result in a $25\%$ reduction in residual standard deviation compared to the stationary model.

The spatial trends of the slope adjustment indicate that, for the same $V_{S30}$, profiles around the Bay Area estuary have gentler slopes (i.e., slower increase in $V_S$ with depth) compared to the global model. In contrast, profiles further inland in the San Francisco Peninsula, San Mateo, and South Bay regions show a faster increase in $V_S$ with depth. A classification of slope adjustments based on surface geology revealed no systematic trends, but greater variability was observed in younger formations.

Along-depth semivariogram analyses determined the correlation length to be approximately $10$ m for both the stationary and spatially varying models.

Developing data-driven interface criteria between the sedimentary models and the USGS SFBA model was unsuccessful. Therefore, it is currently recommended to use the sedimentary models until a shear-wave velocity of $1000$ m/sec is reached, then transition to the USGS SFBA velocity model. To prevent negative impedance in the transition region, the maximum of $1000$ m/sec and the USGS SFBA velocity should be used.

Updating the USGS SFBA velocity model with either the stationary or spatially varying model, using the $V_{S30}$ estimates from \cite{wills2015next} and the aforementioned termination rule, results in a stiffer velocity structure in the South Bay and Livermore Valley. These adjustments align with recent 3D simulations \citep{pinilla2024}, which observed over-amplification of seismic waves in these regions compared to observations.

Lastly, goodness-of-fit (GOF) comparisons with one-dimensional site-response analyses between the actual profiles (treated as reference conditions), the stationary model, the spatially varying model, and the USGS SFBA velocity model demonstrated that the proposed models capture site amplification more accurately, for frequencies up to twice the fundamental frequency of the profiles, as approximated by the quarter-wavelength method. Incorporating along-depth variability expands the GOF range within each profile at medium and high frequencies and reduces over-amplification at high frequencies due to scattering effects.

Future studies should focus on the transition between SVMs and community velocity models, emphasizing the basin edges.  Expanding the dataset of $V_S$ observations and developing a more advanced correlation structure that simultaneously accounts for horizontal and along-depth, as well as within and between geologic unit correlations, is expected to enhance near-surface velocity characterization further. Incorporating refined near-surface characterization and uncertainty in 3D earthquake simulations will provide valuable insights into regional seismic hazard assessments.

\section{\small Acknowledgments}
\small

This material is based upon work supported by the U.S. Geological Survey under Grant No. G21AP10518 and G21AP10448.
The views and conclusions contained in this paper are those of the authors and should not be interpreted as representing the opinions or policies of the U.S. Geological Survey. Mention of trade names or commercial products does not constitute their endorsement by the U.S. Geological Survey.

\section*{\small Data and Resources} \label{sec:data_and_resources}

The regression code and regression datasets are provided in \url{https://github.com/glavrentiadis/svm_sfba}
The statistical regressions were performed using the programming language Python and the statistical software STAN \citep{Stan2023}.
The site-response analyses were performed using the PySeismoSoil package \citep{pyseismosoil}.

\section*{\small Conflict of Interest/Disclosures} 
The authors declared no potential conflicts of interest with respect to the research, authorship, and/or publication of this article.

\bibliography{references_GL, references_other}

\newcommand{\noop}[1]{}
\begin{thebibliography}{39}
\providecommand{\natexlab}[1]{#1}
\providecommand{\url}[1]{\texttt{#1}}
\expandafter\ifx\csname urlstyle\endcsname\relax
  \providecommand{\doi}[1]{doi: #1}\else
  \providecommand{\doi}{doi: \begingroup \urlstyle{rm}\Url}\fi

\bibitem[Olsen et~al.(2006)Olsen, Day, Minster, Cui, Chourasia, Faerman, Moore, Maechling, and Jordan]{olsen2006strong}
KB~Olsen, SM~Day, JB~Minster, Yifeng Cui, Amit Chourasia, Marcio Faerman, Reagan Moore, Philip Maechling, and Thomas Jordan.
\newblock Strong shaking in los angeles expected from southern san andreas earthquake.
\newblock \emph{Geophysical Research Letters}, 33\penalty0 (7), 2006.

\bibitem[Graves and Pitarka(2010)]{graves2010broadband}
Robert~W Graves and Arben Pitarka.
\newblock Broadband ground-motion simulation using a hybrid approach.
\newblock \emph{Bulletin of the Seismological Society of America}, 100\penalty0 (5A):\penalty0 2095--2123, 2010.

\bibitem[Bielak et~al.(2010)Bielak, Graves, Olsen, Taborda, Ram{\'\i}rez-Guzm{\'a}n, Day, Ely, Roten, Jordan, Maechling, et~al.]{bielak2010shakeout}
Jacobo Bielak, Robert~W Graves, Kim~B Olsen, Ricardo Taborda, Leonardo Ram{\'\i}rez-Guzm{\'a}n, Steven~M Day, Geoffrey~P Ely, Daniel Roten, Thomas~H Jordan, Philip~J Maechling, et~al.
\newblock The shakeout earthquake scenario: Verification of three simulation sets.
\newblock \emph{Geophysical Journal International}, 180\penalty0 (1):\penalty0 375--404, 2010.

\bibitem[Taborda and Bielak(2013)]{taborda2013ground}
Ricardo Taborda and Jacobo Bielak.
\newblock Ground-motion simulation and validation of the 2008 chino hills, california, earthquake.
\newblock \emph{Bulletin of the Seismological Society of America}, 103\penalty0 (1):\penalty0 131--156, 2013.

\bibitem[Taborda and Bielak(2014)]{taborda2014ground}
Ricardo Taborda and Jacobo Bielak.
\newblock Ground-motion simulation and validation of the 2008 chino hills, california, earthquake using different velocity models.
\newblock \emph{Bulletin of the Seismological Society of America}, 104\penalty0 (4):\penalty0 1876--1898, 2014.

\bibitem[Bielak et~al.(2016)Bielak, Taborda, Olsen, Graves, Silva, Khoshnevis, Savran, Roten, Shi, Goulet, et~al.]{bielak2016verification}
J~Bielak, R~Taborda, KB~Olsen, RW~Graves, F~Silva, N~Khoshnevis, WH~Savran, D~Roten, Z~Shi, CA~Goulet, et~al.
\newblock Verification and validation of high-frequency (fmax= 5 hz) ground motion simulations of the 2014 m 5.1 la habra, california, earthquake.
\newblock In \emph{AGU Fall Meeting Abstracts}, 2016.

\bibitem[Pinilla-Ramos et~al.(2024)Pinilla-Ramos, Pitarka, McCallen, and Nakata]{pinilla2024}
Camilo Pinilla-Ramos, Arben Pitarka, David McCallen, and Rie Nakata.
\newblock Performance evaluation of the usgs velocity model for the san francisco bay area.
\newblock \emph{Earthquake Spectra}, 2024.

\bibitem[McCallen et~al.(2024)McCallen, Pitarka, Tang, Pankajakshan, Petersson, Miah, and Huang]{mccallen2024regional}
David McCallen, Arben Pitarka, Houjun Tang, Ramesh Pankajakshan, N~Anders Petersson, Mamun Miah, and Junfei Huang.
\newblock Regional-scale fault-to-structure earthquake simulations with the eqsim framework: Workflow maturation and computational performance on gpu-accelerated exascale platforms.
\newblock \emph{Earthquake Spectra}, page 87552930241246235, 2024.

\bibitem[Shi and Day(2013)]{shi2013rupture}
Zheqiang Shi and Steven~M Day.
\newblock Rupture dynamics and ground motion from 3-d rough-fault simulations.
\newblock \emph{Journal of Geophysical Research: Solid Earth}, 118\penalty0 (3):\penalty0 1122--1141, 2013.

\bibitem[Taborda et~al.(2012)Taborda, Bielak, and Restrepo]{taborda2012earthquake}
Ricardo Taborda, Jacobo Bielak, and Doriam Restrepo.
\newblock Earthquake ground-motion simulation including nonlinear soil effects under idealized conditions with application to two case studies.
\newblock \emph{Seismological Research Letters}, 83\penalty0 (6):\penalty0 1047--1060, 2012.

\bibitem[Roten et~al.(2014)Roten, Olsen, Day, Cui, and F{\"a}h]{roten2014expected}
D~Roten, KB~Olsen, SM~Day, Y~Cui, and D~F{\"a}h.
\newblock Expected seismic shaking in los angeles reduced by san andreas fault zone plasticity.
\newblock \emph{Geophysical Research Letters}, 41\penalty0 (8):\penalty0 2769--2777, 2014.

\bibitem[Seylabi et~al.(2019)Seylabi, Stuart, and Asimaki]{elnaz}
Elnaz Seylabi, Andrew Stuart, and Domniki Asimaki.
\newblock Data fusion and assimilation framework for site characterization.
\newblock \emph{To be submitted to Bulletin of the Seismological Society of America}, 2019.

\bibitem[Savran and Olsen(2016)]{savran2016model}
WH~Savran and KB~Olsen.
\newblock Model for small-scale crustal heterogeneity in los angeles basin based on inversion of sonic log data.
\newblock \emph{Geophysical Journal International}, 205\penalty0 (2):\penalty0 856--863, 2016.

\bibitem[Small et~al.(2017)Small, Gill, Maechling, Taborda, Callaghan, Jordan, Olsen, Ely, and Goulet]{small2017scec}
Patrick Small, David Gill, Philip~J Maechling, Ricardo Taborda, Scott Callaghan, Thomas~H Jordan, Kim~B Olsen, Geoffrey~P Ely, and Christine Goulet.
\newblock The scec unified community velocity model software framework.
\newblock \emph{Seismological Research Letters}, 88\penalty0 (6):\penalty0 1539--1552, 2017.

\bibitem[Ely et~al.(2010)Ely, Jordan, Small, and Maechling]{ely2010VS30}
Geoffrey~P Ely, TH~Jordan, Patrick Small, and Philip~J Maechling.
\newblock A vs30-derived nearsurface seismic velocity model.
\newblock In \emph{Abstract S51A-1907, Fall Meeting}. AGU San Francisco, CA, 2010.

\bibitem[Shi and Asimaki(2018)]{Shi2018}
Jian Shi and Domniki Asimaki.
\newblock {A generic velocity profile for basin sediments in California conditioned on VS30}.
\newblock \emph{Seismological Research Letters}, 89\penalty0 (4):\penalty0 1397--1409, 2018.
\newblock ISSN 19382057.
\newblock \doi{10.1785/0220170268}.

\bibitem[Taborda et~al.(2016)Taborda, Azizzadeh-Roodpish, Khoshnevis, and Cheng]{taborda2016}
Ricardo Taborda, Shima Azizzadeh-Roodpish, Naeem Khoshnevis, and Keli Cheng.
\newblock Evaluation of the southern california seismic velocity models through simulation of recorded events.
\newblock \emph{Geophysical Journal International}, 205\penalty0 (3):\penalty0 1342--1364, 2016.

\bibitem[Marafi et~al.(2021)Marafi, Grant, Maurer, Rateria, Eberhard, and Berman]{marafi2021generic}
Nasser~A Marafi, Alex Grant, Brett~W Maurer, Gunjan Rateria, Marc~O Eberhard, and Jeffrey~W Berman.
\newblock A generic soil velocity model that accounts for near-surface conditions and deeper geologic structure.
\newblock \emph{Soil Dynamics and Earthquake Engineering}, 140:\penalty0 106461, 2021.

\bibitem[Aagaard et~al.(2020)Aagaard, Graymer, Thurber, Rodgers, Taira, Catchings, Goulet, and Plesch]{scienceplan}
Brad~T. Aagaard, Russell~W. Graymer, Clifford~H. Thurber, Arthur~J. Rodgers, Taka'aki Taira, Rufus~D. Catchings, Christine~A. Goulet, and Andreas Plesch.
\newblock Science plan for improving three-dimensional seismic velocity models in the san francisco bay region, 2019--24.
\newblock Technical report, 2020.
\newblock URL \url{http://pubs.er.usgs.gov/publication/ofr20201019}.

\bibitem[Aagaard et~al.(2008{\natexlab{a}})Aagaard, Brocher, Dolenc, Dreger, Graves, Harmsen, Hartzell, Larsen, McCandless, Nilsson, et~al.]{aagaard2008ground_a}
Brad~T Aagaard, Thomas~M Brocher, David Dolenc, Douglas Dreger, Robert~W Graves, Stephen Harmsen, Stephen Hartzell, Shawn Larsen, Kathleen McCandless, Stefan Nilsson, et~al.
\newblock Ground-motion modeling of the 1906 san francisco earthquake, part ii: Ground-motion estimates for the 1906 earthquake and scenario eventsground-motion modeling of the 1906 san francisco earthquake, part ii.
\newblock \emph{Bulletin of the Seismological Society of America}, 98\penalty0 (2):\penalty0 1012--1046, 2008{\natexlab{a}}.

\bibitem[Aagaard et~al.(2008{\natexlab{b}})Aagaard, Brocher, Dolenc, Dreger, Graves, Harmsen, Hartzell, Larsen, and Zoback]{aagaard2008ground_b}
Brad~T Aagaard, Thomas~M Brocher, David Dolenc, Douglas Dreger, Robert~W Graves, Stephen Harmsen, Stephen Hartzell, Shawn Larsen, and Mary~Lou Zoback.
\newblock Ground-motion modeling of the 1906 san francisco earthquake, part i: Validation using the 1989 loma prieta earthquake.
\newblock \emph{Bulletin of the Seismological Society of America}, 98\penalty0 (2):\penalty0 989--1011, 2008{\natexlab{b}}.

\bibitem[Hirakawa and Aagaard(2022)]{hirakawa2022evaluation}
Evan Hirakawa and Brad Aagaard.
\newblock Evaluation and updates for the usgs san francisco bay region 3d seismic velocity model in the east and north bay portions.
\newblock \emph{Bulletin of the Seismological Society of America}, 112\penalty0 (4):\penalty0 2070--2096, 2022.

\bibitem[Tehrani et~al.(2023)Tehrani, Lavrentiadis, Seylabi, McCallen, and Asimaki]{tehrani2023}
Hesam Tehrani, Grigorios Lavrentiadis, Elnaz Seylabi, David McCallen, and Domniki Asimaki.
\newblock Final technical report (2021-2022) towards a three-dimensional geotechnical layer model for northern california collaborative research with the university of nevada reno and california institute of technology.
\newblock 2023.

\bibitem[Yong et~al.(2013)Yong, Martin, Stokoe, and Diehl]{yong2013arra}
Alan Yong, Antony Martin, Kenneth Stokoe, and John Diehl.
\newblock Arra-funded v s30 measurements using multi-technique approach at strong-motion stations in california and central-eastern united states.
\newblock Technical report, US Geological Survey, 2013.

\bibitem[Boore(2003)]{boore2003compendium}
David~M Boore.
\newblock A compendium of p-and s-wave velocities from surface-to-borehole logging; summary and reanalysis of previously published data and analysis of unpublished data.
\newblock Technical report, US Geological Survey, 2003.

\bibitem[Kwak et~al.(2021)Kwak, Ahdi, Wang, Zimmaro, Brandenberg, Stewart, et~al.]{kwak2021web}
Dongyoup Kwak, Sean~K Ahdi, Pengfei Wang, Paolo Zimmaro, Scott~J Brandenberg, Jonathan~P Stewart, et~al.
\newblock Web portal for shear wave velocity and hvsr databases in support of site response research and applications.
\newblock 2021.

\bibitem[Boore and Brown(1998)]{boore1998comparing}
David~M Boore and Leo~T Brown.
\newblock Comparing shear-wave velocity profiles from inversion of surface-wave phase velocities with downhole measurements: systematic differences between the cxw method and downhole measurements at six usc strong-motion sites.
\newblock \emph{Seismological Research Letters}, 69:\penalty0 222--229, 1998.

\bibitem[Boore and Asten(2008)]{boore2008comparisons}
David~M Boore and Michael~W Asten.
\newblock Comparisons of shear-wave slowness in the santa clara valley, california, using blind interpretations of data from invasive and noninvasive methods.
\newblock \emph{Bulletin of the Seismological Society of America}, 98\penalty0 (4):\penalty0 1983--2003, 2008.

\bibitem[Brown et~al.(2002)Brown, Boore, and Stokoe]{brown2002comparison}
Leo~T Brown, David~M Boore, and Kenneth~H Stokoe.
\newblock Comparison of shear-wave slowness profiles at 10 strong-motion sites from noninvasive sasw measurements and measurements made in boreholes.
\newblock \emph{Bulletin of the Seismological Society of America}, 92\penalty0 (8):\penalty0 3116--3133, 2002.

\bibitem[Rix et~al.(2002)Rix, Hebeler, and Orozco]{rix2002near}
Glenn~J Rix, Gregory~L Hebeler, and M~Catalina Orozco.
\newblock Near-surface vs profiling in the new madrid seismic zone using surface-wave methods.
\newblock \emph{Seismological Research Letters}, 73\penalty0 (3):\penalty0 380--392, 2002.

\bibitem[Stephenson et~al.(2005)Stephenson, Louie, Pullammanappallil, Williams, and Odum]{stephenson2005blind}
William~J Stephenson, John~N Louie, S~Pullammanappallil, RA~Williams, and Jackson~K Odum.
\newblock Blind shear-wave velocity comparison of remi and masw results with boreholes to 200 m in santa clara valley: implications for earthquake ground-motion assessment.
\newblock \emph{Bulletin of the Seismological Society of America}, 95\penalty0 (6):\penalty0 2506--2516, 2005.

\bibitem[Bas et~al.(2022)Bas, Seylabi, Yong, Tehrani, and Asimaki]{bas2022p}
Elif~Ecem Bas, Elnaz Seylabi, Alan Yong, Hesam Tehrani, and Domniki Asimaki.
\newblock P-and s-wave velocity estimation by ensemble kalman inversion of dispersion data for strong motion stations in california.
\newblock \emph{Geophysical Journal International}, 231\penalty0 (1):\penalty0 536--551, 2022.

\bibitem[{Stan Development Team}(2023)]{Stan2023}
{Stan Development Team}.
\newblock {The Stan Core Library}, 2023.
\newblock URL \url{http://mc-stan.org/}.
\newblock Version 2.33.0.

\bibitem[Assimaki et~al.(2014)Assimaki, Shi, and Yong]{asimaki2014}
D.~Assimaki, J.~Shi, and A.~Yong.
\newblock Site-specific amplification factors at strong motion stations in western us: Collaborative effort between gatech and the usgs pasadena office, {Final Report NEHRP Grant Award G11AP20052:02/01/2011-07/31/2012}.
\newblock Technical report, 2014.

\bibitem[Simpson et~al.(2017)Simpson, Rue, Riebler, Martins, and S{\o}rbye]{Simpson2017}
Daniel Simpson, Håvard Rue, Andrea Riebler, Thiago~G. Martins, and Sigrunn~H. S{\o}rbye.
\newblock {Penalising model component complexity: A principled, practical approach to constructing priors}.
\newblock \emph{Statistical Science}, 32\penalty0 (1):\penalty0 1--28, 2017.
\newblock ISSN 08834237.
\newblock \doi{10.1214/16-STS576}.

\bibitem[Lavrentiadis et~al.(2021)Lavrentiadis, Abrahamson, and Kuehn]{Lavrentiadis2021}
Grigorios Lavrentiadis, Norman~A. Abrahamson, and Nicolas~M. Kuehn.
\newblock {A non-ergodic effective amplitude ground-motion model for California}.
\newblock \emph{Bulletin of Earthquake Engineering}, \penalty0 (0123456789), 9 2021.
\newblock ISSN 1570-761X.
\newblock \doi{10.1007/s10518-021-01206-w}.
\newblock URL \url{https://link.springer.com/10.1007/s10518-021-01206-w}.

\bibitem[Wills et~al.(2015)Wills, Gutierrez, Perez, and Branum]{wills2015next}
CJ~Wills, CI~Gutierrez, FG~Perez, and DM~Branum.
\newblock A next generation vs30 map for california based on geology and topographya next generation vs30 map for california based on geology and topography.
\newblock \emph{Bulletin of the Seismological Society of America}, 105\penalty0 (6):\penalty0 3083--3091, 2015.

\bibitem[Shi and Asimaki(2017)]{shi2017stiffness}
Jian Shi and Domniki Asimaki.
\newblock From stiffness to strength: Formulation and validation of a hybrid hyperbolic nonlinear soil model for site-response analyses.
\newblock \emph{Bulletin of the Seismological Society of America}, 107\penalty0 (3):\penalty0 1336--1355, 2017.

\bibitem[Shi et~al.(2024)Shi, Asimaki, and Li]{pyseismosoil}
Jian Shi, Domniki Asimaki, and Wei Li.
\newblock Caltech-geoquake/pyseismosoil: v0.5.4, February 2024.
\newblock URL \url{https://doi.org/10.5281/zenodo.10652844}.

\end{thebibliography}

\end{document}


\maketitle

\begin{figure}[htbp!]
    \centering
    \begin{subfigure}[t]{0.32\textwidth}
        \caption{}
        \includegraphics[height = 1\textwidth]{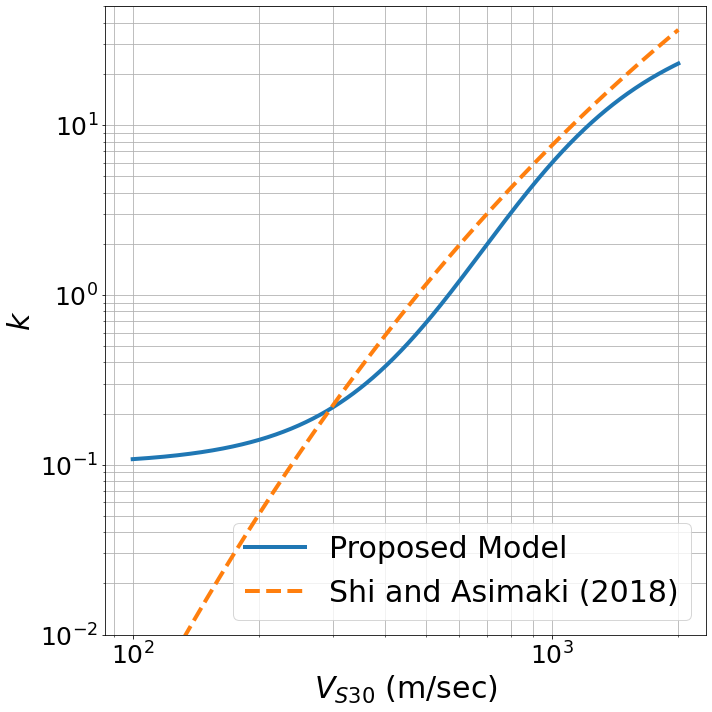}
    \end{subfigure}  
    \hfill
    \begin{subfigure}[t]{0.32\textwidth}
        \caption{}
        \includegraphics[height = 1\textwidth]{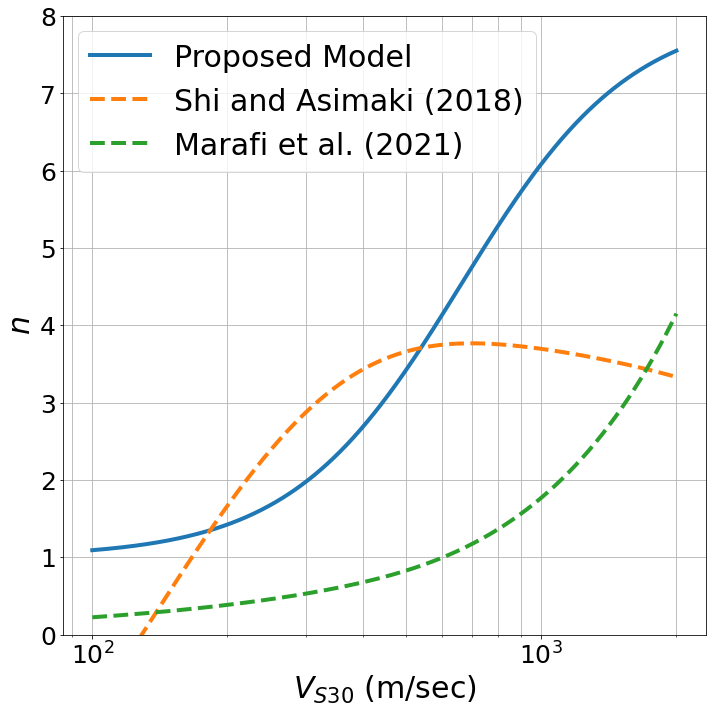}
    \end{subfigure} 
    \hfill
    \begin{subfigure}[t]{0.32\textwidth}
        \caption{}
        \includegraphics[height = 1\textwidth]{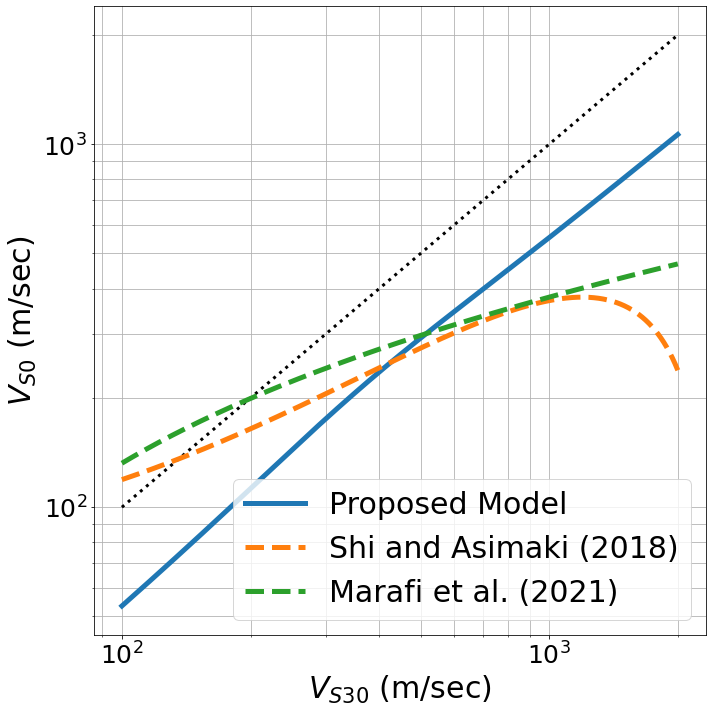}
    \end{subfigure}  
    \caption{Comparison of $k$, $n$, and $V_{S0}$ scaling between the proposed model, Shi and Asimaki (2018) and Marafi et al. (2021); the black dotted line in subfigure (c) corresponds to the one-to-one line; the region above this line corresponds to $V_{S0}$ greater than $V_{S30}$.}
    \label{esupp:fig:scl_comparision}
\end{figure}

Figure \ref{esupp:fig:scl_comparision} compares the $k$, $n$, and $V_{S0}$ scaling between the proposed model, Shi and Asimaki (2018), and Marafi et al. (2021). In Subfigure (c), the black dotted line represents the one-to-one line; the region above this line corresponds to $V_{S0}$ values greater than $V_{S30}$, indicating unphysical scaling.

\begin{figure}[htbp!]
    \centering
    \begin{subfigure}[t]{0.49\textwidth}
        \caption{}
        \includegraphics[width = 1\textwidth]{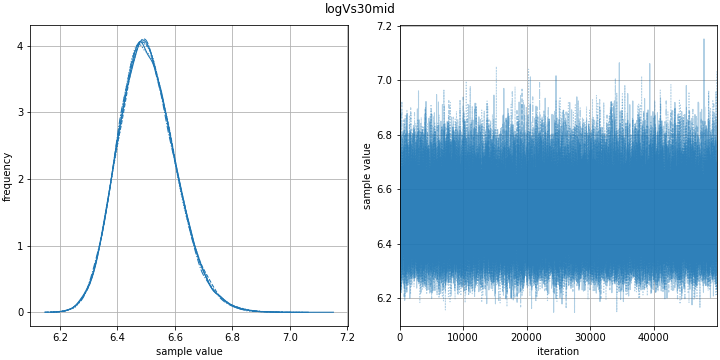}
    \end{subfigure}
    \begin{subfigure}[t]{0.49\textwidth}
        \caption{}
        \includegraphics[width = 1\textwidth]{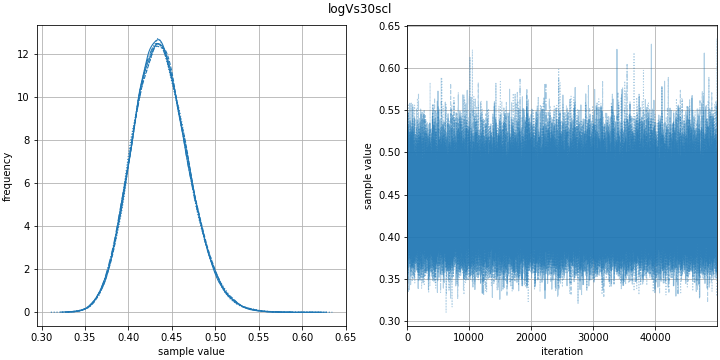}
    \end{subfigure}  
    \begin{subfigure}[t]{0.49\textwidth}
        \caption{}
        \includegraphics[width = 1\textwidth]{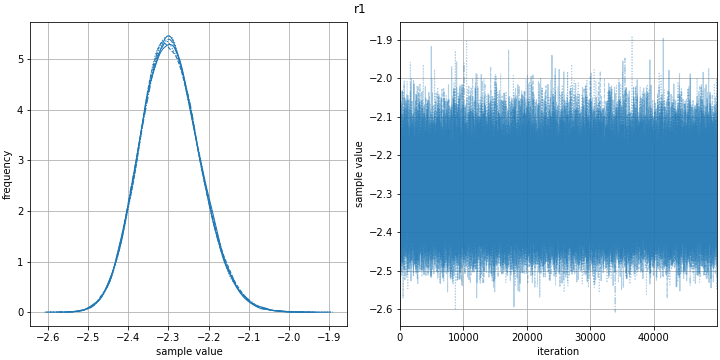}
    \end{subfigure}  
    \begin{subfigure}[t]{0.49\textwidth}
        \caption{}
        \includegraphics[width = 1\textwidth]{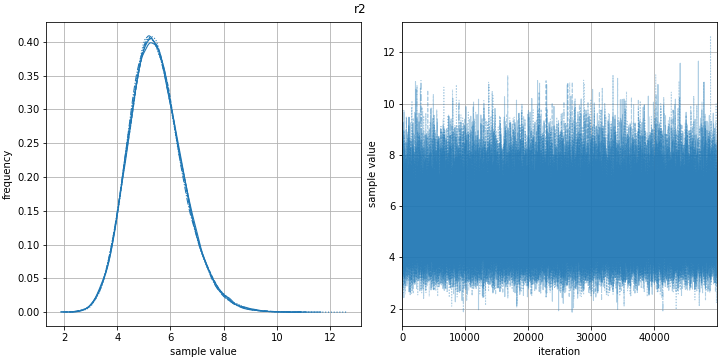}
    \end{subfigure} 
    \begin{subfigure}[t]{0.49\textwidth}
        \caption{}
        \includegraphics[width = 1\textwidth]{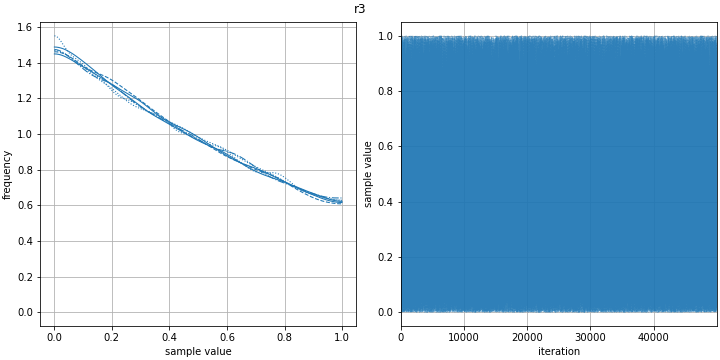}
    \end{subfigure}
    \begin{subfigure}[t]{0.49\textwidth}
        \caption{}
        \includegraphics[width = 1\textwidth]{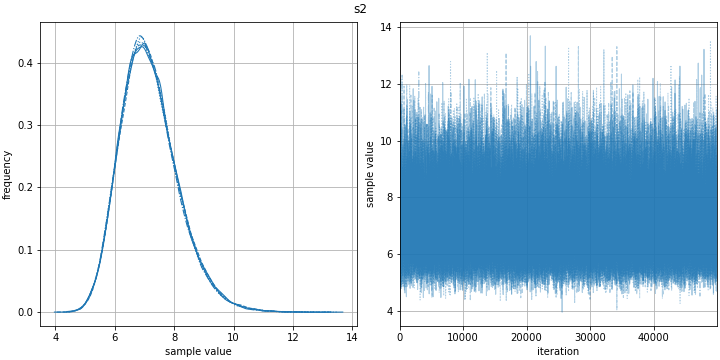}
    \end{subfigure} 
    \begin{subfigure}[t]{0.49\textwidth}
        \caption{}
        \includegraphics[width = 1\textwidth]{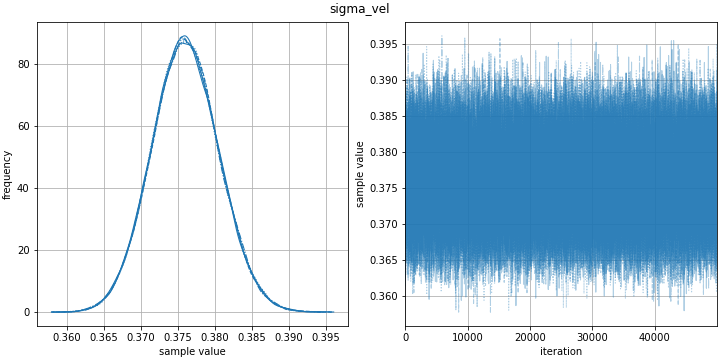}
    \end{subfigure}
    \caption{Monte Carlo Markov Chain for posterior distributions of stationary model for:\\
    (a) $V_{S30ref}$,
    (b) $V_{S30w}$,
    (c) $r_1$,
    (d) $r_2$, 
    (e) $r_3$,
    (f) $s_2$, and
    (g) $\sigma_{st}$.}
    \label{esupp:fig:st_mcmc_chains}
\end{figure}

Figure \ref{esupp:fig:st_mcmc_chains} displays the Markov-chain Monte Carlo (MCMC) trace plots and posterior distributions for the hyperparameters of the stationary model.

\begin{figure}[htbp!]
    \centering
    \begin{subfigure}[t]{0.32\textwidth}
        \caption{}
        \includegraphics[height = 1\textwidth]{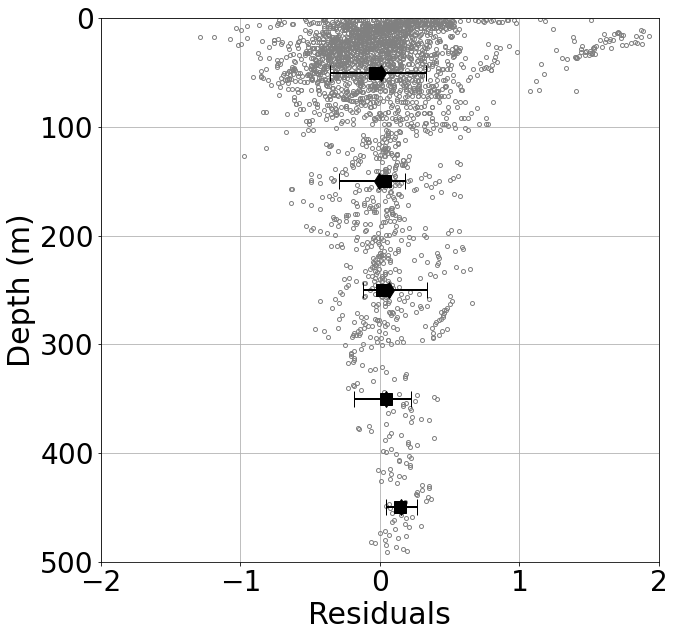}
    \end{subfigure}  
    \hfill
    \begin{subfigure}[t]{0.32\textwidth}
        \caption{}
        \includegraphics[height = 1\textwidth]{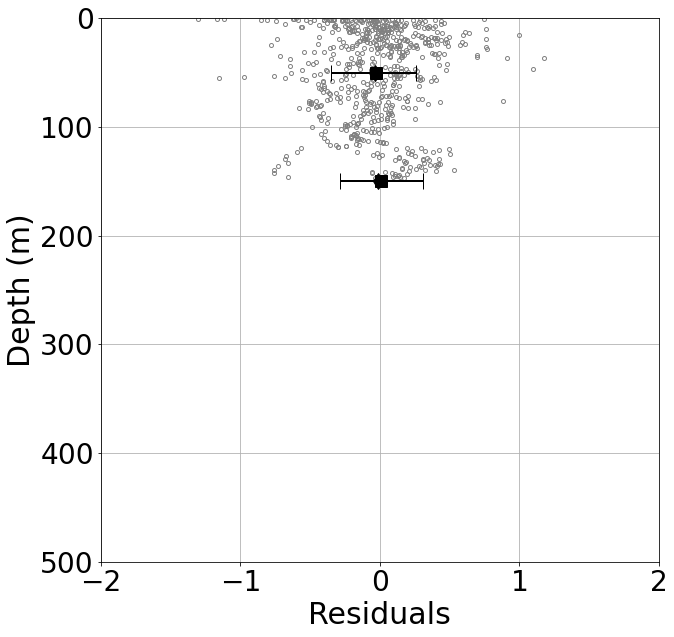}
    \end{subfigure} 
    \hfill
    \begin{subfigure}[t]{0.32\textwidth}
        \caption{}
        \includegraphics[height = 1\textwidth]{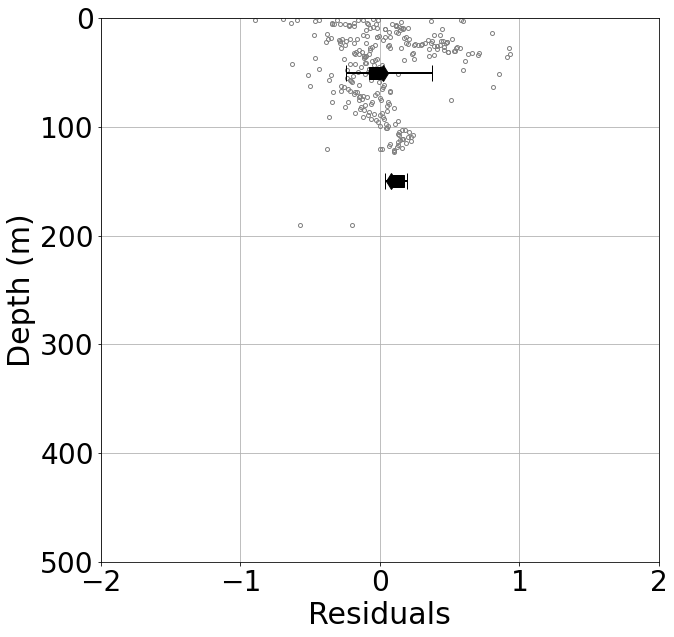}
    \end{subfigure}  
    \caption{Stationary model residuals versus depth for different $V_{S30}$ bins. 
    (a) $V_{S30} \in [100, 400)~\text{m/sec}$,
    (b) $V_{S30} \in [400, 800)~\text{m/sec}$,
    (c) $V_{S30} \in [800, 3000)~\text{m/sec}$.}
    \label{esupp:fig:st_scl_vs30_bin_res}
\end{figure}

\begin{figure}[htbp!]
    \centering
    \begin{subfigure}[t]{0.32\textwidth}
        \caption{}
        \includegraphics[height = 1\textwidth]{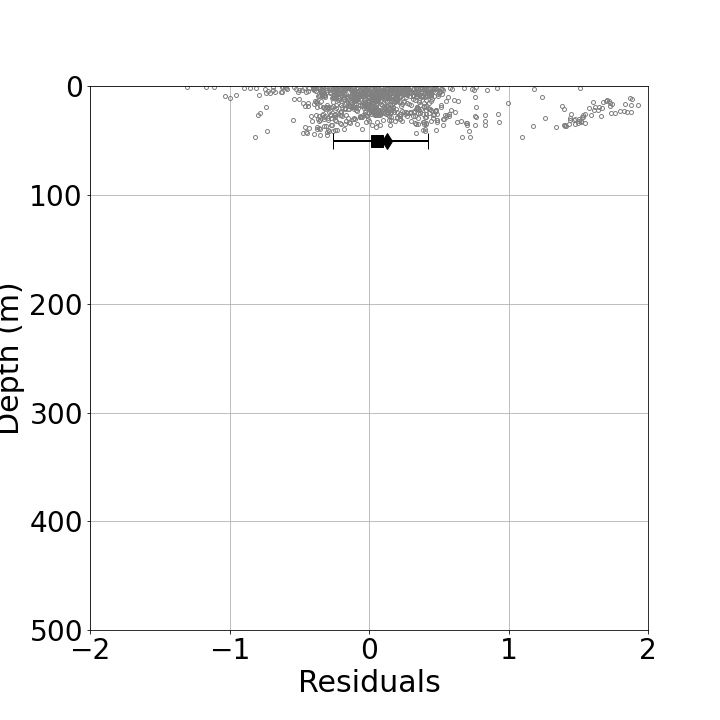}
    \end{subfigure}  
    \begin{subfigure}[t]{0.32\textwidth}
        \caption{}
        \includegraphics[height = 1\textwidth]{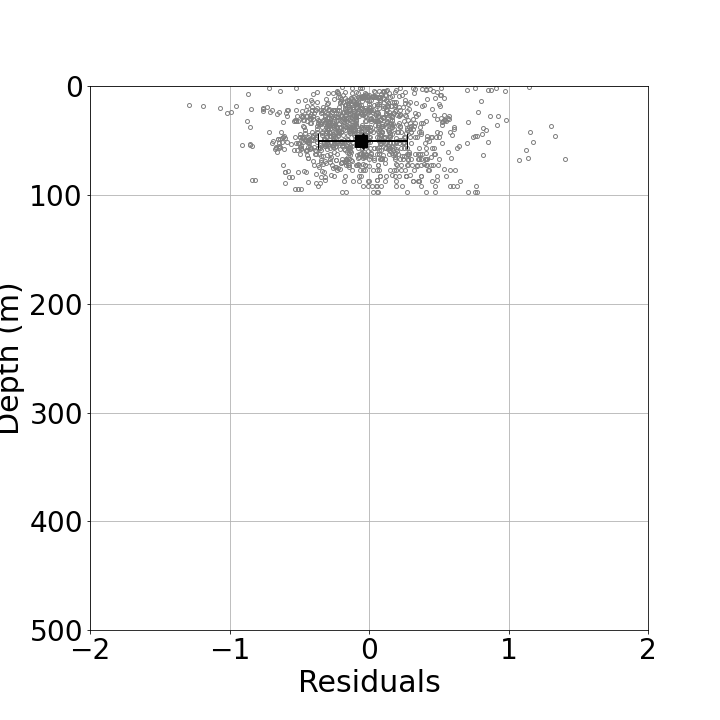}
    \end{subfigure} 
    \\
    \begin{subfigure}[t]{0.32\textwidth}
        \caption{}
        \includegraphics[height = 1\textwidth]{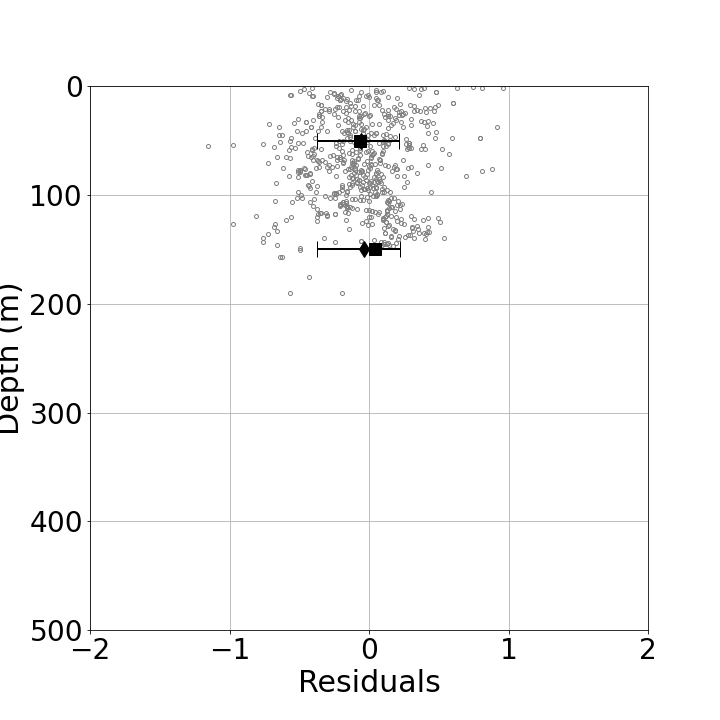}
    \end{subfigure}  
    \begin{subfigure}[t]{0.32\textwidth}
        \caption{}
        \includegraphics[height = 1\textwidth]{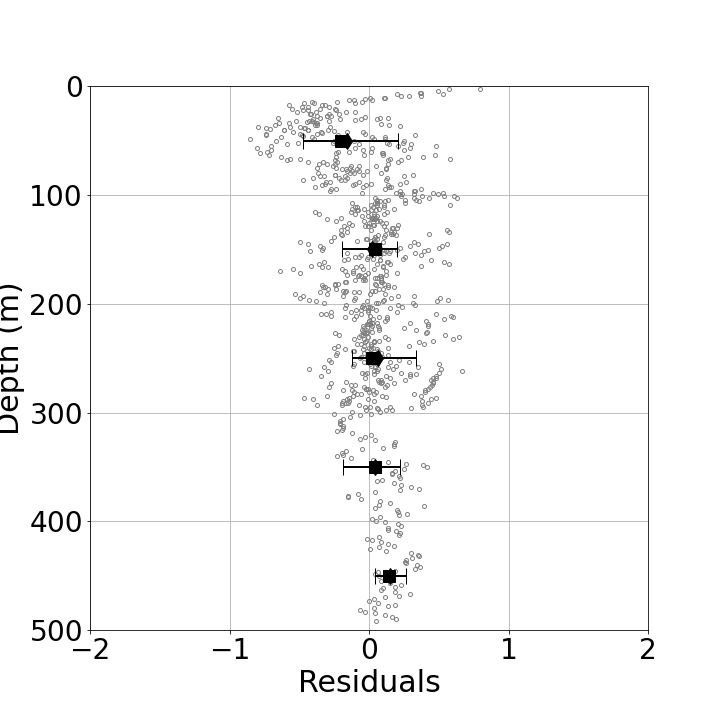}
    \end{subfigure}  
    \caption{Stationary model residuals versus depth for different velocity profile depths ($z_{\max}$) bins. 
    (a) $z_{\max} \in [0, 50)~\text{m/sec}$,
    (b) $z_{\max} \in [50, 100)~\text{m/sec}$,
    (c) $z_{\max} \in [100, 250)~\text{m/sec}$
    (d) $z_{\max} \in [250, 500)~\text{m/sec}$.}
    \label{esupp:fig:st_scl_depth_bin_res}
\end{figure}

The along-depth residuals of the stationary model are shown for different $V_{S30}$ bins in Figure \ref{esupp:fig:st_scl_vs30_bin_res}, and for various maximum profile depths in Figure \ref{esupp:fig:st_scl_depth_bin_res}.

\begin{figure}[htbp!]
    \centering
    \begin{subfigure}[t]{0.49\textwidth}
        \caption{}
        \includegraphics[width = 1\textwidth]{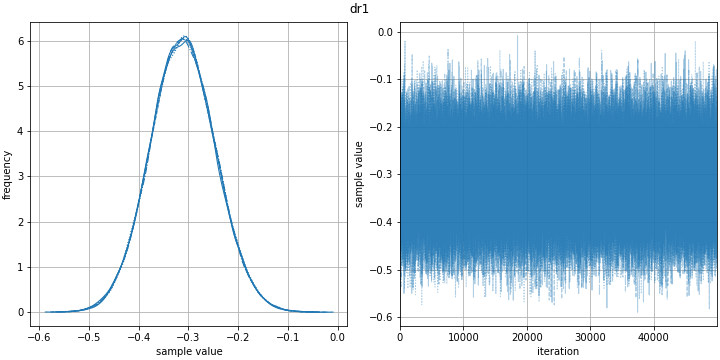}
    \end{subfigure}  
    \begin{subfigure}[t]{0.49\textwidth}
        \caption{}
        \includegraphics[width = 1\textwidth]{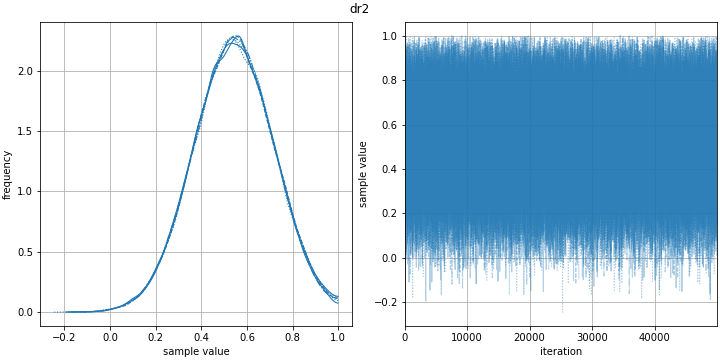}
    \end{subfigure} 
    \begin{subfigure}[t]{0.49\textwidth}
        \caption{}
        \includegraphics[width = 1\textwidth]{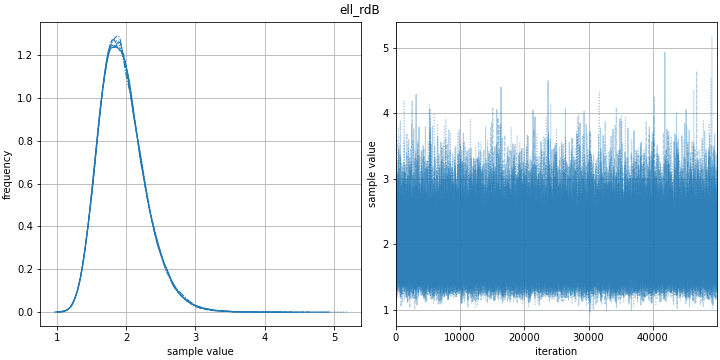}
    \end{subfigure}
    \begin{subfigure}[t]{0.49\textwidth}
        \caption{}
        \includegraphics[width = 1\textwidth]{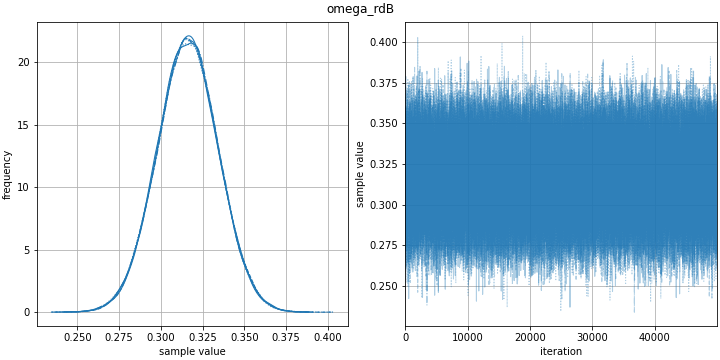}
    \end{subfigure} 
    \begin{subfigure}[t]{0.49\textwidth}
        \caption{}
        \includegraphics[width = 1\textwidth]{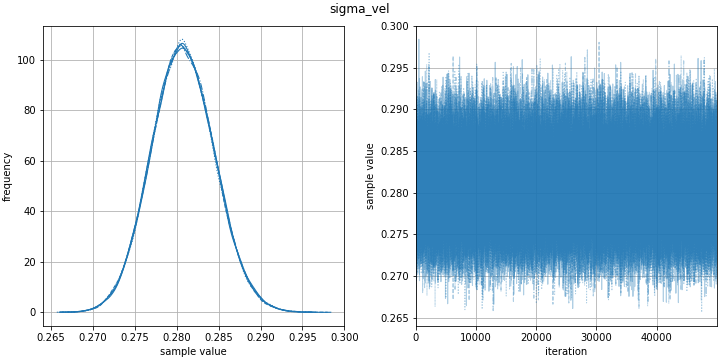}
    \end{subfigure}
    \caption{Monte Carlo Markov Chain for posterior distributions of spatially varying model for:\\
    (a) $\delta r_1$,
    (b) $\delta r_2$, 
    (c) $\ell_{\delta Br}$ 
    (d) $\omega_{\delta Br}$
    (g) $\sigma_{svar}$.}
    \label{esupp:fig:sv_mcmc_chains}
\end{figure}

Similarly, Figure \ref{esupp:fig:sv_mcmc_chains} presents the MCMC trace plots and posterior distributions for the hyperparameters of the spatially varying model.

\begin{figure}[htbp!]
    \centering
    \begin{subfigure}[t]{0.32\textwidth}
        \caption{}
        \includegraphics[height = 1\textwidth]{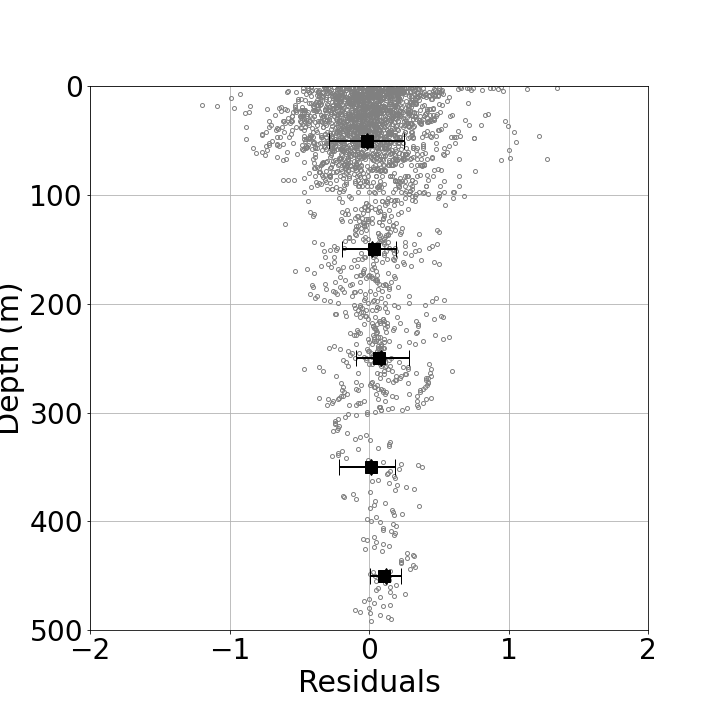}
    \end{subfigure}  
    \hfill
    \begin{subfigure}[t]{0.32\textwidth}
        \caption{}
        \includegraphics[height = 1\textwidth]{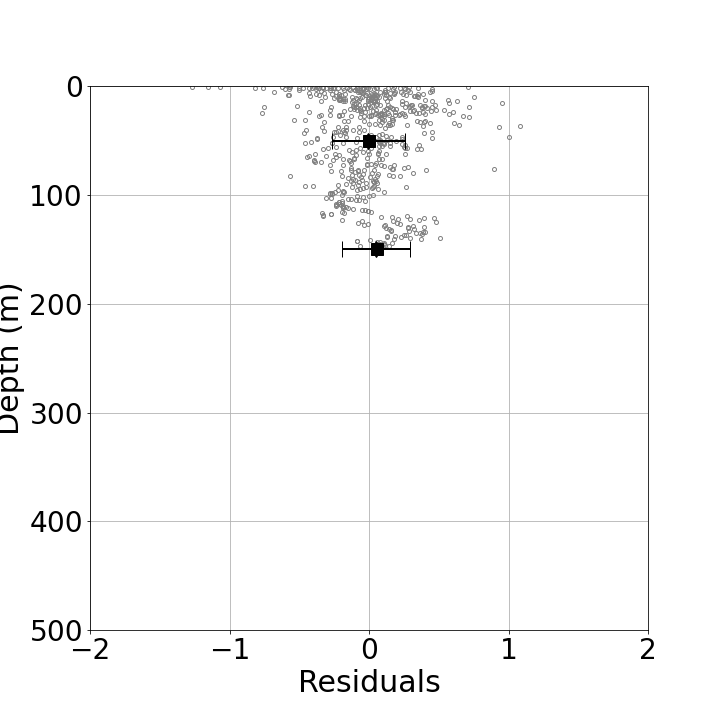}
    \end{subfigure} 
    \hfill
    \begin{subfigure}[t]{0.32\textwidth}
        \caption{}
        \includegraphics[height = 1\textwidth]{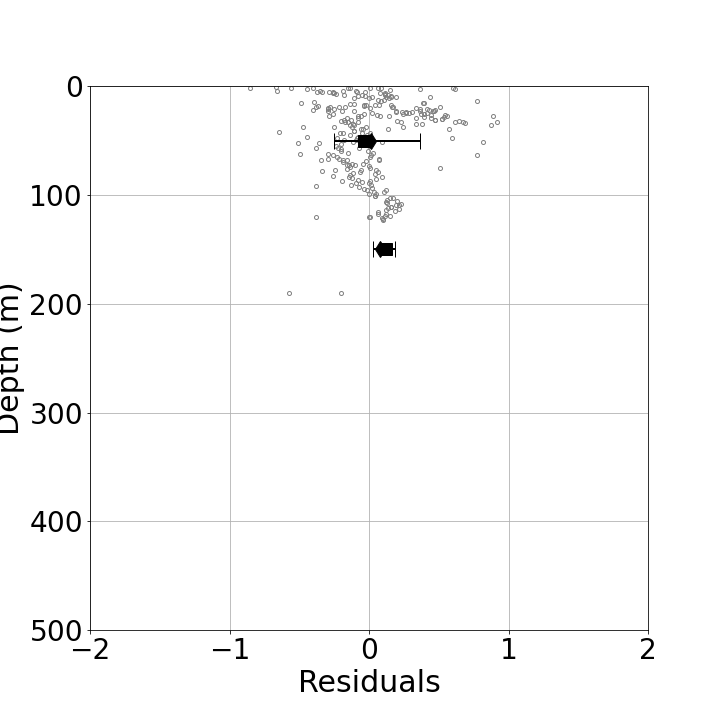}
    \end{subfigure}  
    \caption{Spatially varying model residuals versus depth for different $V_{S30}$ bins. 
    (a) $V_{S30} \in [100, 400)~\text{m/sec}$,
    (b) $V_{S30} \in [400, 800)~\text{m/sec}$,
    (c) $V_{S30} \in [800, 3000)~\text{m/sec}$.}
    \label{esupp:fig:sv_scl_vs30_bin_res}
\end{figure}

\begin{figure}[htbp!]
    \centering
    \begin{subfigure}[t]{0.32\textwidth}
        \caption{}
        \includegraphics[height = 1\textwidth]{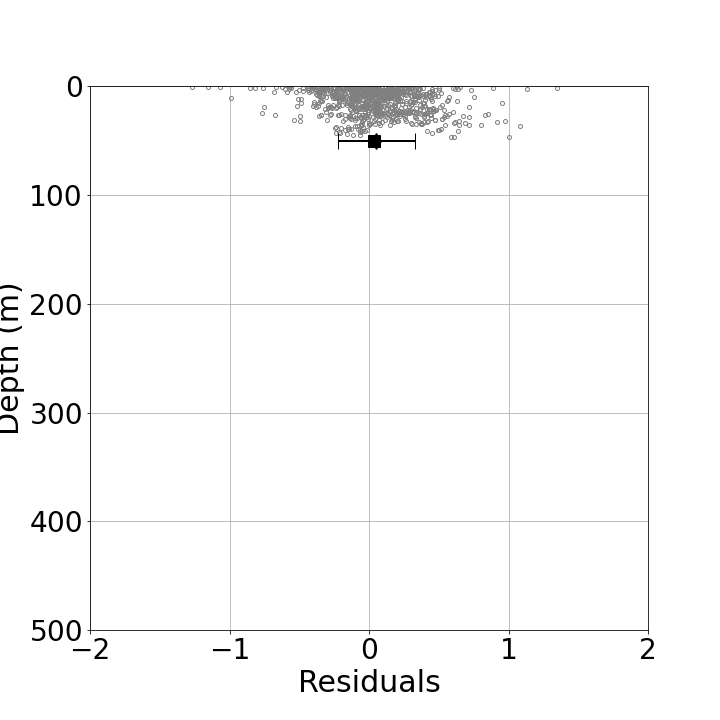}
    \end{subfigure} 
    \qquad
    \begin{subfigure}[t]{0.32\textwidth}
        \caption{}
        \includegraphics[height = 1\textwidth]{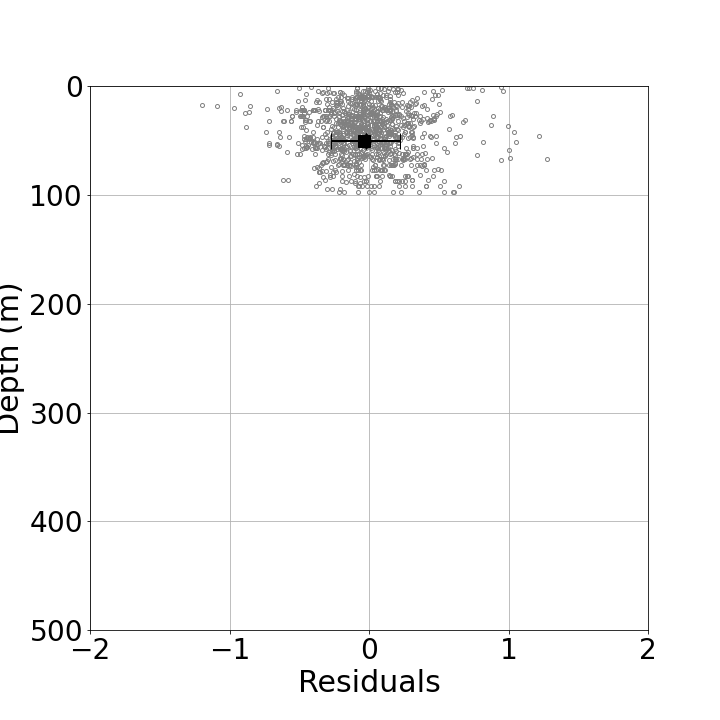}
    \end{subfigure} 
    \\
    \begin{subfigure}[t]{0.32\textwidth}
        \caption{}
        \includegraphics[height = 1\textwidth]{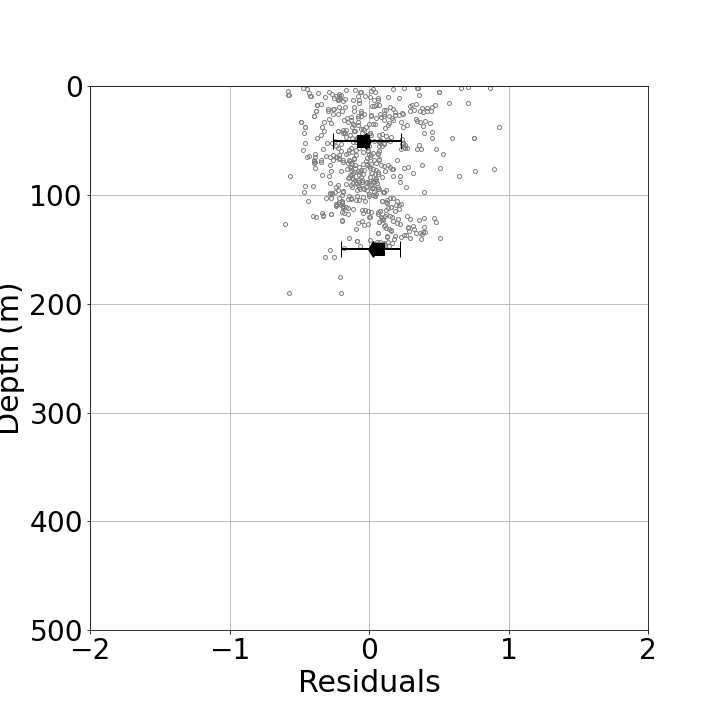}
    \end{subfigure}  
    \qquad
    \begin{subfigure}[t]{0.32\textwidth}
        \caption{}
        \includegraphics[height = 1\textwidth]{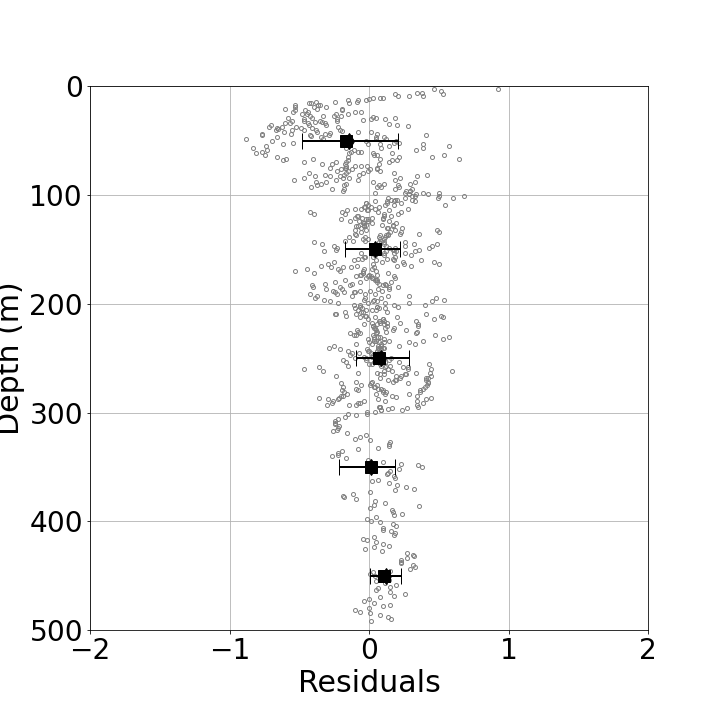}
    \end{subfigure}  
    \caption{Spatially varying model residuals versus depth for different velocity profile depths ($z_{max}$) bins. 
    (a) $z_{max} \in [0, 50)~\text{m/sec}$,
    (b) $z_{max} \in [50, 100)~\text{m/sec}$,
    (c) $z_{max} \in [100, 250)~\text{m/sec}$
    (d) $z_{max} \in [250, 500)~\text{m/sec}$.}
    \label{esupp:fig:sv_scl_depth_bin_res}
\end{figure}

The along-depth residuals of the spatially varying model are shown in Figure \ref{esupp:fig:sv_scl_vs30_bin_res} for different $V_{S30}$ bins and in Figure \ref{esupp:fig:sv_scl_depth_bin_res} for different maximum profile depths.

\begin{figure} 
    \centering
    \includegraphics[width= 0.85\textwidth]{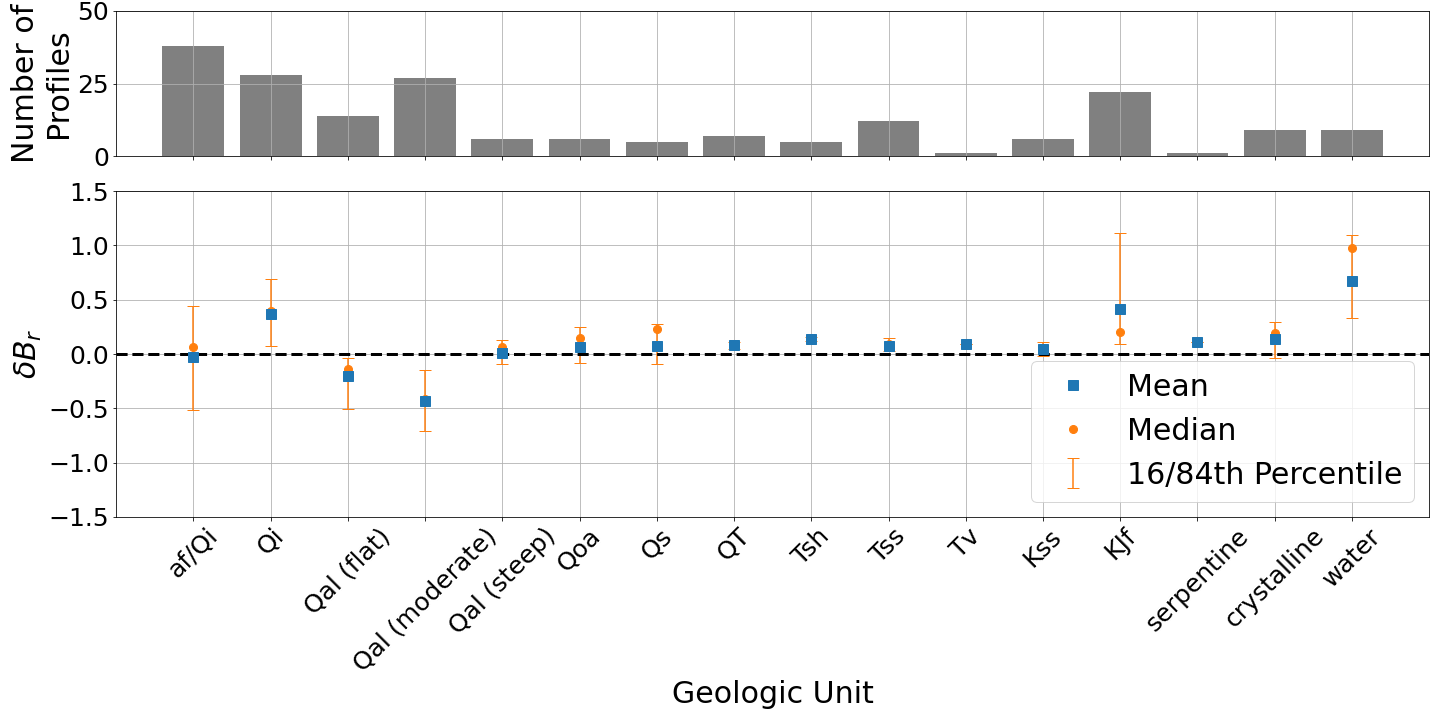}
    \caption{Distribution of the spatially varying slope adjustment term, $\delta B_r$, binned by the geologic unit; the square marker indicates the mean value, the circular marker represents the median value, and the error bars show the $16^{th}$ and $84^{th}$ percentiles.}
    \label{esupp:fig:sv_model_dBr_geol}
\end{figure}

Figure \ref{esupp:fig:sv_model_dBr_geol} illustrates the distribution of the spatially varying slope adjustment term, $\delta B_r$, categorized by geologic unit.

\begin{figure} 
    \centering
    \begin{subfigure}[t]{0.32\textwidth}
        \caption{}
        \includegraphics[height = 1\textwidth]{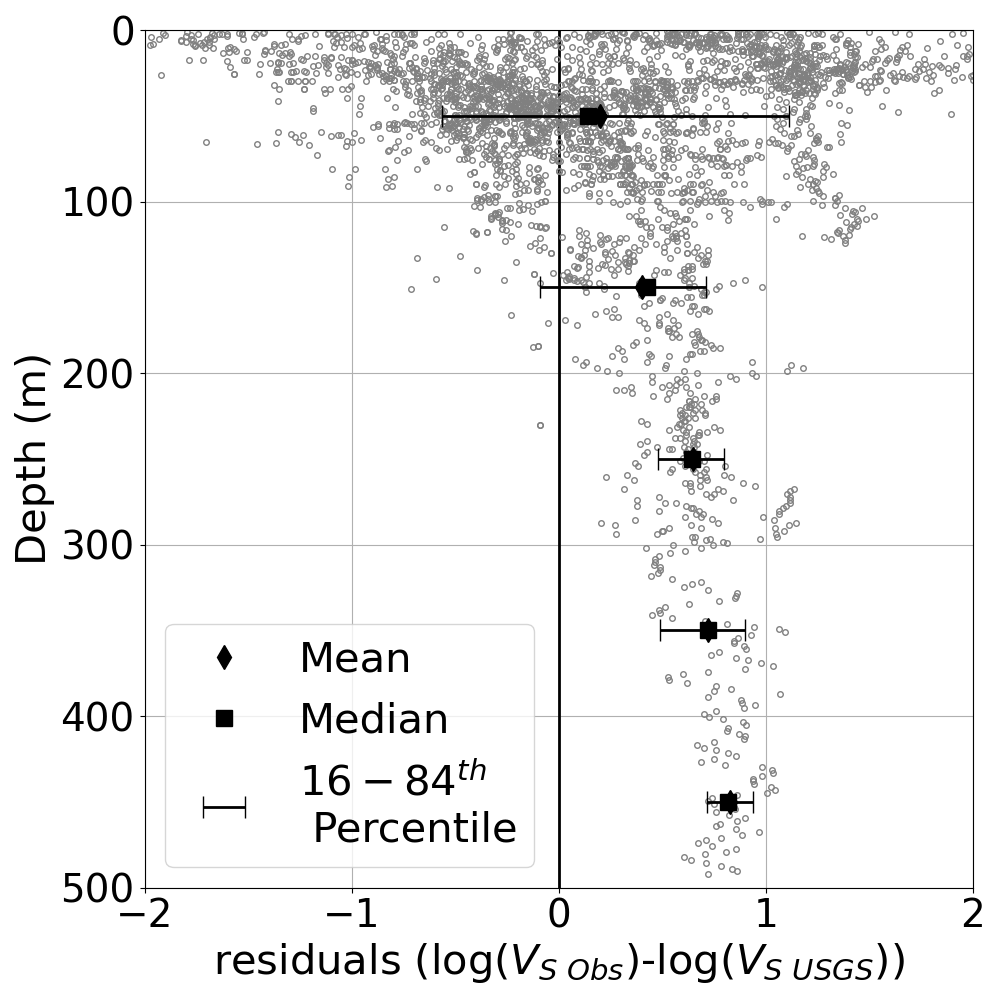}
    \end{subfigure}  
    \qquad
    \begin{subfigure}[t]{0.32\textwidth}
        \caption{}
        \includegraphics[height = 1\textwidth]{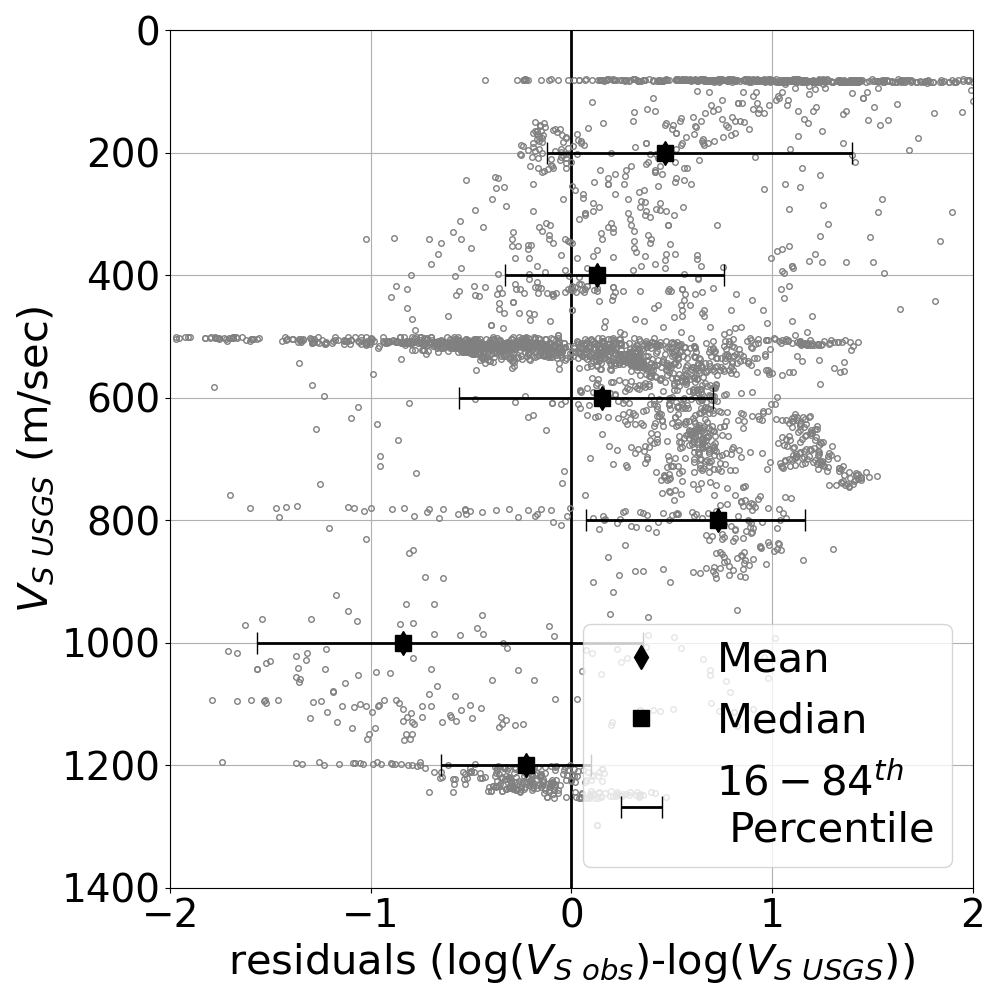}
    \end{subfigure} 
    \caption{Missfit between the $V_{S}$ observations and the USGS SFBA community velocity model; (a) observation - USGS misfit versus depth and (b) observation - USGS misfit versus layer's shear-wave velocity.}
    \label{esupp:fig:cmp_emp_usgs}
\end{figure}

\begin{figure}[htbp!]
    \centering
    \begin{subfigure}[t]{0.32\textwidth}
        \caption{}
        \includegraphics[height = 1\textwidth]{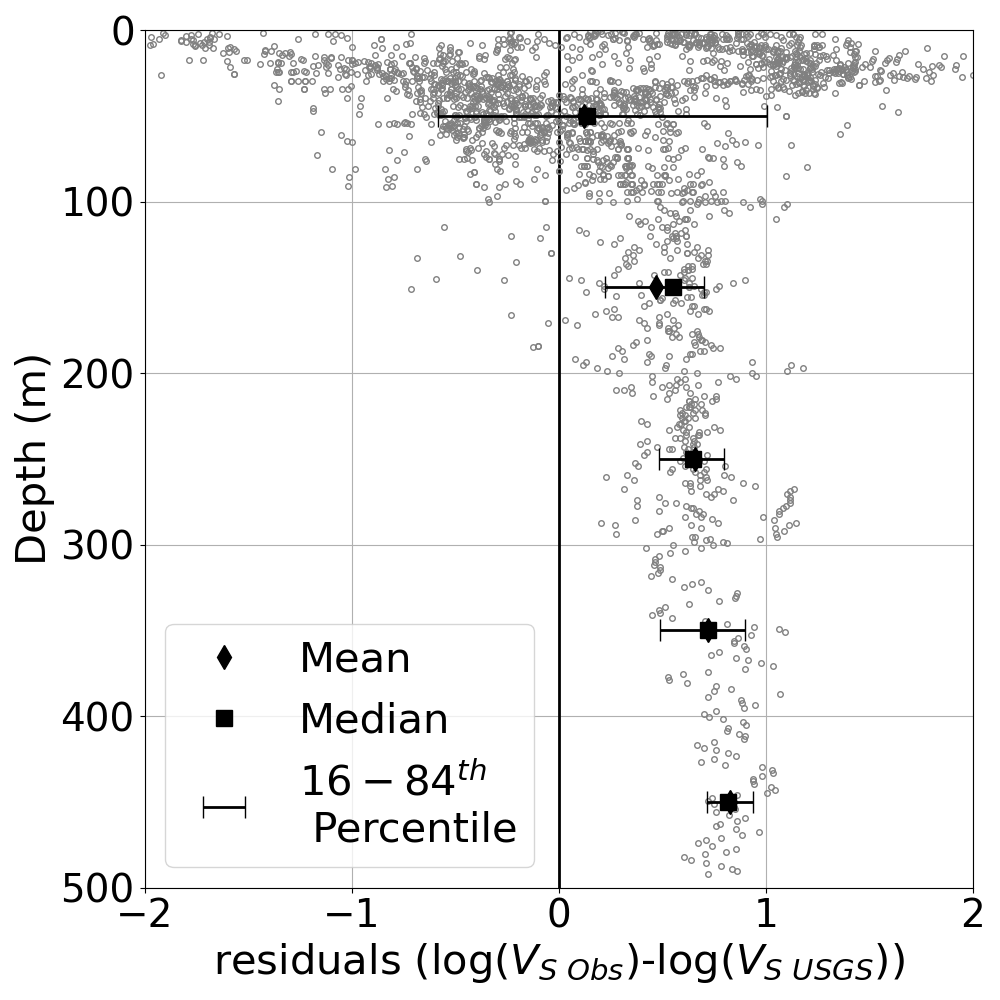}
    \end{subfigure} 
    \hfill
    \begin{subfigure}[t]{0.32\textwidth}
        \caption{}
        \includegraphics[height = 1\textwidth]{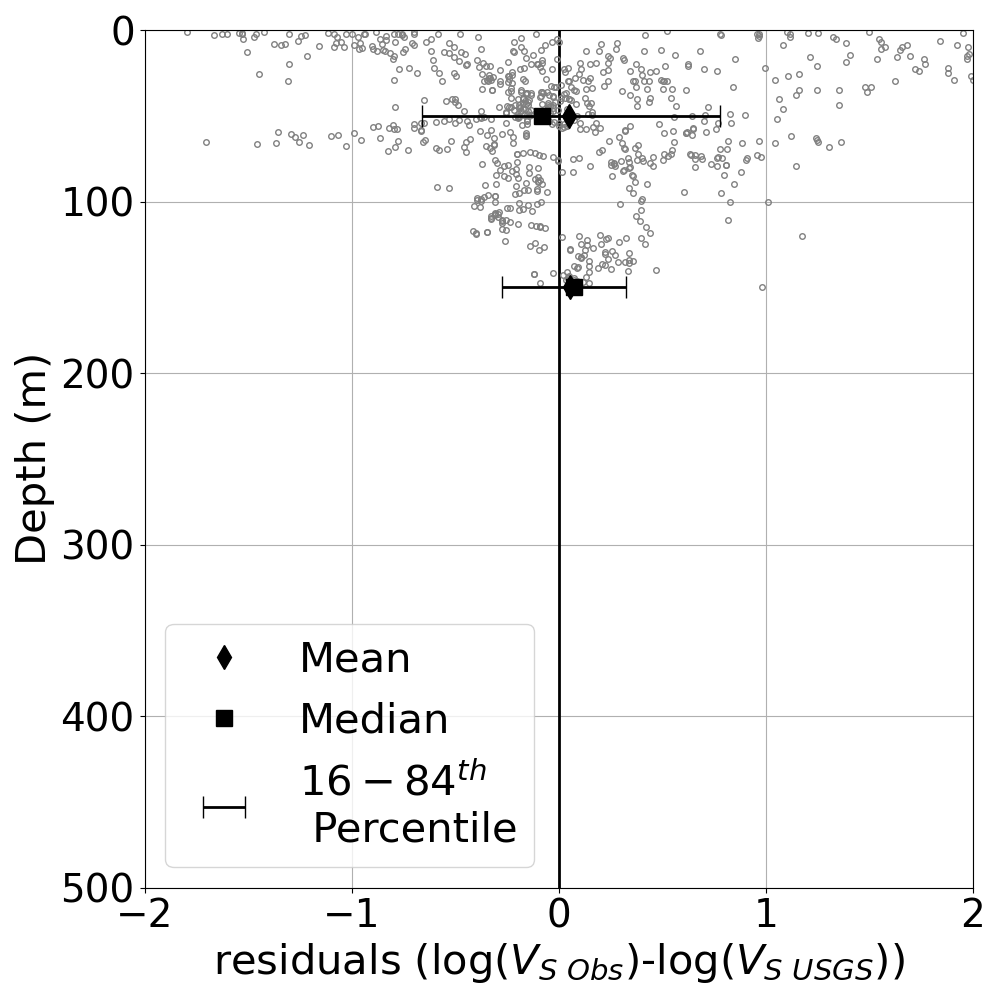}
    \end{subfigure}  
    \hfill
    \begin{subfigure}[t]{0.32\textwidth}
        \caption{}
        \includegraphics[height = 1\textwidth]{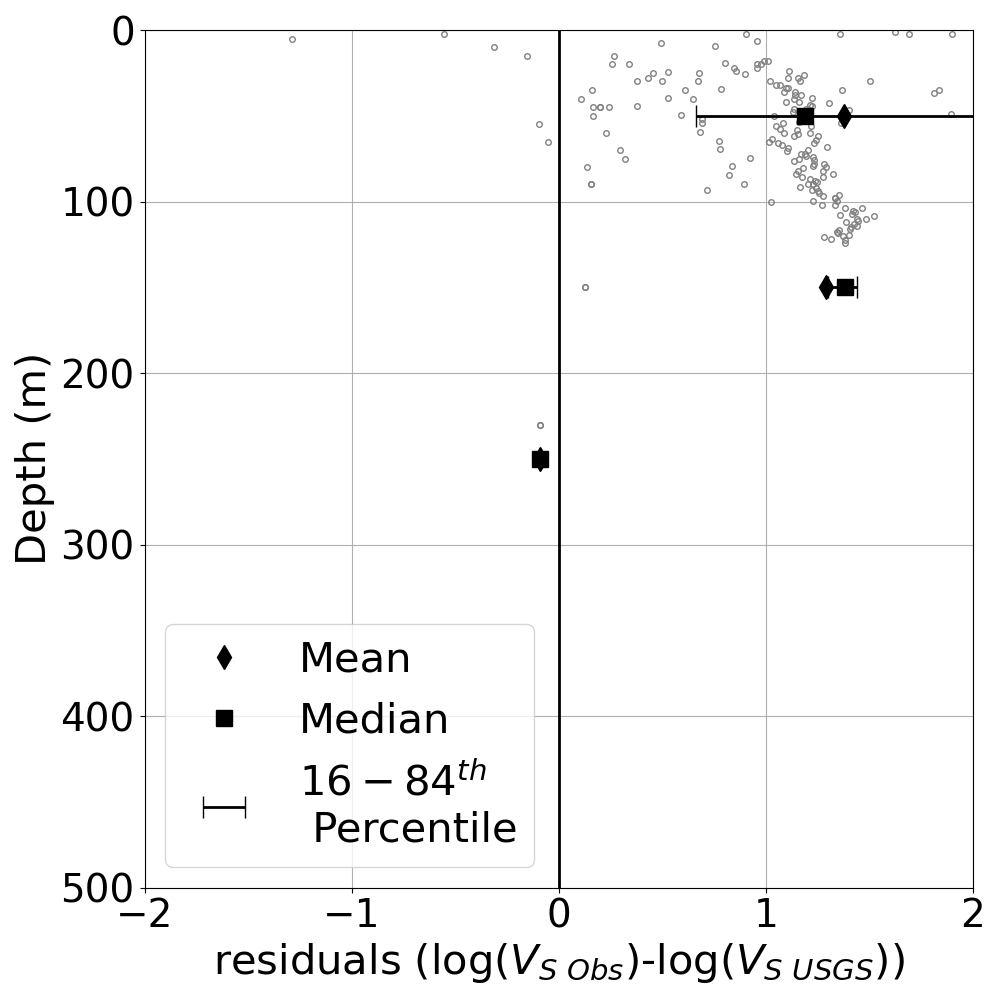}
    \end{subfigure}  
    \caption{Missfit between the $V_{S}$ observations and the USGS SFBA community velocity model for different $V_{S30}$ bins; 
    (a) $V_{S30} \in [100, 400)~\text{m/sec}$,
    (b) $V_{S30} \in [400, 800)~\text{m/sec}$,
    (c) $V_{S30} \in [800, 3000)~\text{m/sec}$.}
    \label{esupp:fig:cmp_emp_usgs_binned}
\end{figure}

Figure \ref{esupp:fig:cmp_emp_usgs} presents the misfit between the direct $V_S$ observations and the USGS SFBA community velocity model as a function of depth and measured shear-wave velocity in Subfigures (a) and (b), respectively. Figure \ref{esupp:fig:cmp_emp_usgs_binned} compares the misfit between the $V_S$ observations and the USGS model, grouped by $V_{S30}$.

Figures \ref{esupp:fig:vel_model_10m} and \ref{esupp:fig:vel_model_100m} provide comparisons of the different velocity models at depths of $10$ and $100~\text{m}$. Subfigure (a) shows the cross-section of the stationary model; Subfigures (b) and (c) display the median value and uncertainty of the spatially varying model conditioned on the available data; Subfigures (d) and (e) show the median value and uncertainty of the spatially varying model without conditioning on any data; and Subfigure (f) presents the cross-section of the USGS SFBA community velocity model. 

\begin{figure}[htbp!]
    \centering
    \begin{subfigure}[t]{0.28\textwidth}
        \caption{}
        \includegraphics[height = 1\textwidth]{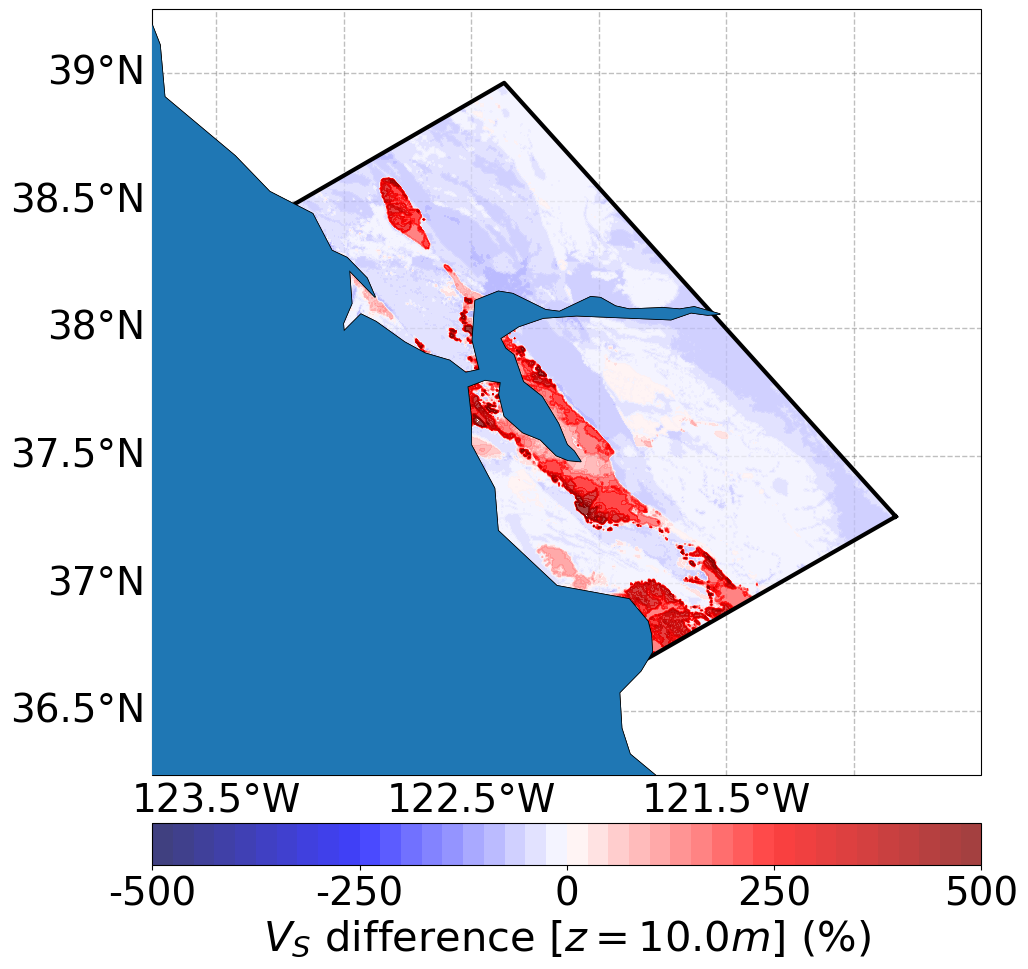}
    \end{subfigure}  
    \hfill
    \begin{subfigure}[t]{0.28\textwidth}
        \caption{}
        \includegraphics[height = 1\textwidth]{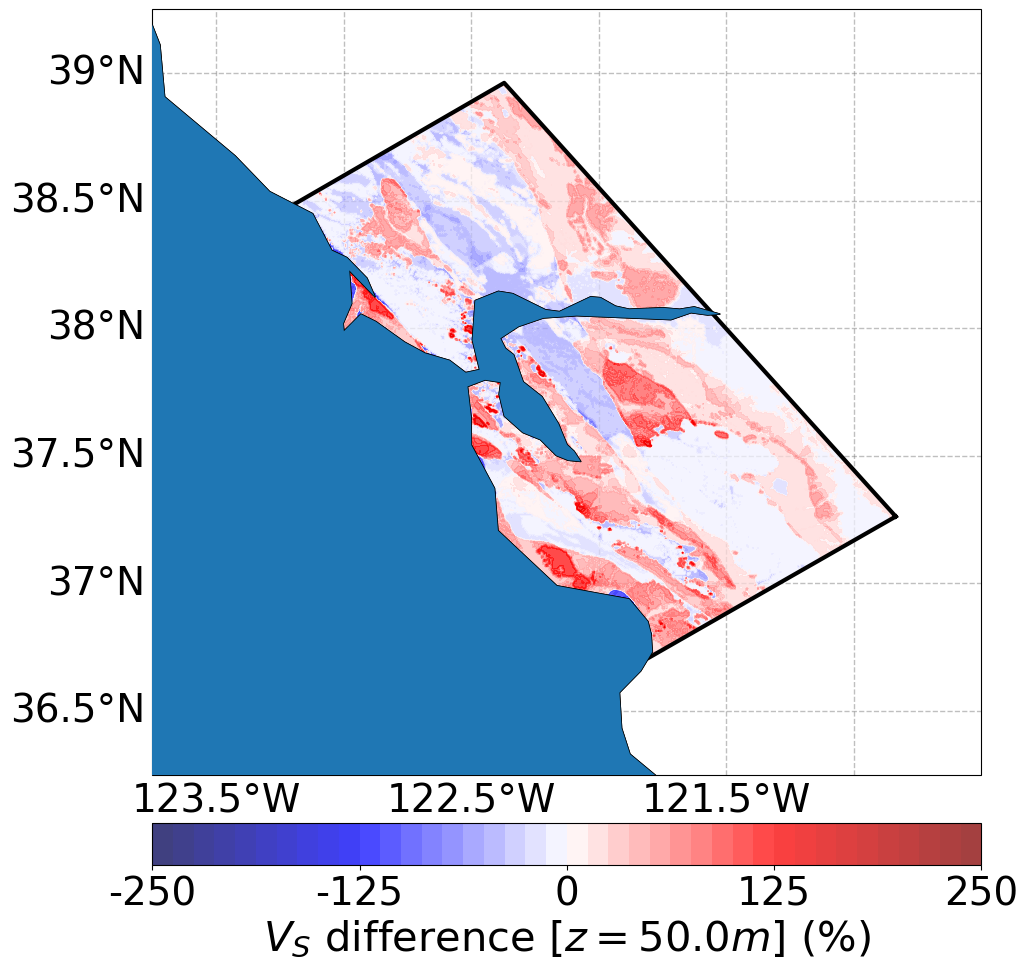}
    \end{subfigure} 
    \hfill
    \begin{subfigure}[t]{0.28\textwidth}
        \caption{}
        \includegraphics[height = 1\textwidth]{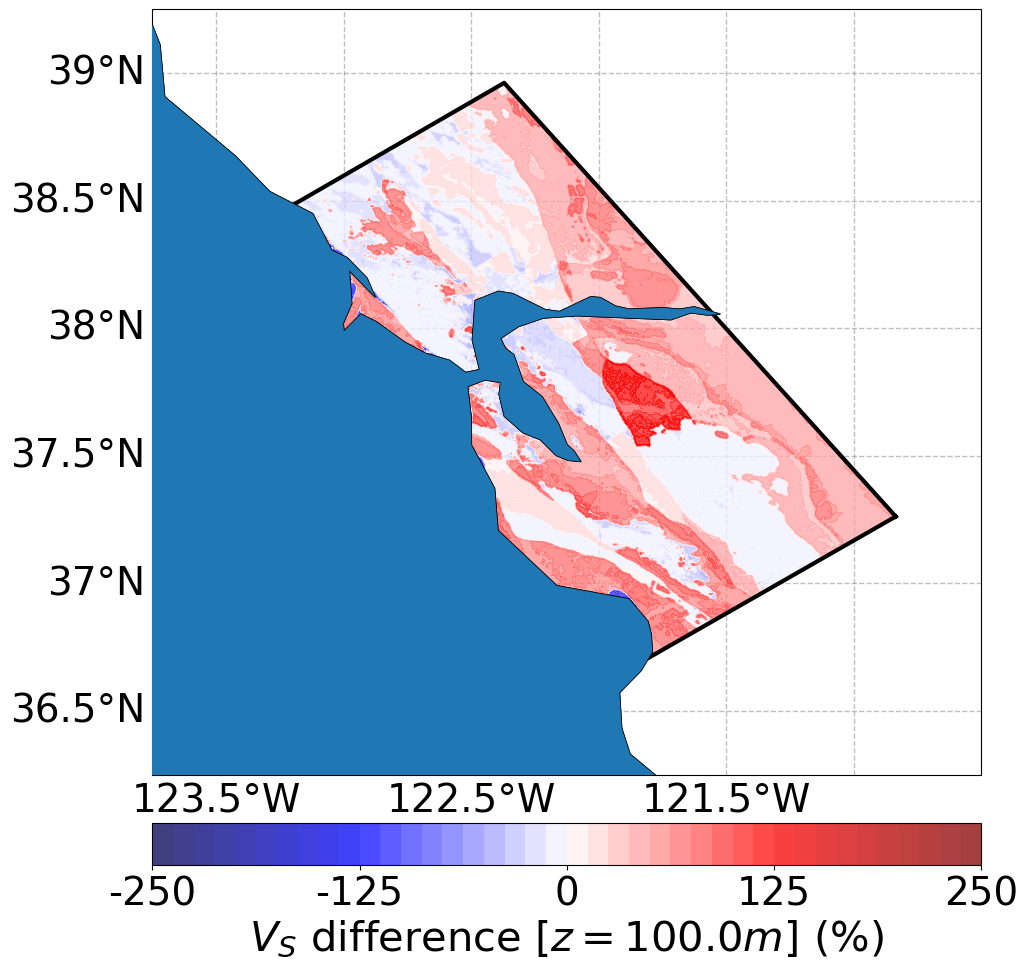}
    \end{subfigure}  
    \caption{Percent difference between the stationary and USGS SFBA velocity models ($(V_{S,\texttt{stat}} - V_{S,\texttt{USGS}})/V_{S,\texttt{USGS}}$). Positive values indicate higher $V_S$ in the stationary model, while negative values indicate higher $V_S$ in the USGS SFBA model. Map views at (a) 10 m, (b) 50 m, and (c) 100 m depth.}
    \label{esupp:fig:vel_stat_usgs_diff}
\end{figure}

\begin{figure}[htbp!]
    \centering
    \begin{subfigure}[t]{0.28\textwidth}
        \caption{}
        \includegraphics[height = 1\textwidth]{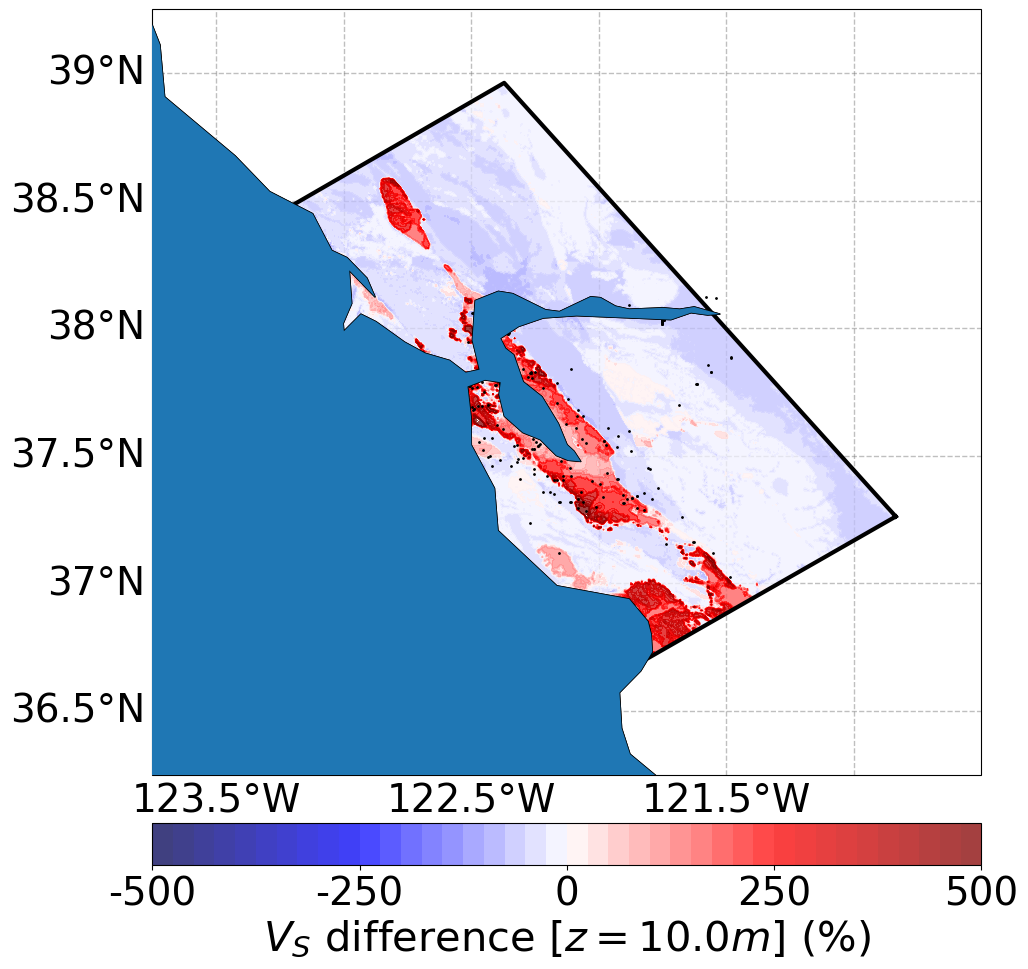}
    \end{subfigure}  
    \hfill
    \begin{subfigure}[t]{0.28\textwidth}
        \caption{}
        \includegraphics[height = 1\textwidth]{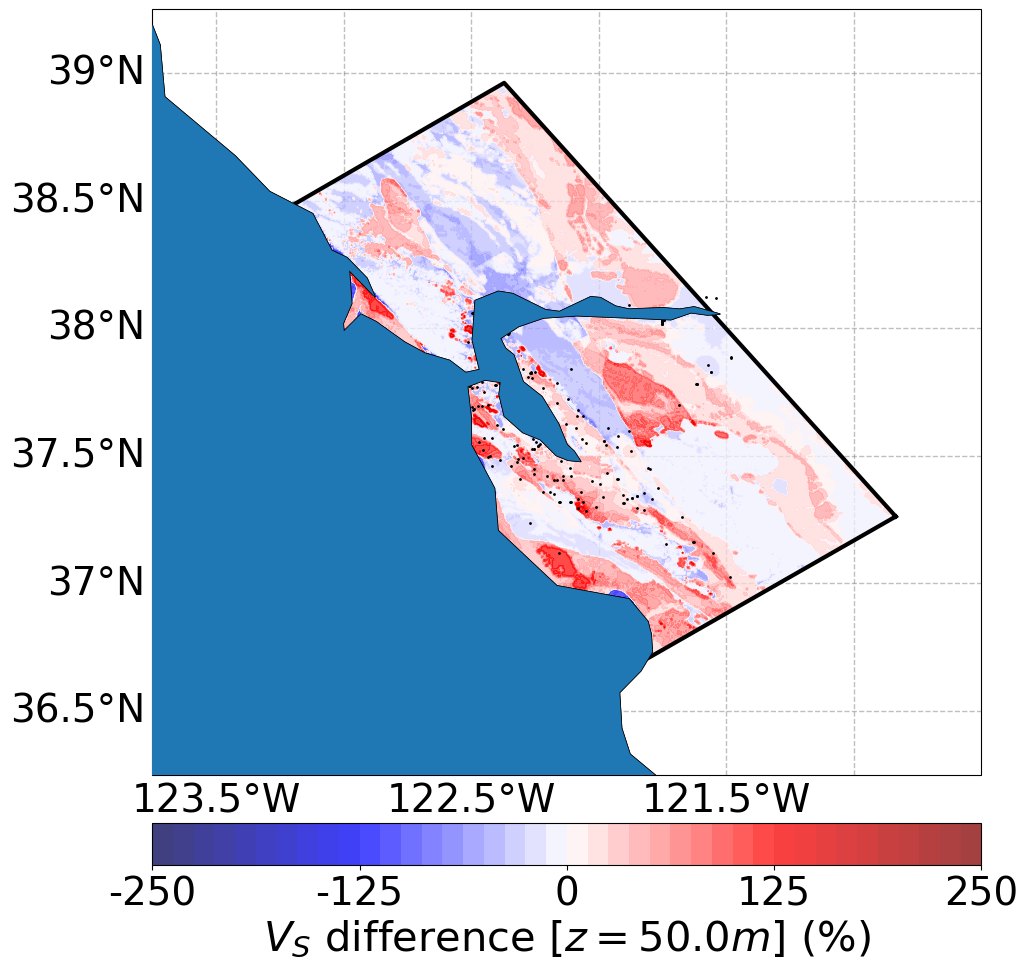}
    \end{subfigure} 
    \hfill
    \begin{subfigure}[t]{0.28\textwidth}
        \caption{}
        \includegraphics[height = 1\textwidth]{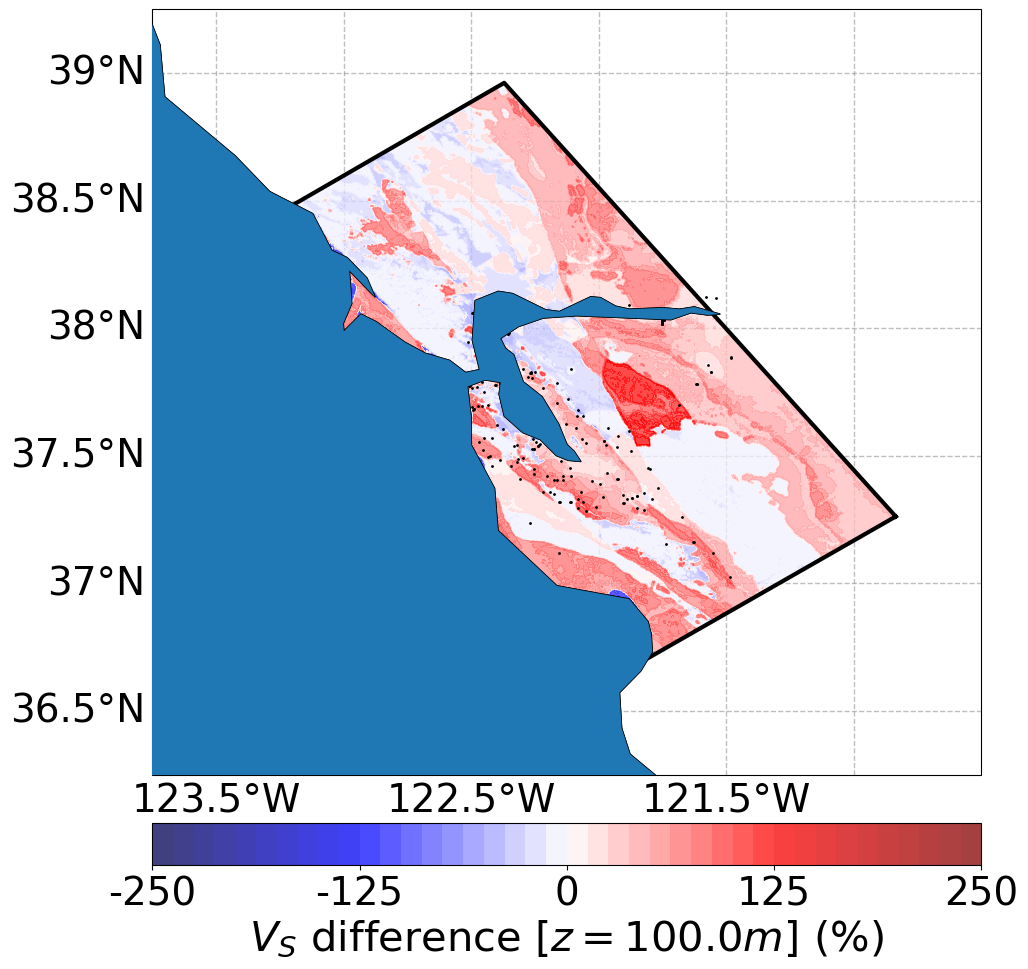}
    \end{subfigure}  
    \caption{Percent difference between the spatially varying and USGS SFBA velocity models ($(V_{S,\texttt{svar}} - V_{S,\texttt{USGS}})/V_{S,\texttt{USGS}}$). Positive values indicate higher $V_S$ in the spatially varying model, while negative values indicate higher $V_S$ in the USGS SFBA model. Map views at (a) 10 m, (b) 50 m, and (c) 100 m depth.}
\label{esupp:fig:vel_svar_usgs_diff}
\end{figure}

\begin{figure}[htbp!]
    \centering
    \begin{subfigure}[t]{0.28\textwidth}
        \caption{}
        \includegraphics[height = 1\textwidth]{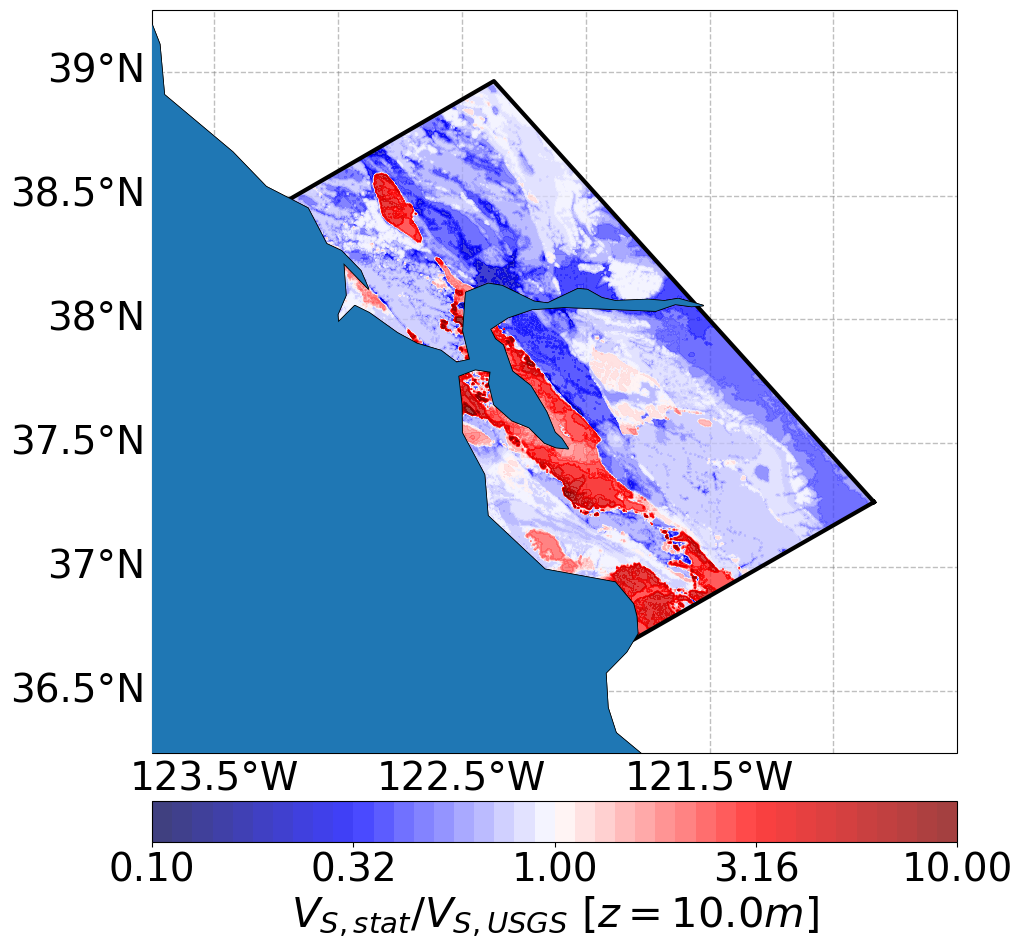}
    \end{subfigure}  
    \hfill
    \begin{subfigure}[t]{0.28\textwidth}
        \caption{}
        \includegraphics[height = 1\textwidth]{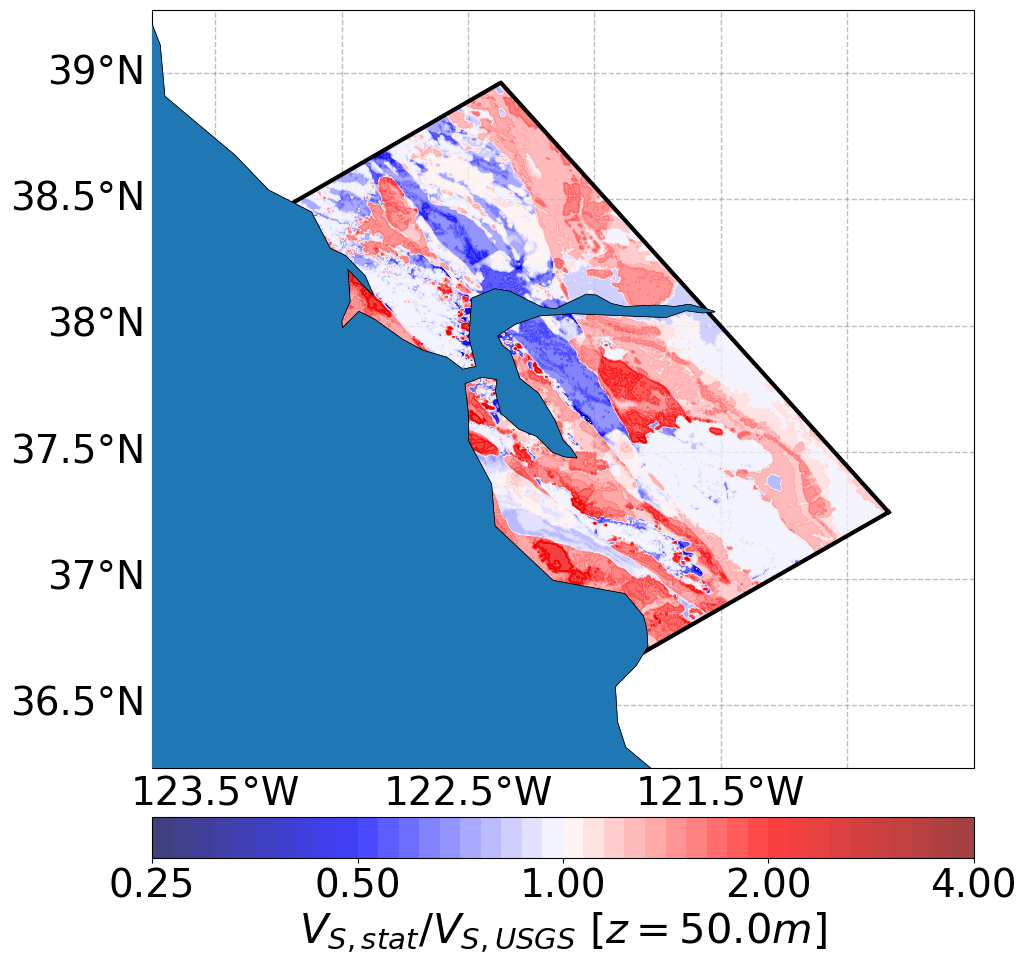}
    \end{subfigure} 
    \hfill
    \begin{subfigure}[t]{0.28\textwidth}
        \caption{}
        \includegraphics[height = 1\textwidth]{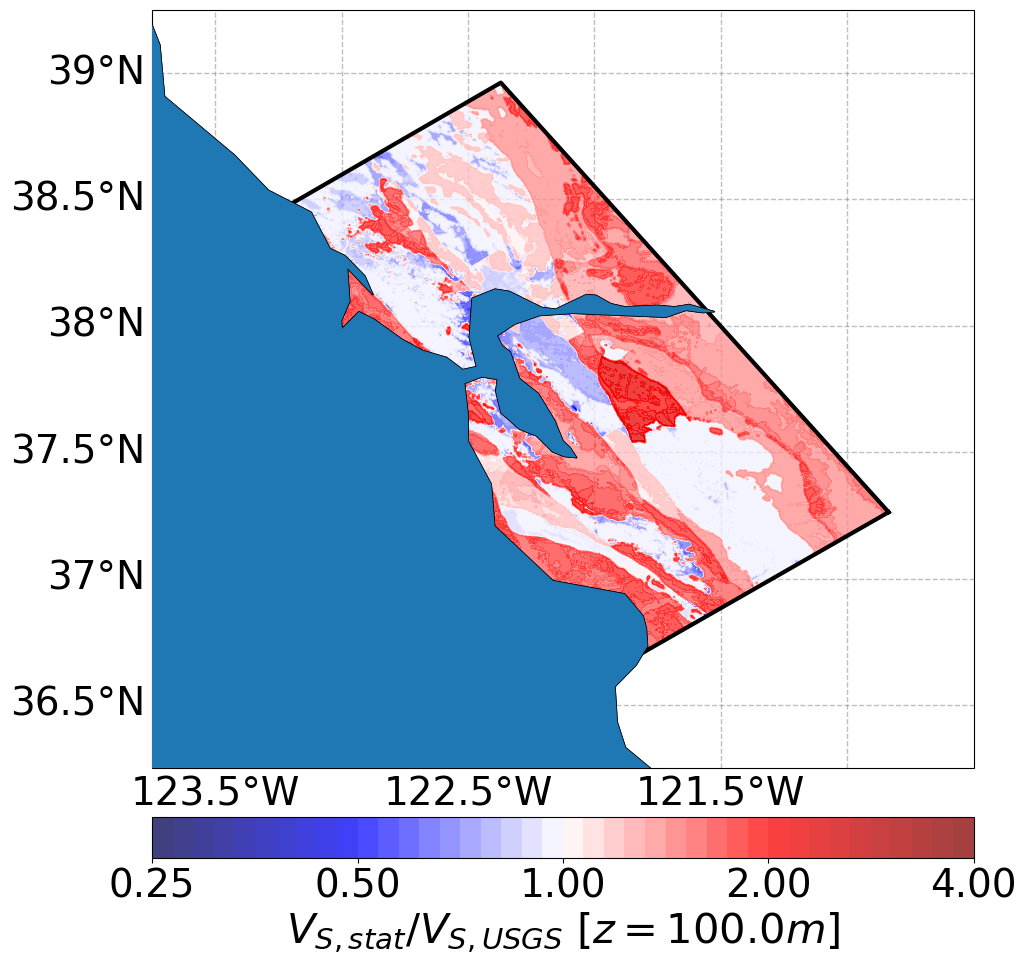}
    \end{subfigure}  
    \caption{Shear-wave velocity ratio between the stationary and USGS SFBA velocity models ($V_{S,\texttt{stat}}/V_{S,\texttt{USGS}}$). Ratios greater than one indicate higher $V_S$ in the stationary model, while values below one indicate higher $V_S$ in the USGS SFBA model. Map views are shown at depths of (a) 10 m, (b) 50 m, and (c) 100 m.}
    \label{esupp:fig:vel_stat_usgs_ratio}
\end{figure}

\begin{figure}[htbp!]
    \centering
    \begin{subfigure}[t]{0.28\textwidth}
        \caption{}
        \includegraphics[height = 1\textwidth]{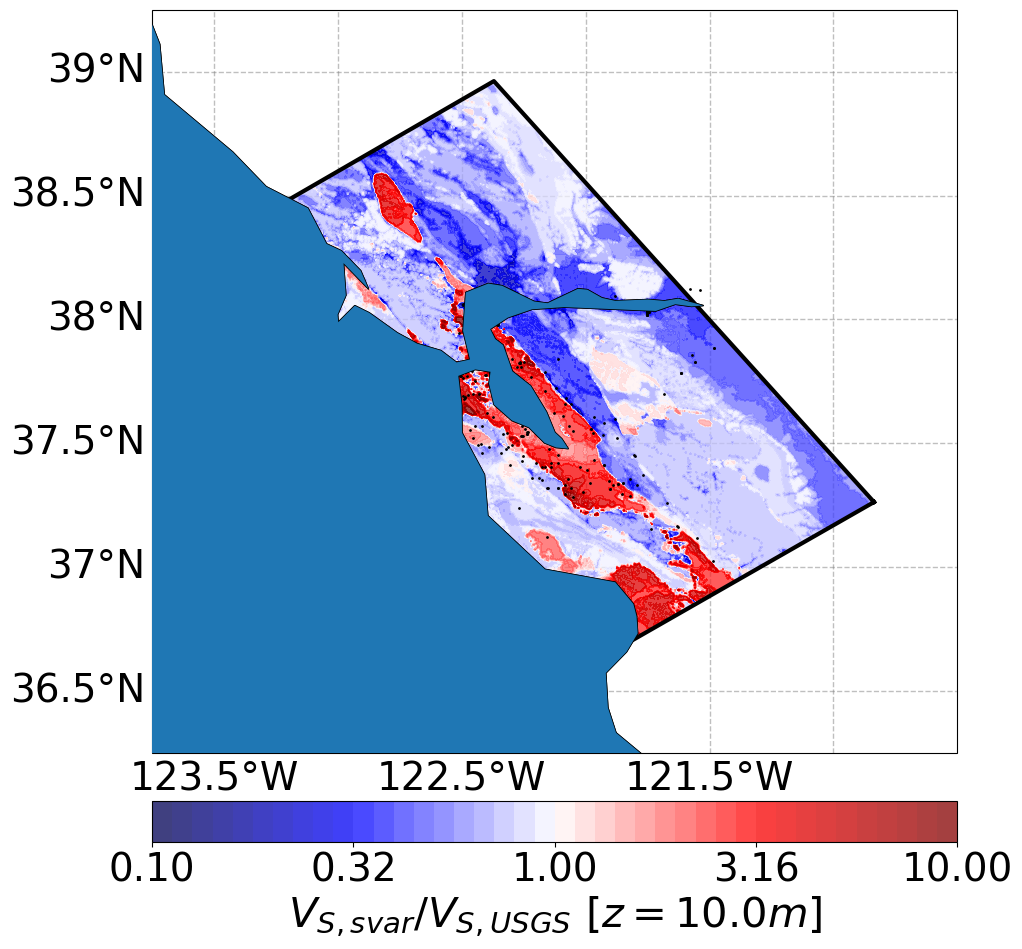}
    \end{subfigure}  
    \hfill
    \begin{subfigure}[t]{0.28\textwidth}
        \caption{}
        \includegraphics[height = 1\textwidth]{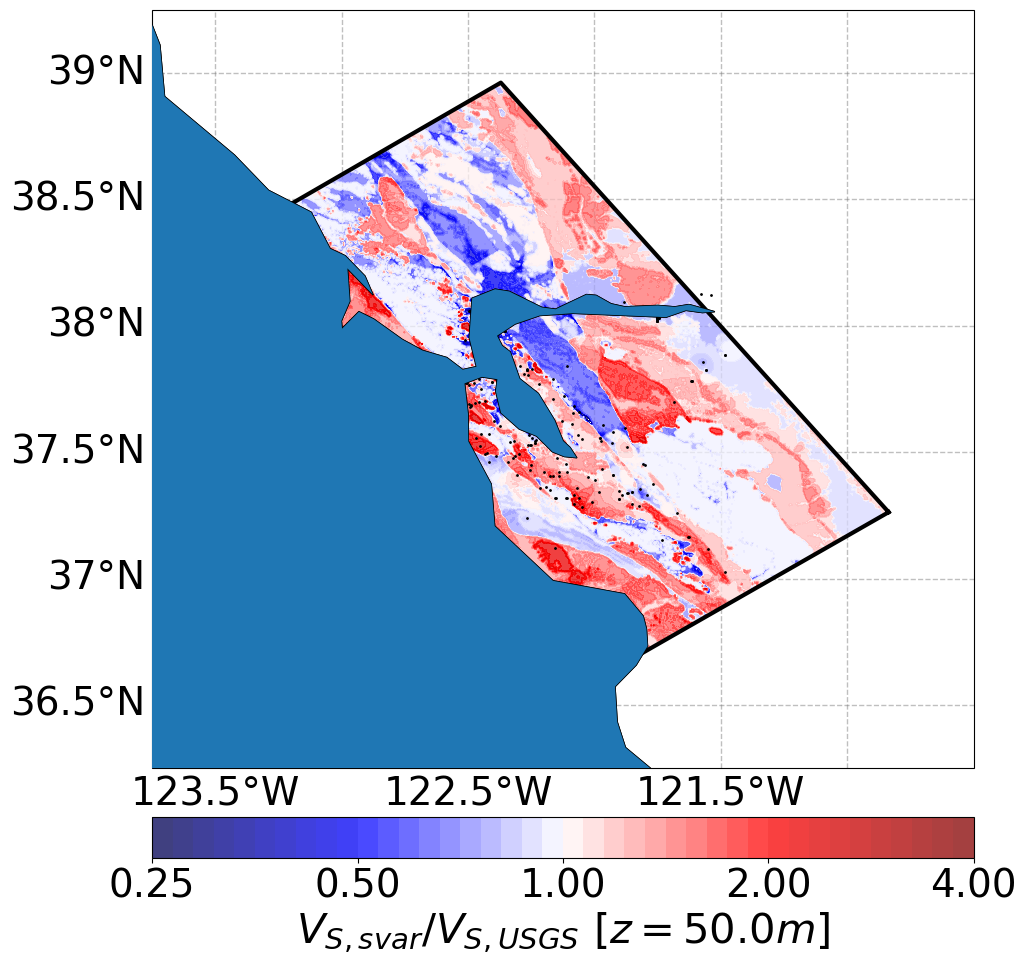}
    \end{subfigure} 
    \hfill
    \begin{subfigure}[t]{0.28\textwidth}
        \caption{}
        \includegraphics[height = 1\textwidth]{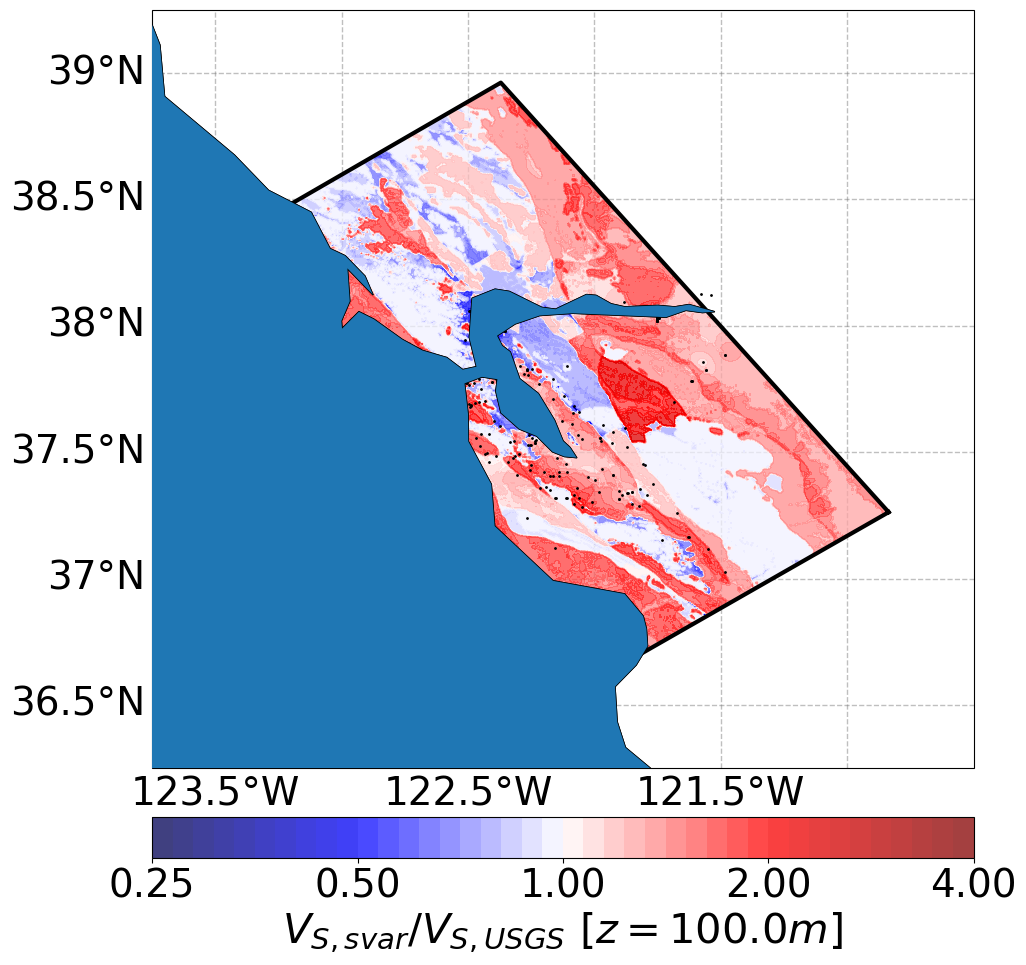}
    \end{subfigure}  
    \caption{Shear-wave velocity ratio between the spatially varying and USGS SFBA velocity models ($V_{S,\texttt{stat}}/V_{S,\texttt{USGS}}$). Ratios greater than one indicate higher $V_S$ in the spatially varying model, while values below one indicate higher $V_S$ in the USGS SFBA model. Map views are shown at depths of (a) 10 m, (b) 50 m, and (c) 100 m.}    \label{esupp:fig:vel_svar_usgs_ratio}
\end{figure}

Figures \ref{esupp:fig:vel_stat_usgs_diff} and \ref{esupp:fig:vel_svar_usgs_diff} present the percentage difference between the stationary and spatially varying models compared to the USGS SFBA velocity model at depths of 10, 50, and 100 meters. Similarly, Figures \ref{esupp:fig:vel_stat_usgs_ratio} and \ref{esupp:fig:vel_svar_usgs_ratio} display the shear-wave velocity ratio between the stationary and spatially varying models and the USGS SFBA model at the same depths.

\begin{figure} 
    \centering
    \includegraphics[width= 0.40\textwidth]{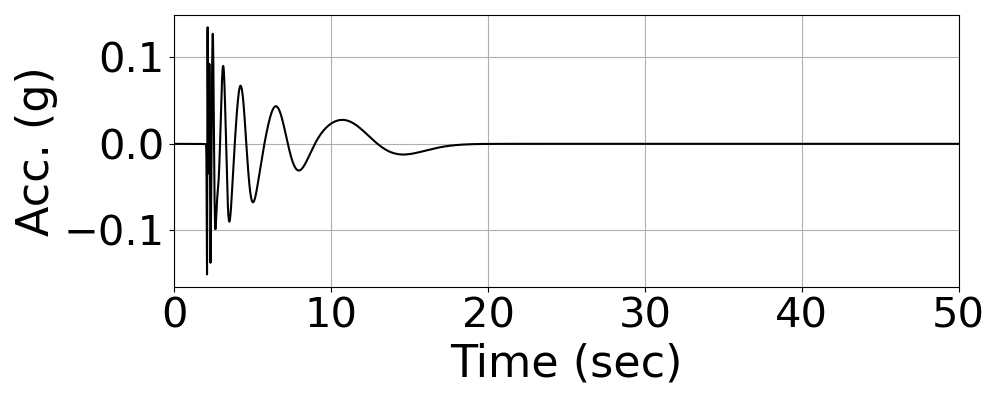}
    \caption{Ricker wavelet ensemble used as input for site-response analysis.}
    \label{esupp:fig:sra_time_history}
\end{figure}

\begin{figure} 
    \centering
    \includegraphics[width= 0.40\textwidth]{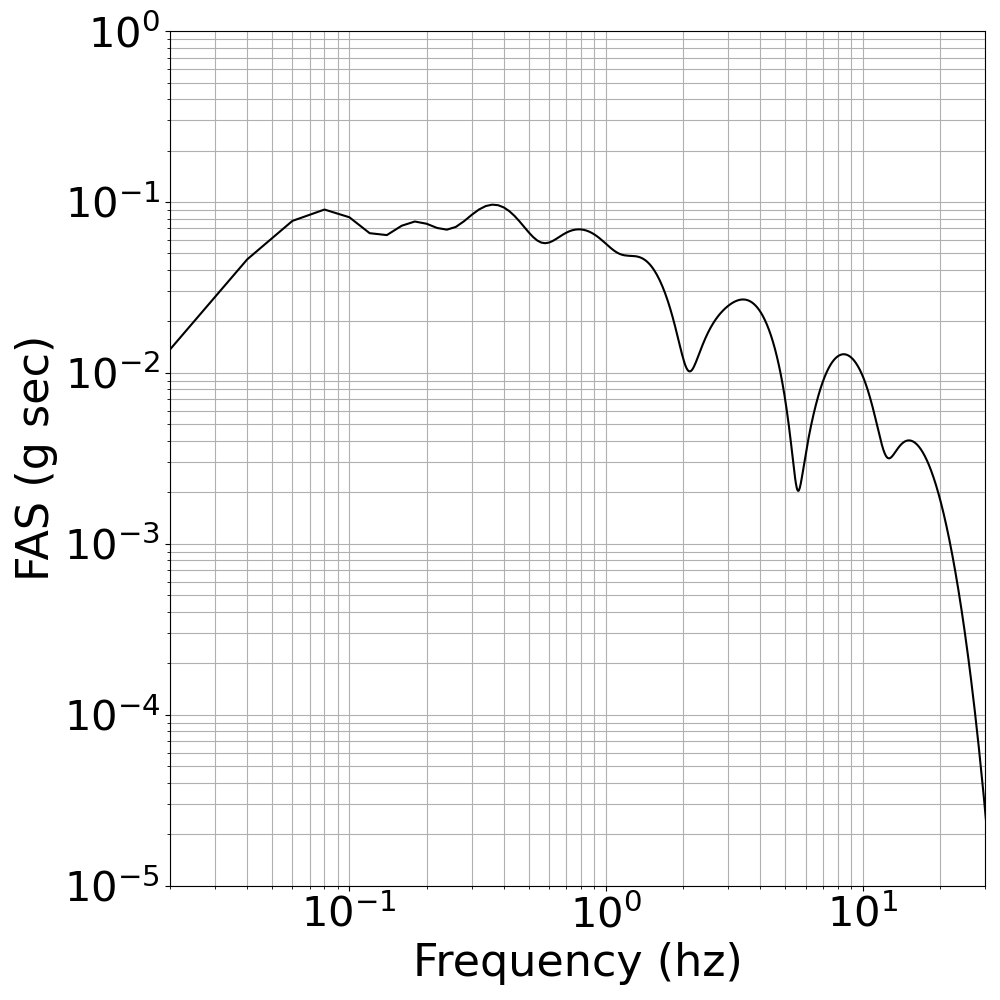}
    \caption{Frequency content of input motion for site-response analysis.}
    \label{esupp:fig:sra_time_history_fas}
\end{figure}

Figures \ref{esupp:fig:sra_time_history} and \ref{esupp:fig:sra_time_history_fas} show the input time history used for the site-response analysis and its corresponding frequency content.

\begin{figure}[htbp!]
    \centering
    \begin{subfigure}[t]{0.32\textwidth}
        \caption{}
        \includegraphics[width = 1\textwidth]{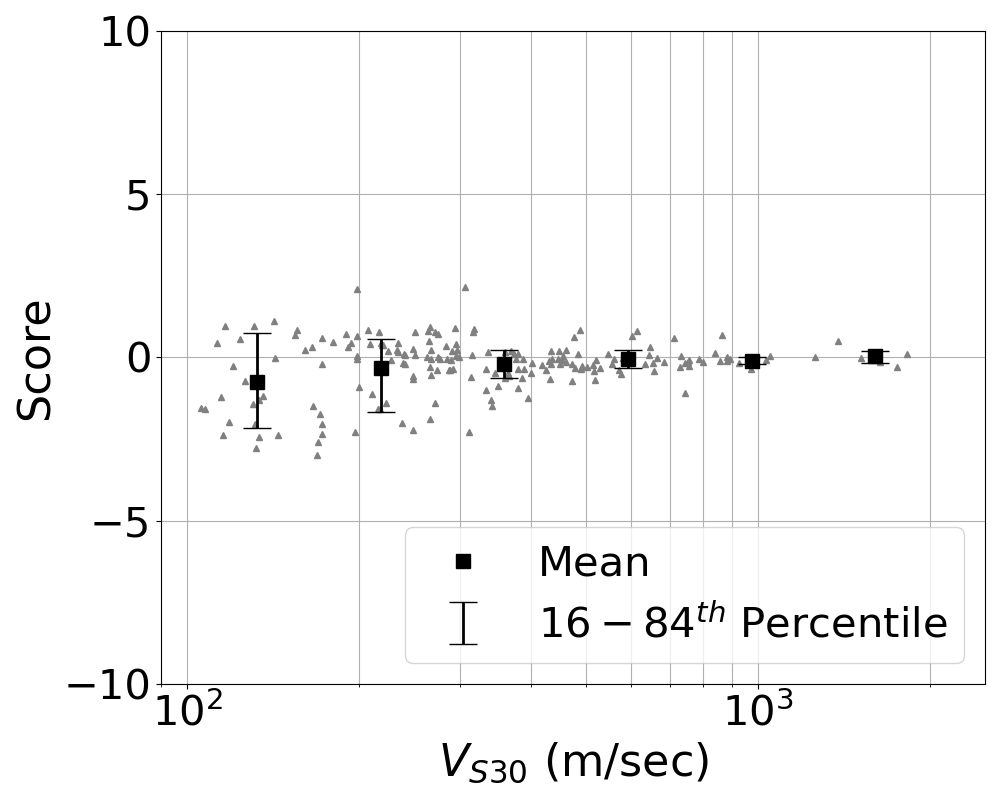}
    \end{subfigure} 
    \hfill
    \begin{subfigure}[t]{0.32\textwidth}
        \caption{}
        \includegraphics[width = 1\textwidth]{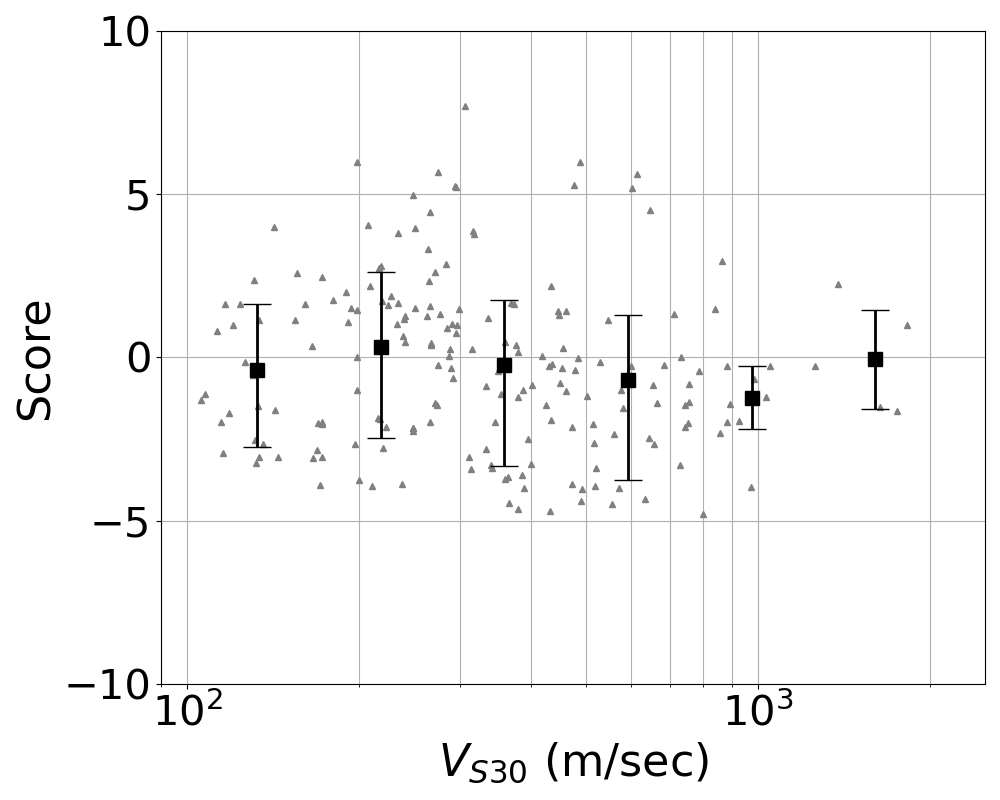}
    \end{subfigure}  
    \hfill
    \begin{subfigure}[t]{0.32\textwidth}
        \caption{}
        \includegraphics[width = 1\textwidth]{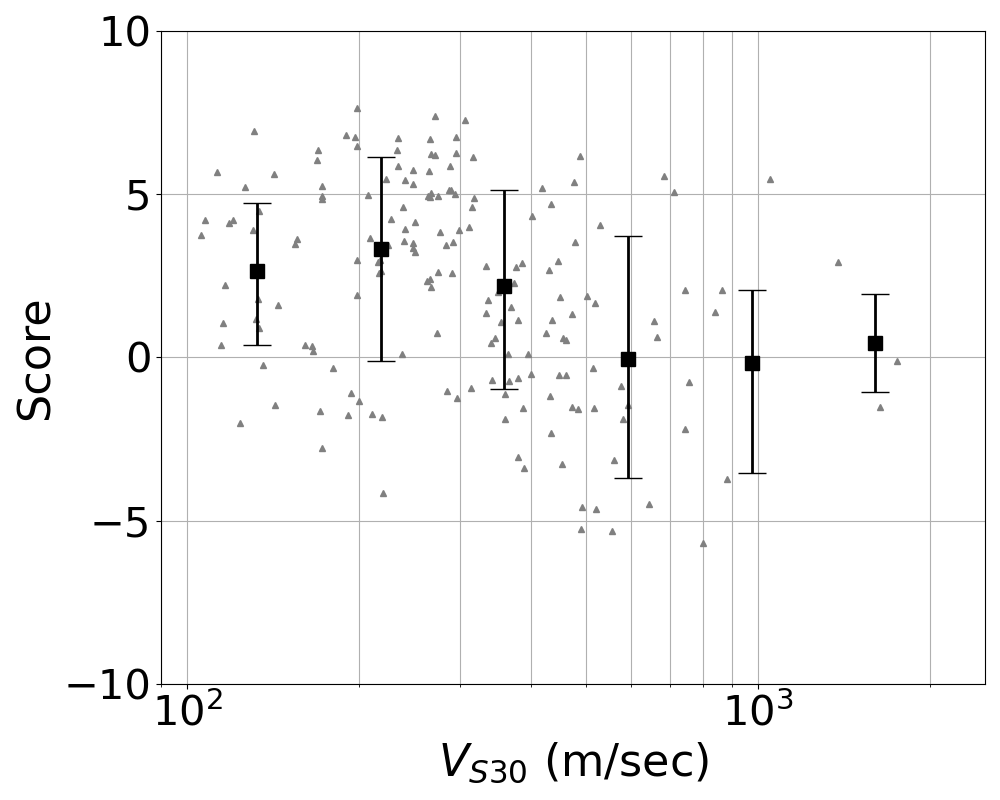}
    \end{subfigure}  
    \caption{Average goodness-of-fit (GOF) scores for the stationary model mean profiles; gray dots represent the GOF scores of individual profiles, black squares indicate the mean GOF values for each $V_{S30}$ bin, and error bars show the 16th to 84th percentile range; 
    (a) Frequency bin: $f \in [0.01, f_P)~\text{Hz}$,
    (b) Frequency bin: $f \in [f_P, 2 f_P)~\text{Hz}$,
    (c) Frequency bin: $f \in [2 f_P, 10)~\text{Hz}$.}
    \label{esupp:fig:sra_gof_stat}
\end{figure}

\begin{figure}[htbp!]
    \centering
    \begin{subfigure}[t]{0.32\textwidth}
        \caption{}
        \includegraphics[width = 1\textwidth]{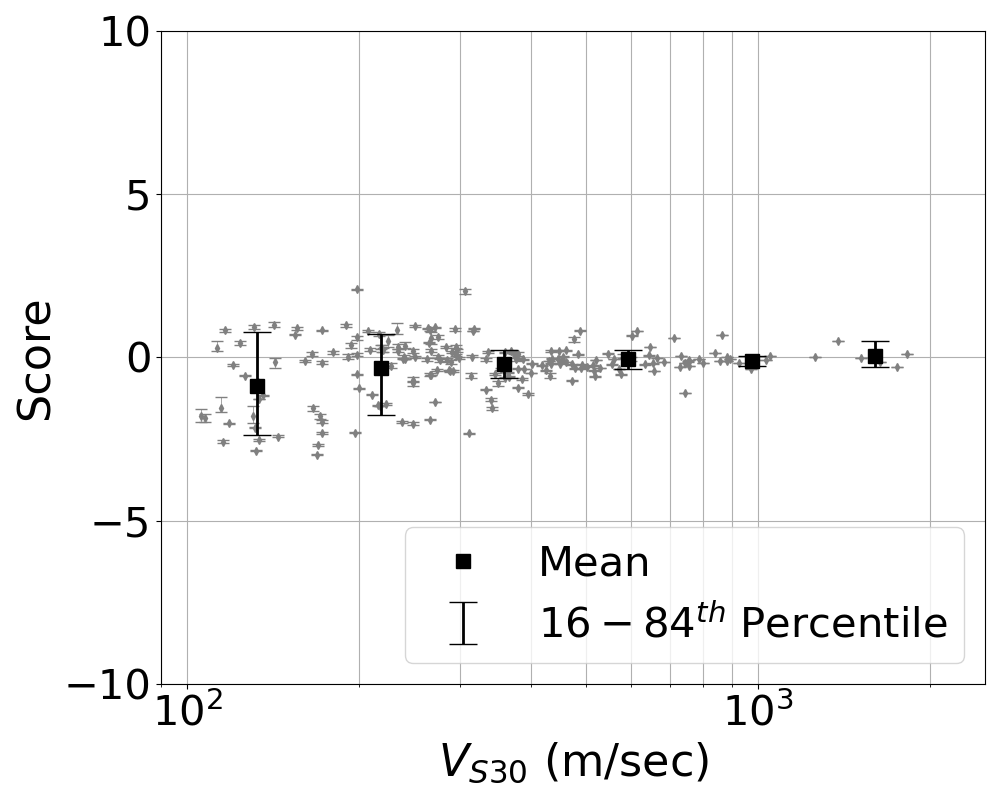}
    \end{subfigure} 
    \hfill
    \begin{subfigure}[t]{0.32\textwidth}
        \caption{}
        \includegraphics[width = 1\textwidth]{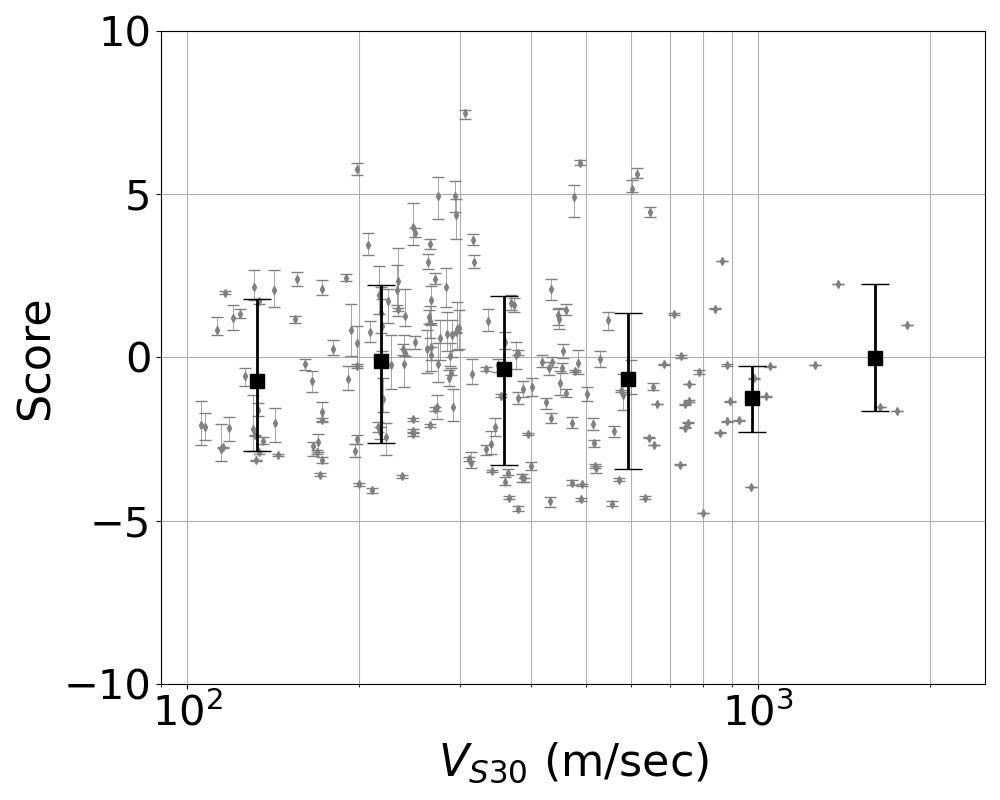}
    \end{subfigure}  
    \hfill
    \begin{subfigure}[t]{0.32\textwidth}
        \caption{}
        \includegraphics[width = 1\textwidth]{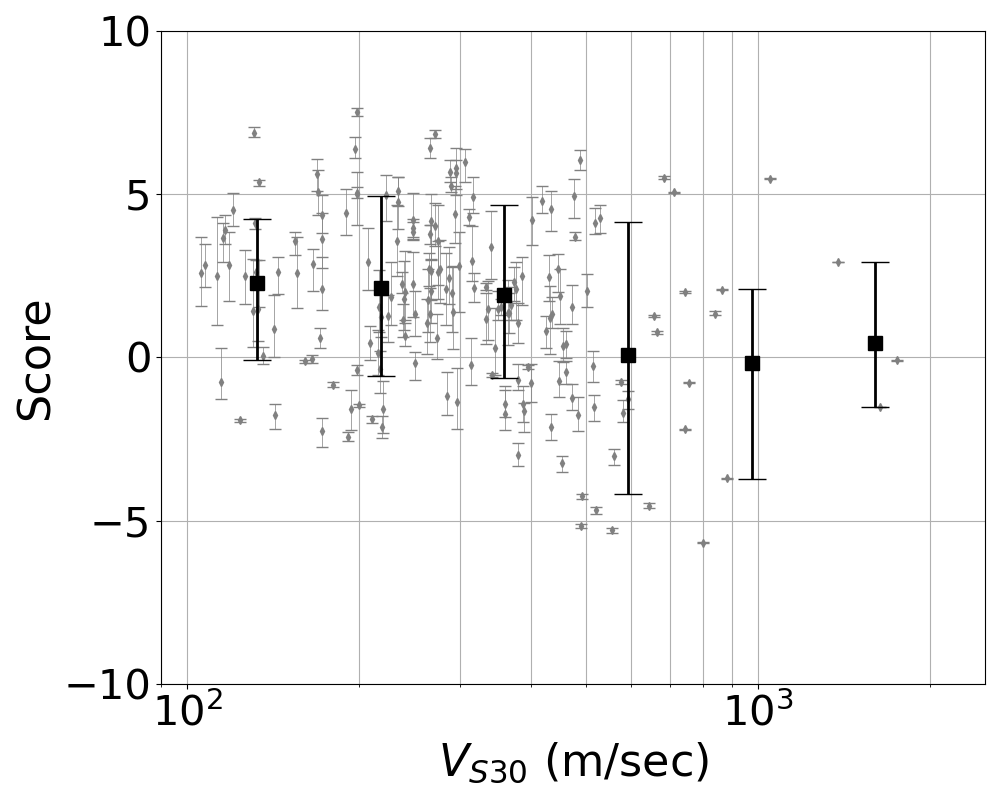}
    \end{subfigure}  
    \caption{Average goodness-of-fit (GOF) scores for the spatially varying model mean profiles; gray dots represent the GOF scores of individual profiles, black squares indicate the mean GOF values for each $V_{S30}$ bin, and error bars show the 16th to 84th percentile range; 
    (a) Frequency bin: $f \in [0.01, f_P)~\text{Hz}$,
    (b) Frequency bin: $f \in [f_P, 2 f_P)~\text{Hz}$,
    (c) Frequency bin: $f \in [2 f_P, 10)~\text{Hz}$.}
    \label{esupp:fig:sra_gof_svar}
\end{figure}

Figure \ref{esupp:fig:sra_gof_stat} presents the goodness-of-fit (GOF) score for the stationary model without along-depth variability. Figure \ref{esupp:fig:sra_gof_svar} presents the GOF score for the spatially varying model without variability.

\begin{landscape}

\begin{figure}
    \centering
    \begin{minipage}[c]{0.3\textwidth}
        \begin{subfigure}[t]{1\textwidth}
            \caption{}
            \includegraphics[height = 0.95\textwidth]{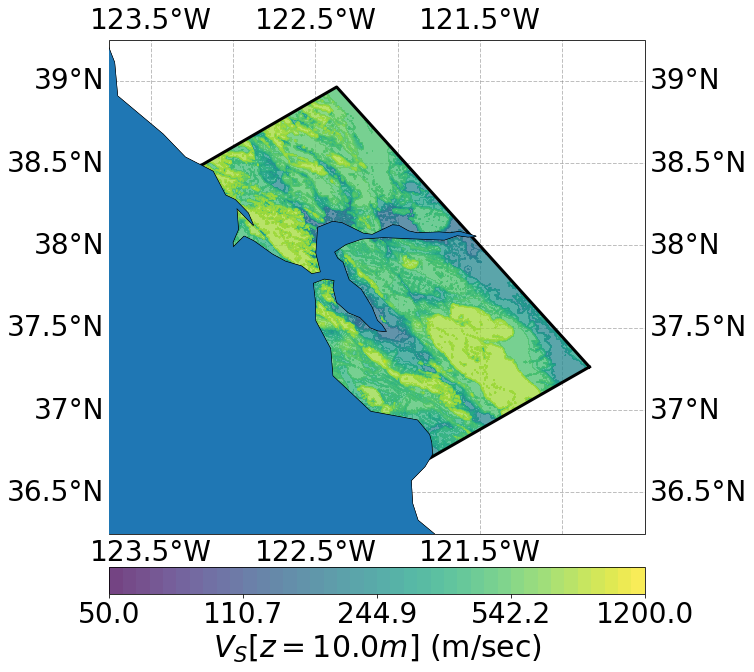} 
        \end{subfigure} 
    \end{minipage}
    \begin{minipage}[b]{0.3\textwidth}
        \begin{subfigure}[t]{1\textwidth}
            \caption{}
            \includegraphics[height = 0.95\textwidth]{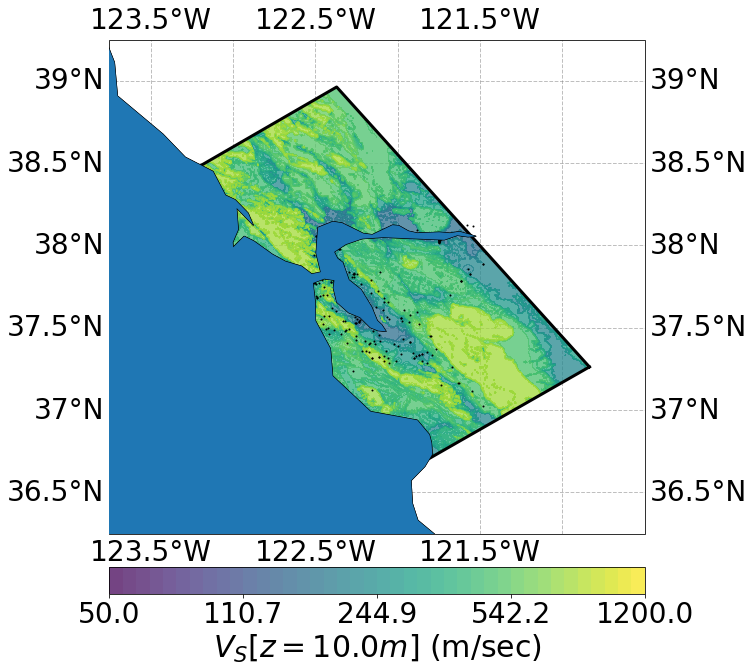}
        \end{subfigure} \\
        \begin{subfigure}[t]{1\textwidth}
            \caption{}
            \includegraphics[height = 0.95\textwidth]{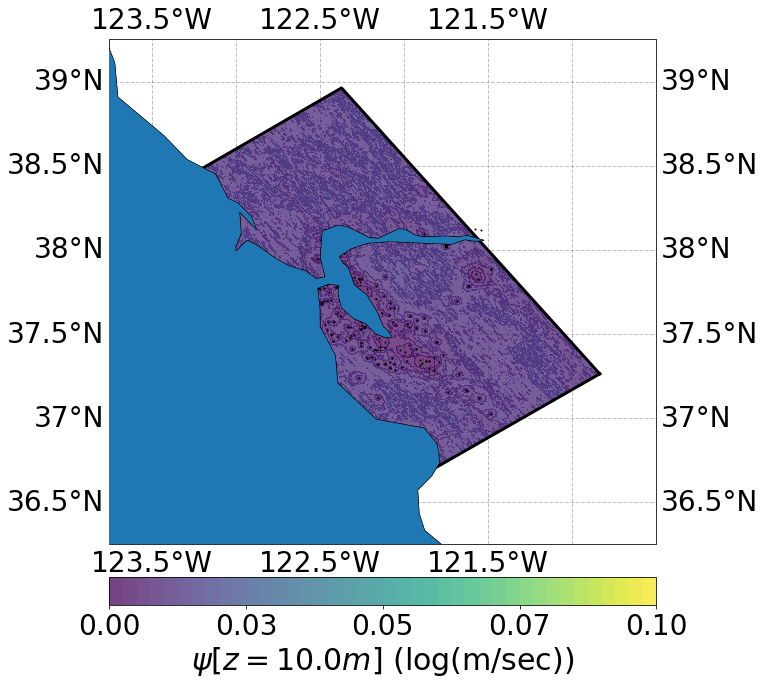}
        \end{subfigure}  
    \end{minipage}
    \begin{minipage}[b]{0.3\textwidth}
        \begin{subfigure}[t]{1\textwidth}
            \caption{}
            \includegraphics[height = 0.95\textwidth]{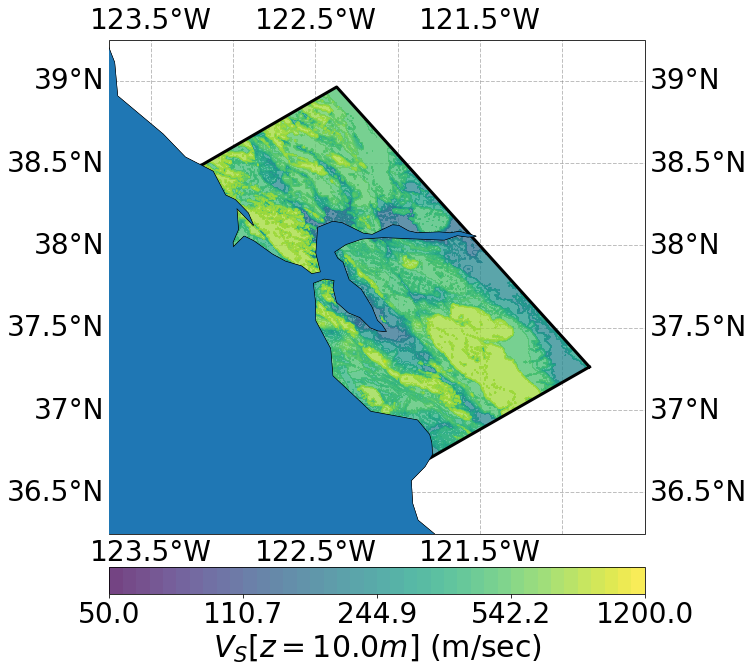}
        \end{subfigure} \\
        \begin{subfigure}[t]{1\textwidth}
            \caption{}
            \includegraphics[height = 0.95\textwidth]{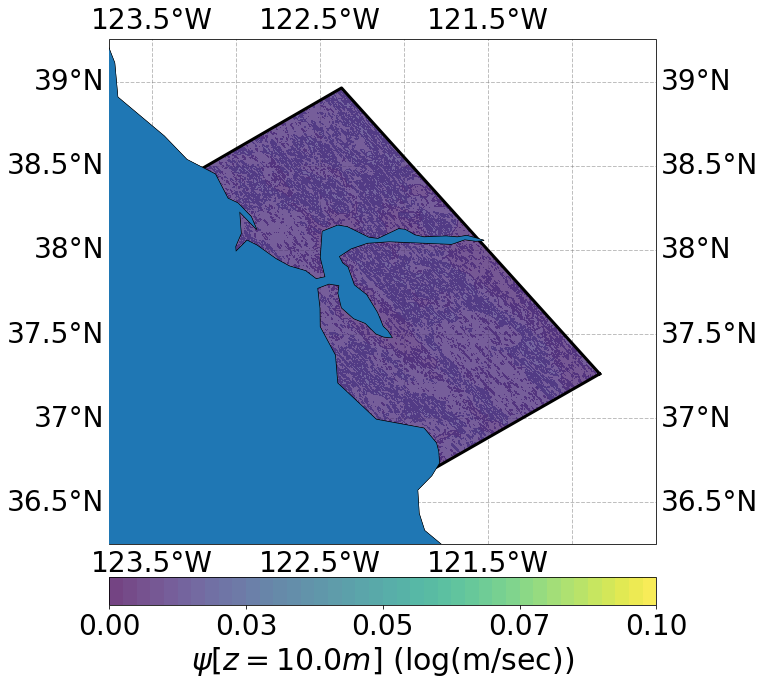}
        \end{subfigure}  
    \end{minipage}
    \begin{minipage}[c]{0.3\textwidth}
        \begin{subfigure}[t]{1\textwidth}
            \caption{}
            \includegraphics[height = 0.95\textwidth]{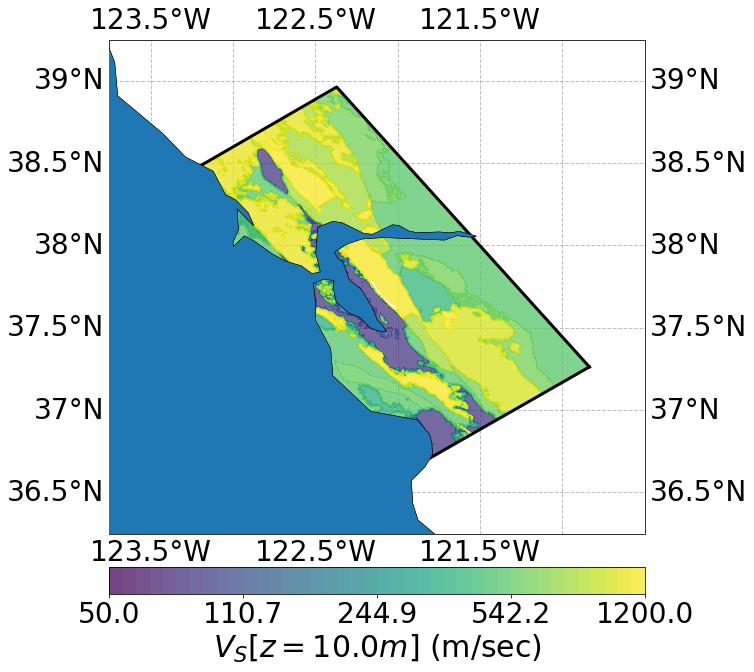}
        \end{subfigure} 
    \end{minipage}
    \caption{Velocity model cross session at depth of $10~\text{m}$: 
    (a) mean $V_S$ of the stationary model, 
    (b) mean $V_S$ of spatially varying model conditioned on available velocity profiles, 
    (c) $V_S$ uncertainty of spatially varying model conditioned on available velocity profiles, 
    (d) mean $V_S$ of spatially varying model unconditional on any data, 
    (e) $V_S$ uncertainty of spatially varying model unconditional on any data, 
    (d) $V_S$ of USGS San-Francisco Bay Area velocity model.}
    \label{esupp:fig:vel_model_10m}
\end{figure}

\begin{figure}
    \centering
    \begin{minipage}[c]{0.3\textwidth}
        \begin{subfigure}[t]{1\textwidth}
            \caption{}
            \includegraphics[height = 0.95\textwidth]{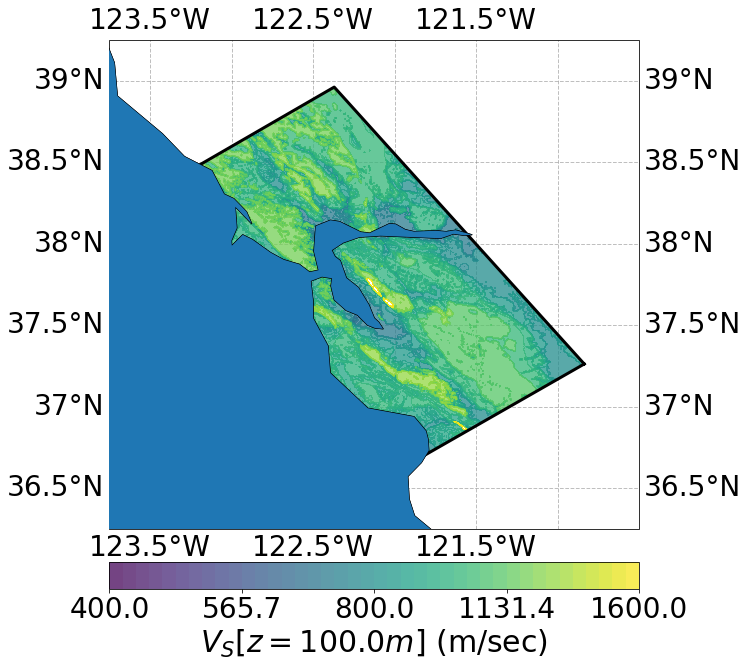} 
        \end{subfigure} 
    \end{minipage}
    \begin{minipage}[b]{0.3\textwidth}
        \begin{subfigure}[t]{1\textwidth}
            \caption{}
            \includegraphics[height = 0.95\textwidth]{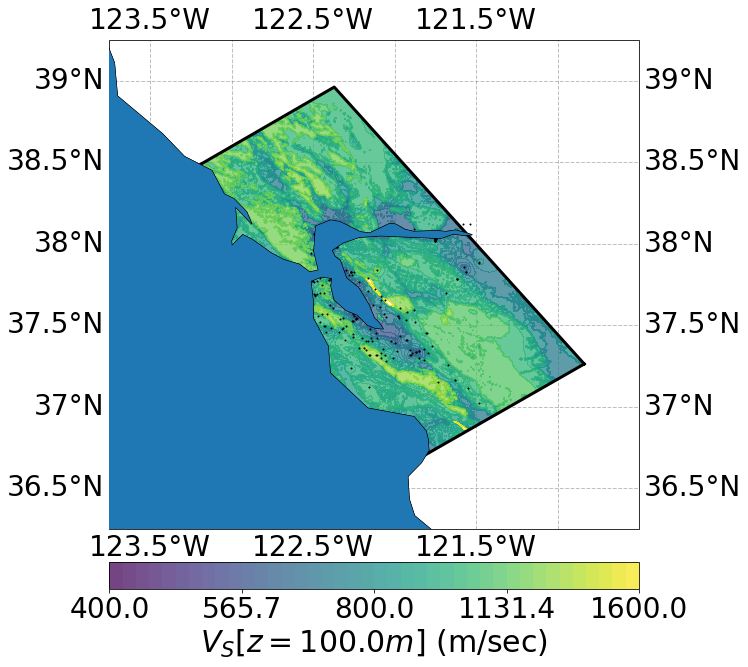}
        \end{subfigure} \\
        \begin{subfigure}[t]{1\textwidth}
            \caption{}
            \includegraphics[height = 0.95\textwidth]{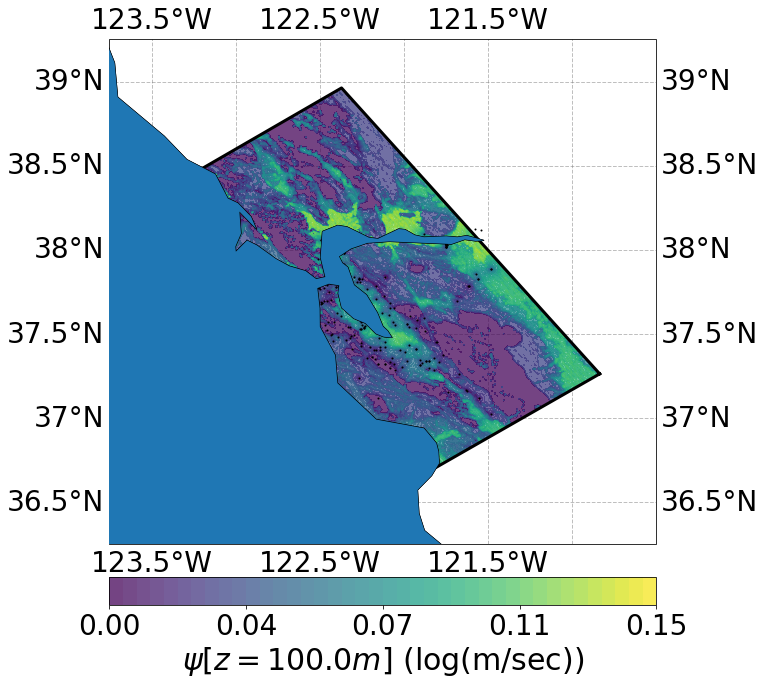}
        \end{subfigure}  
    \end{minipage}
    \begin{minipage}[b]{0.3\textwidth}
        \begin{subfigure}[t]{1\textwidth}
            \caption{}
            \includegraphics[height = 0.95\textwidth]{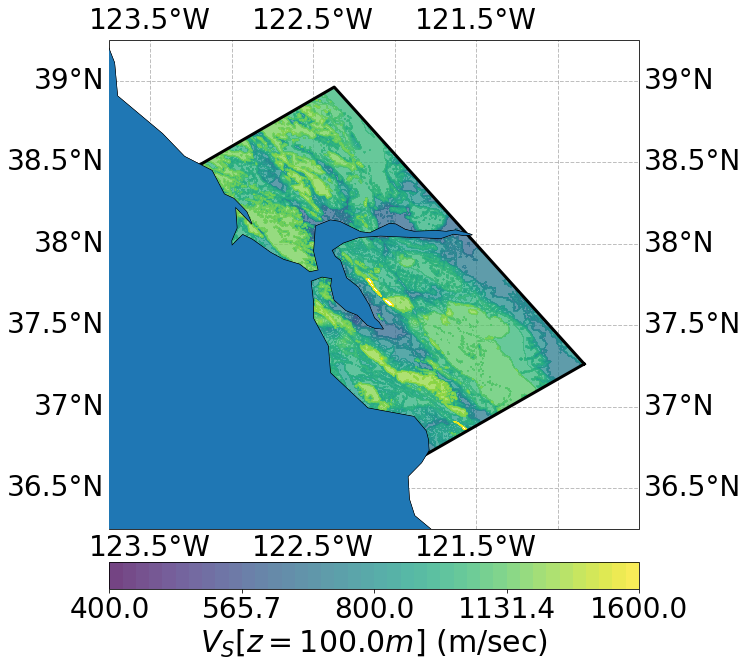}
        \end{subfigure} \\
        \begin{subfigure}[t]{1\textwidth}
            \caption{}
            \includegraphics[height = 0.95\textwidth]{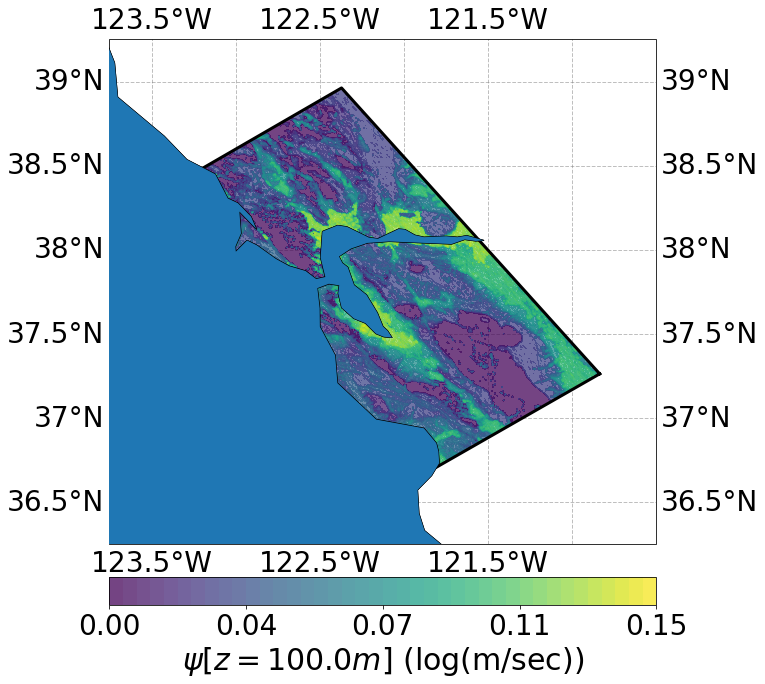}
        \end{subfigure}  
    \end{minipage}
    \begin{minipage}[c]{0.3\textwidth}
        \begin{subfigure}[t]{1\textwidth}
            \caption{}
            \includegraphics[height = 0.95\textwidth]{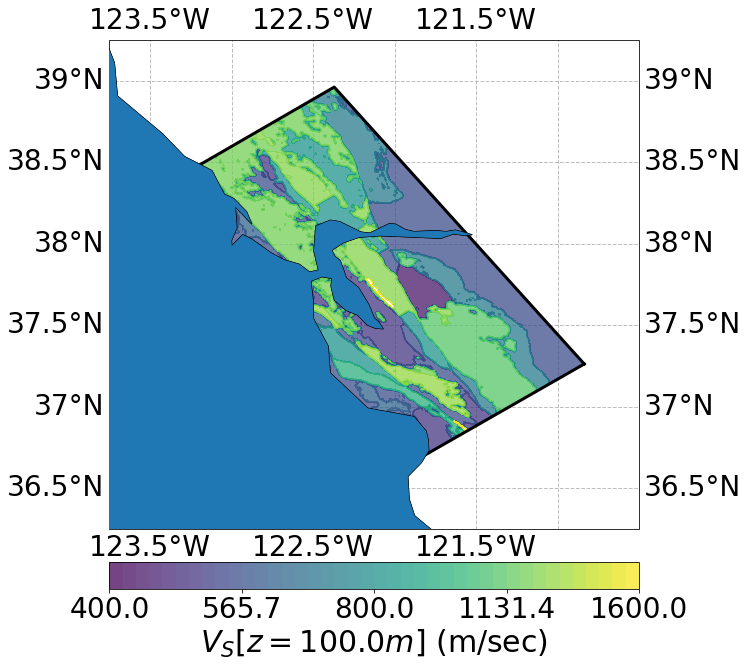}
        \end{subfigure} 
    \end{minipage}
    \caption{Velocity model cross session at depth of $100~\text{m}$: 
    (a) mean $V_S$ of the stationary model, 
    (b) mean $V_S$ of spatially varying model conditioned on available velocity profiles, 
    (c) $V_S$ uncertainty of spatially varying model conditioned on available velocity profiles, 
    (d) mean $V_S$ of spatially varying model unconditional on any data, 
    (e) $V_S$ uncertainty of spatially varying model unconditional on any data, 
    (d) $V_S$ of USGS San-Francisco Bay Area velocity model.}
    \label{esupp:fig:vel_model_100m}
\end{figure}

\end{landscape}

\begin{figure}[htbp!]
    \centering
    \begin{subfigure}[t]{0.32\textwidth}
        \caption{}
        \includegraphics[height = 1\textwidth]{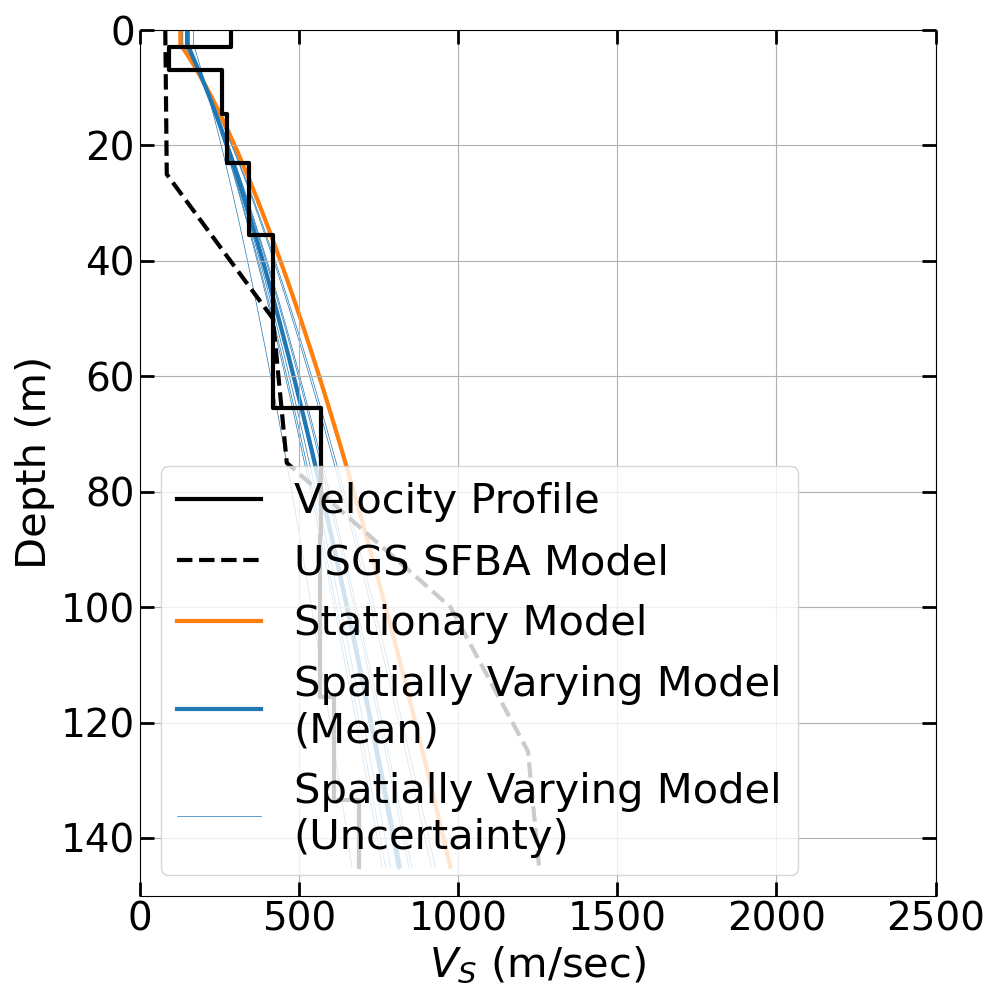}
    \end{subfigure} 
    \hfill
    \begin{subfigure}[t]{0.32\textwidth}
        \caption{}
        \includegraphics[height = 1\textwidth]{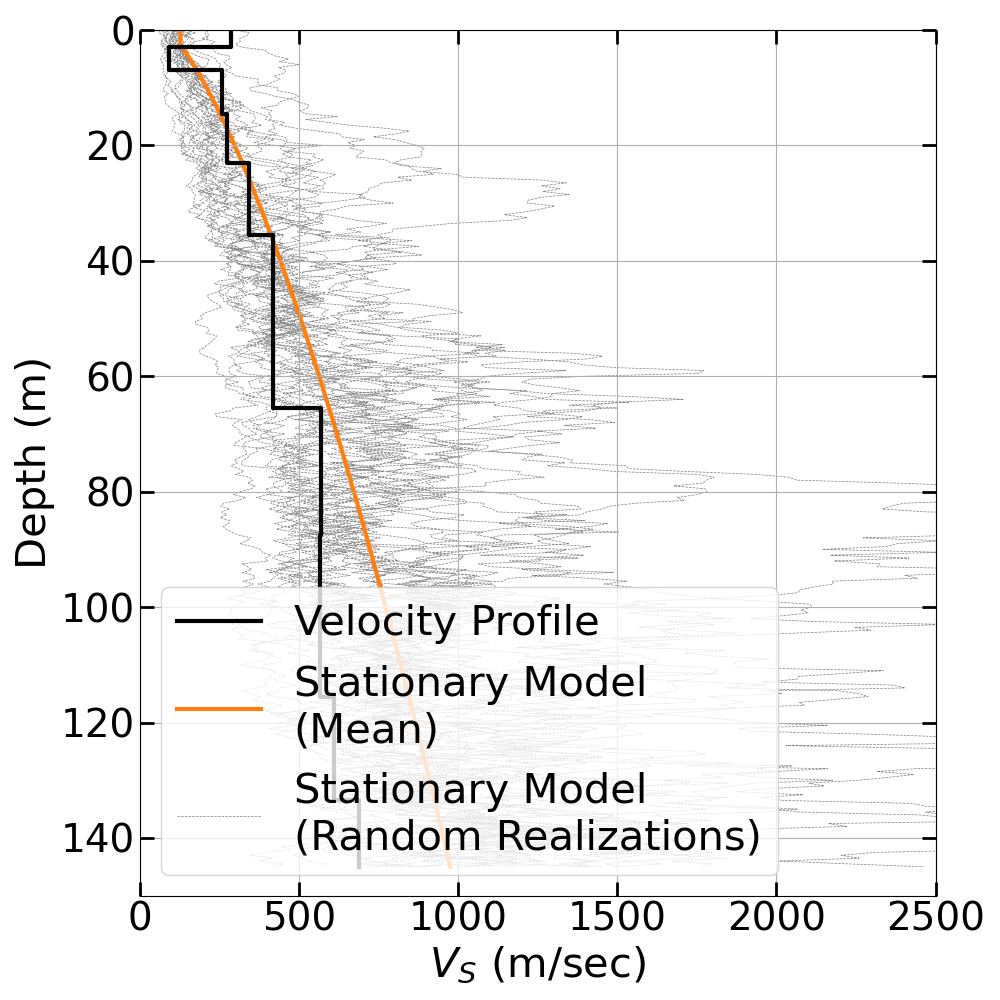}
    \end{subfigure}  
    \hfill
    \begin{subfigure}[t]{0.32\textwidth}
        \caption{}
        \includegraphics[height = 1\textwidth]{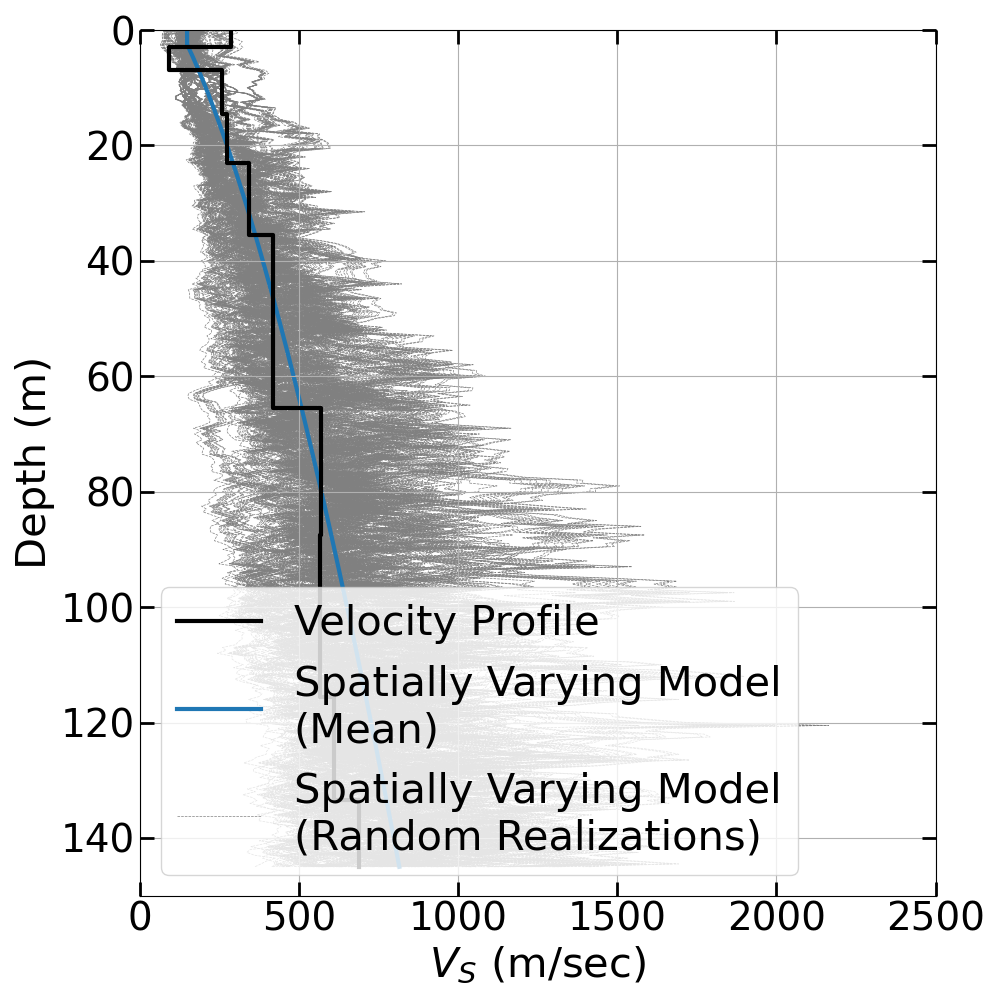}
    \end{subfigure}  
    \caption{Example velocity profiles used in the site-response analysis; The solid black line represents the measured velocity profile, the dashed black line represents the USGS SFBA model at the same location, the orange line indicates the mean stationary model and the blue line shows the spatially varying model; 
    (a) comparison of the measured profile with the USGS SFBA model, stationary model, and spatially varying model, including uncertainty due to the slope adjustment ($\delta B_r$),
    (b) Comparison of the measured profile with the stationary model, incorporating along-depth variability,
    (c) Comparison of the measured profile with the spatially varying model, including both $\delta B_r$ uncertainty and along-depth variability.}
    \label{esupp:fig:vprof_ex}
\end{figure}

